\documentclass[a4paper,fleqn]{cas-sc}

\usepackage[numbers]{natbib}
\usepackage{subcaption}
\usepackage{tikz}
\usetikzlibrary{arrows.meta}

\usepackage{pdfpages}
\usepackage{placeins}

\def\tsc#1{\csdef{#1}{\textsc{\lowercase{#1}}\xspace}}
\tsc{WGM}
\tsc{QE}

\begin{document}
\let\WriteBookmarks\relax
\def\floatpagepagefraction{1}
\def\textpagefraction{.001}

\shorttitle{A dynamic pore-network model with discrete microbubble transport}

\shortauthors{K. Skjelanger et al.} 

\title [mode = title]{A dynamic pore-network model with discrete microbubble transport}  




\author[1]{Kristoffer Skjelanger}[orcid=0009-0008-1872-9381]

\cormark[1]

\ead{kristoffer.skjelanger@hvl.no}

\affiliation[1]{organization={Western Norway University of Applied Sciences},
            addressline={Inndalsveien 28}, 
            city={5063 Bergen},
            citysep={}, % Uncomment if no comma needed between city and postcode
            %postcode={5063}, 
            state={},
            country={Norway}}

% Kristoffer
\credit{Conceptualization, Data curation, Formal Analysis, Methodology, Investigation, Software, Validation, Visualization, Writing - Original Draft}

\author[2]{Anna Mareike Kostelecky}%[]
\credit{Conceptualization, Methodology, Writing - Review \& Editing}

\author[2]{Maziar Veyskarami}%[]
\credit{Conceptualization, Methodology, Writing - Review \& Editing}

\author[2]{Timo Koch}%[]
\credit{Conceptualization, Formal analysis, Methodology, Supervision, Writing - Review \& Editing}

% Footnote of the second author

% Email id of the second author
% \ead{timo.koch@iws.uni-stuttgart.de}


\affiliation[2]{organization={Department of Hydromechanics and Modelling of Hydrosystems, University of Stuttgart},
            addressline={Pfaffenwaldring 61}, 
            city={Stuttgart},
%          citysep={}, % Uncomment if no comma needed between city and postcode
            postcode={70569}, 
            %state={},
            country={Germany}}

\cortext[1]{Corresponding author}


\begin{abstract}
We develop a hybrid pore-network model (PNM) for discrete microbubble transport coupled to connected two-phase flow in porous media. We apply the model to simulate flow in the porous transport layer of a rotating proton exchange membrane (PEM) electrolyser. At sufficiently large apparent gravitational forces, bubbles growing at the catalyst layer can detach at sizes smaller than the characteristic pore size and move through liquid-filled pores via buoyancy and advection.

The model couples a dynamic two-phase PNM to a Lagrangian bubble tracker. Pore-scale Stokes simulations in square throats provide closures for the bubble buoyancy velocity, the hydraulic resistance of bubble-occupied throats, and the reciprocal coupling between liquid flow and bubble motion. Verification tests assess conservation, pressure-flow coupling, bubble growth, advection, merging, trapping, and transfer to the connected gas phase. 

The network simulations show that increasing apparent gravitational forces shifts a larger fraction of the produced gas toward discrete-bubble transport. The increase mostly comes from a reduced detachment radius, which decreases bubble size and lets them move more freely through the network. 
\end{abstract}


% \begin{highlights}
% \item 
% \item 
% \item 
% \end{highlights}

% \includegraphics[width=\linewidth]{pictures/GraphicalAbstract.png}

% \includepdf[pages=1]{pdfs/cover_letter.pdf}
% \includepdf[pages=1]{pdfs/highlights.pdf}

% Keywords
% Each keyword is separated by \sep
\begin{keywords}
 Pore-network modelling\sep Bubble transport in enhanced gravity\sep Two-phase flow in porous media\sep Bubbles in a square capillary\sep Rotating PEM electrolyser
\end{keywords}

\maketitle

% Main text
\section{Introduction}
Gas bubbles generated by degassing, biological activity, or electrochemical reactions can strongly affect transport in porous media \citep{swiegersProspectsDevelopingHighly2021}. In proton exchange membrane (PEM) electrolysers, bubbles formed at the catalyst surface obstruct water supply and reduce performance \citep{lvMechanismAnalysesOptimization2023}. Experiments report improved performance under centrifugation, presumably due to accelerated bubble detachment and enhanced gas removal \citep{wangWaterElectrolysisEnhanced2010}. Under stationary conditions, oxygen transport through the porous transport layer (PTL) is commonly dominated by connected gas pathways formed by capillary invasion \citep{zlobinskiElucidationFluidStreamlining2021,zhangReviewGasliquidFlow2025}. However, at strong apparent gravitational forces, the gas transport pattern undergoes a gradual transition from a continuous finger flow to a discontinuous bubbly flow \citep{chenGasTransportSaturated2025}. It remains unclear how strongly apparent gravitational forces shift the balance between connected-phase and discrete-bubble transport. 

Interface-resolving methods such as the volume-of-fluid method \citep{raeiniModellingTwophaseFlow2012,shamsNumericalModelTwophase2018} have been used to simulate capillary-dominated flow and bubble detachment in the PTL of a PEM electrolyser \citep{scheuflerAccurateEfficientSurface2019, schmidtModelingPoreScaleCapillaryDominated2024}. Free-surface lattice-Boltzmann models have likewise been applied to oxygen-bubble transport in the same domain \citep{liuMultiphaseLatticeBoltzmann2016, jeonOxygenBubbleTransport2023}. These methods resolve individual interfaces but become computationally infeasible for network-scale simulations involving many bubbles. Pore-network models (PNM) provide a complementary reduced description by assigning storage and primary variables to pore bodies and fluxes to throats. Therefore, this work develops a hybrid PNM that retains a dynamic two-phase description while tracking individual mobile microbubbles.

A PNM approximates a porous medium as a network of geometrically idealised pore bodies interconnected by throats, capturing the general pore-scale behaviour at much lower computational cost and with reasonable accuracy compared to methods that fully resolve the pore space. Quasi-static PNMs, based on the invasion percolation concept, capture the invasion pattern at capillary equilibrium state by comparing local capillary thresholds with an externally applied global capillary pressure and using pore connectivity. Such models have been used to study the influence of centrifugal forces \citep{holtInfluenceCentrifugalForces2003}, predict permeability and gas saturation within a sintered titanium PTL \citep{leePoreNetworkModelling2019}, and model electrochemical reactions and charge transport \citep{elhannachPoreNetworkModel2012}. Dynamic PNMs resolve transient phase displacements by solving the pressure field and phase fluxes, where the storage terms and primary variables (e.g., phase saturation and pressure) are associated with the pore bodies and the fluxes are solved across the throats \citep{weishauptModelConceptsCoupling2020}. While dynamic PNMs have been used for fuel cells \citep{qinWaterTransportGas2015}, a review by \citet{sunSystematicReviewWaterOxygen2026} suggests that future electrolyser modelling should prioritise dynamic two-phase models. \citet{weishauptEfficientCouplingFree2019} coupled a dynamic PNM with free flow to capture interface dynamics between the porous and free-flow domains, and \citet{michalkowskiTwophaseFlowDynamics2022} extended this work to include bubble growth and detachment at the interface between the PTL and the channel of a PEM fuel cell. 

Conventional dynamic two-phase PNMs describe connected phase occupancy and invasion rather than individual mobile bubbles. Previous network models have investigated transport of long bubbles in a network of capillary tubes \citep{starkMotionLongBubbles2000}, tracked the migration and trapping of gas bubbles \citep{mahabadiGasBubbleMigration2018}, buoyancy-driven motion of bubbles \citep{ghasemianBubbleMigrationVelocity2019}, and represented Ostwald ripening of immobile bubbles \citep{mehmaniPorenetworkModelingOstwald2022}. These studies address related bubble-scale mechanisms, but they do not couple a discrete-bubble tracker to a dynamic two-phase PNM with conservative transfer between the discrete and connected gas representations and hydrodynamic feedback on the liquid flow.

To develop a PNM with discrete bubble tracking, local rules for the bubble movement through the porous medium must be established. The rising velocity of bubbles in circular capillaries is a well-studied classic problem, with early work by \citet{brethertonMotionLongBubbles1961} and \citet{whiteVelocityRiseSingle1962}. The work of Bretherton was extended by \citet{ratulowskiTransportGasBubbles1989} to higher capillary numbers and rectangular capillaries. Bretherton's results predict that bubbles are stationary at low Bond numbers, while \citet{bicoRiseLiquidsBubbles2002} showed that gravity can keep bubbles mobile in square channels due to corner flow. Here, we consider pore throats with a square cross-section, allowing for corner flow around a bubble in the throat. Additionally, the gas phase may exist as a continuous phase within the system. We assume a small capillary number and a Bond number slightly less than unity; therefore, the bubble shape is determined entirely by the bubble volume and throat geometry. We therefore develop local rules for bubble movement due to advection and buoyancy in our system. We obtain these from numerical Stokes simulations of the flow around bubbles confined in channels with square cross-sections and use them as the reference for bubble movement in the hybrid PNM.

In this work, we develop a hybrid PNM that couples a dynamic two-phase pore-network model to a discrete bubble tracker. The model distinguishes continuous liquid, connected gas, and discrete gas bubbles. Discrete bubbles reduce the available pore volume, modify throat conductance, generate a pressure contribution when driven by buoyancy, and exchange mass with the connected gas phase. Pore-scale Stokes simulations provide three independent closure functions: a buoyancy mobility, a conductance correction, and a reciprocal pressure-advection coupling coefficient. To our knowledge, this is the first conservative network formulation that combines these discrete and connected gas descriptions within one dynamic two-phase model.

The paper is structured as follows. We first discuss the influence of apparent gravitational forces, the existence of discrete microbubbles, and their properties in \autoref{sec:problem}. We then present the governing equations of the hybrid PNM in \autoref{sec:pnm_model}, with the closure relations obtained from the Stokes simulations in \autoref{sec:Stokes_models}. Next, we present the hybrid PNM numerical algorithm in \autoref{sec:numerical-algorithm}, then verify the model internally in \autoref{sec:verification}. Finally, we present the model application and predictions in \autoref{sec:application}, followed by the final conclusion in \autoref{sec:conclusion}. 

\section{Problem setup and assumptions}\label{sec:problem}
In this section, we present the general settings and assumptions for the rotating electrolyser system described in this study. First, we define the terminology related to the system's apparent gravitational forces. Next, we motivate the presence of microbubbles within the porous medium, describe their motion through it, and state the assumptions used to model bubble motion.

The system is assumed to be isothermal, with fluid properties fixed to the constant values given by \autoref{tab:parameters}. 

\begin{table}[pos=ht!p]
    \centering
    \caption{Constant material properties used for the rotating PEM electrolyser system. The values are chosen to represent a typical PEM electrolyser titanium PTL, with fluid properties at $60^\circ~\mathrm{C}$, and will be used throughout the paper for all developed models.}
    \label{tab:parameters}
    \begin{tabular}{lll}
        \toprule
        \textbf{Parameter} & \textbf{Symbol} & \textbf{Value} \\
        \midrule
        %Temperature & $T$ & $333 ~\mathrm{K}$ \\
        Contact angle & $\theta$ & $0^\circ$\\
        Contact radius & $r_c$ & $0.1 ~\mathrm{\mu m}$\\
        Density of water ($60^\circ ~\mathrm{C}$) & $\rho_w$ & $983.19 ~\mathrm{kg/m^3}$ \\
        Density of oxygen ($60^\circ ~\mathrm{C}$, $1\cdot 10^{5}~\mathrm{Pa}$) & $\rho_n$ & $1.156 ~\mathrm{kg/m^3}$ \\
        Dynamic viscosity of water ($60^\circ ~\mathrm{C}$) & $\mu_w$ & $4.656 \cdot 10^{-4} ~\mathrm{Pa\;s}$ \\
        Dynamic viscosity of oxygen ($60^\circ ~\mathrm{C}$) & $\mu_n$ & $2.253 \cdot 10^{-5} ~\mathrm{Pa s}$ \\
        Surface tension of water $(60^\circ ~\mathrm{C})$ & $\sigma_w$ & $0.0662 ~\mathrm{N/m}$ \\
        Gravitational acceleration & $g_0$ & $9.81 ~\mathrm{m/s^2}$ \\
        Maximum angular velocity & $\omega_{\text{max}}$ & $220 ~\mathrm{rad/s}$ \\ % Corresponds to 295G in our system, 0.06m from the centre of rotation at 2100 RPM.
        Maximum G-level & $\Gamma_{\mathrm{max}}$ & $300$ \\ 
        Distance to axis of rotation & $R_{0}$ & $60 ~\mathrm{mm}$ \\
        Region extent & $L$ & $1\text{-}5 ~\mathrm{mm}$ \\
        Mean pore inscribed radius & $\bar r_p$ & $70~\mathrm{\mu m}$ \\
        Typical throat inscribed radius & $\bar r_{ij}$ & $45~\mathrm{\mu m}$ \\
        Typical throat length & $\bar l_{ij}$ & $200~\mathrm{\mu m}$ \\
        Typical liquid phase velocity & $\bar u_{w}$ & $1~\mathrm{mm/s}$ \\
        Typical bubble velocity with $\bar u_w=1~\mathrm{mm/s}$ and $\Gamma=1$ & $\bar u_{b}$ & $0.001 ~\mathrm{m/s}$ \\ % stokes velocity of a bubble with the detachment radius at 1G, using our velocity model.
        \hphantom{Typical bubble velocity with $\bar u_w=1~\mathrm{mm/s}$ and} $\Gamma=300$ & $ $ & $0.27 ~\mathrm{m/s}$ \\ % stokes velocity of a bubble with the detachment radius at 300G, using our velocity model.
        Wetting film thickness & $d_{ij}$ & $0.01~r_{ij}$ \\
        Wall correction coefficient for circular tube & $k_c$ & $2.10443$ \\
        \hphantom{Wall correction coefficient for} square channel & $k_s$ & $1.903266$ \\ 
        \bottomrule
    \end{tabular}
\end{table}

\subsection{Apparent gravitational forces}\label{sec:gravity}
Let \(\boldsymbol\omega\) denote the constant angular-velocity vector and \(\boldsymbol R\) the position relative to the rotation axis. In an inertial frame, a point fixed in the rotating device has the inward centripetal acceleration
\(\boldsymbol a_{\mathrm{cp}} =\boldsymbol\omega\times(\boldsymbol\omega\times\boldsymbol R).\)
In the co-rotating frame used here, the same kinematic effect is represented by the outward centrifugal acceleration
\(\boldsymbol a_{\mathrm{cf}} =-\boldsymbol\omega\times(\boldsymbol\omega\times\boldsymbol R) = \|\boldsymbol\omega\|^2 \boldsymbol R.\)
Neglecting the ordinary gravitational field, the acceleration in the co-rotating frame can be expressed as a multiple of the gravitational acceleration $g_0$, hereafter denoted by the dimensionless G-level, $\Gamma$,
\begin{equation}
    \Gamma=\frac{\|\boldsymbol a_{\mathrm{cf}}\|}{g_0}.
\end{equation}

A value of \(\Gamma=1\) corresponds to a stationary system experiencing an acceleration equal to the local gravitational field, while \(\Gamma>1\) indicates a stronger effective gravitational force due to rotation.

To model a porous region of extent \(L\) centred at radial distance \(R_0\), where \(L \ll R_0\), we evaluate \(\boldsymbol a_{\mathrm{cf}}\) at \(R_0\) and treat it as uniform and constant within the pore network. We therefore retain the centrifugal contribution while neglecting its variation across the PTL.
At constant \(\boldsymbol\omega\), the Euler acceleration vanishes. The remaining Coriolis acceleration is \(\boldsymbol a_{\mathrm{cor}}=-2\boldsymbol\omega\times\boldsymbol u\), and its magnitude relative to the centrifugal acceleration is bounded by
\begin{equation*}
    \frac{\|\boldsymbol a_{\mathrm{cor}}\|}{\|\boldsymbol a_{\mathrm{cf}}\|} \leq 2 \bar{u}_b /(\omega_\text{max} R_0) = 0.04 \ll1 \,,
\end{equation*}
using parameter values of \autoref{tab:parameters}. The Coriolis force is therefore small compared to the centrifugal acceleration.
For local viscous flow on the bubble scale, the ratio of Coriolis force and viscous forces is given by
$\mathcal C_\Omega = 2\rho_w\omega \bar r_{ij}^2 \mu_w^{-1} \approx 2$. The Coriolis force can therefore be of the same order of magnitude as viscous forces for large $r_{ij}$ and $\Gamma$. For simplicity, we nevertheless neglect its possible effects on the bubble-drag-dependent closure relations derived subsequently from throat-local Stokes simulations. At the network scale, since the Coriolis force acts perpendicular to the velocity, i.e., perpendicular to the throat direction, it only displaces moving bubbles laterally. This effect is assumed negligible.
We assume that \(\boldsymbol a_{\mathrm{cf}}\) is normal to the catalyst surface of the electrolyser, and express the apparent gravitational forces in the rotating system through $\Gamma$.

\subsection{A buoyancy-driven bubble flow regime}
\begin{figure}[pos=ht!,width=0.47\linewidth]
    \centering
    \begin{minipage}[t]{0.47\linewidth}
        \vspace{0pt}
        \centering
        \includegraphics[width=\linewidth]{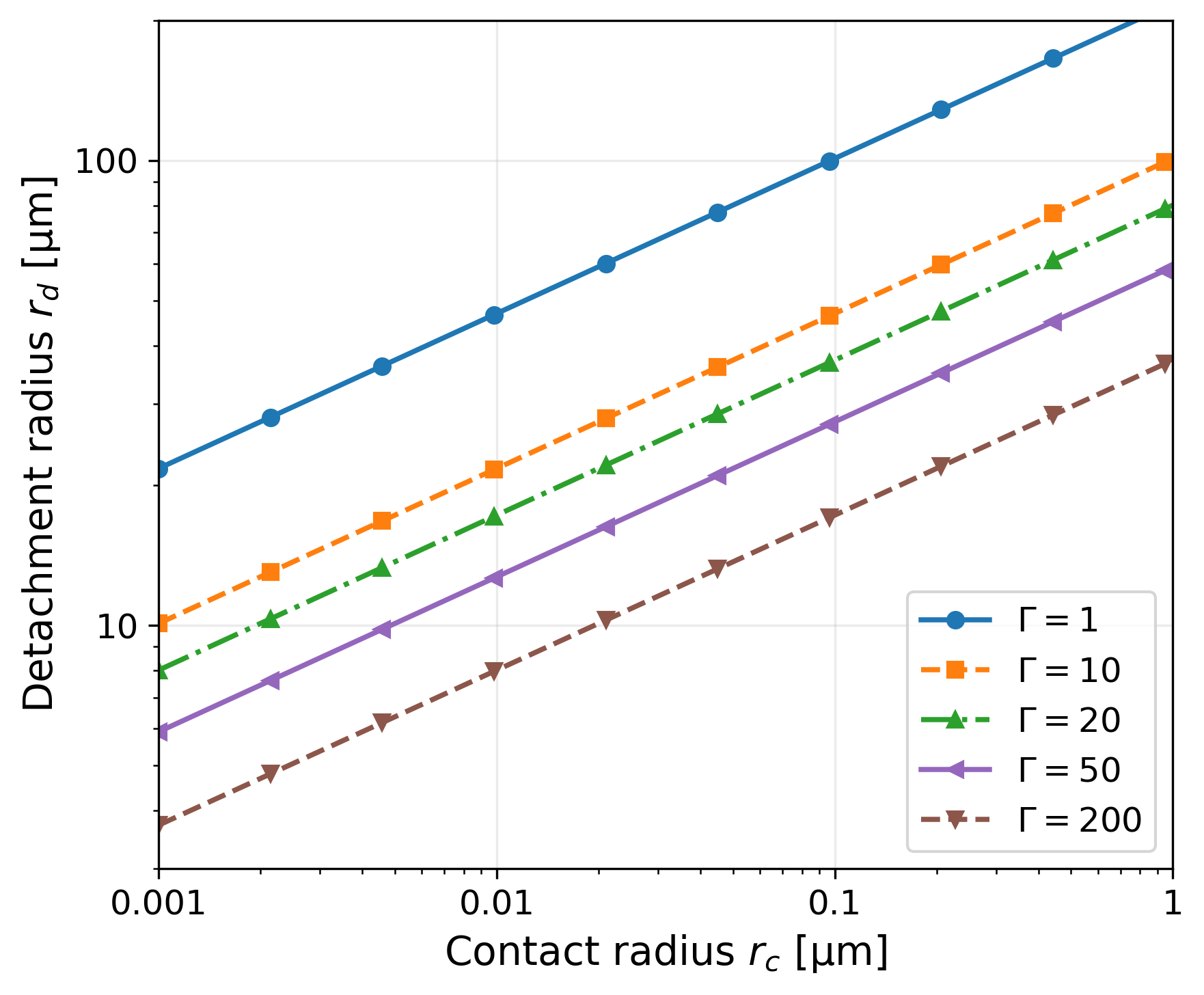}
        \caption{Plot of \autoref{eqn:detachment_radius}, giving the bubble detachment radius \(r_d\) as a function of contact radius \(r_c\) for different effective gravitational forces $\Gamma$.}
        \label{fig:detachment-radius}
    \end{minipage}
    \hfill
    \begin{minipage}[t]{0.47\linewidth}
        \vspace{0pt}
        \centering
        \includegraphics[width=\linewidth]{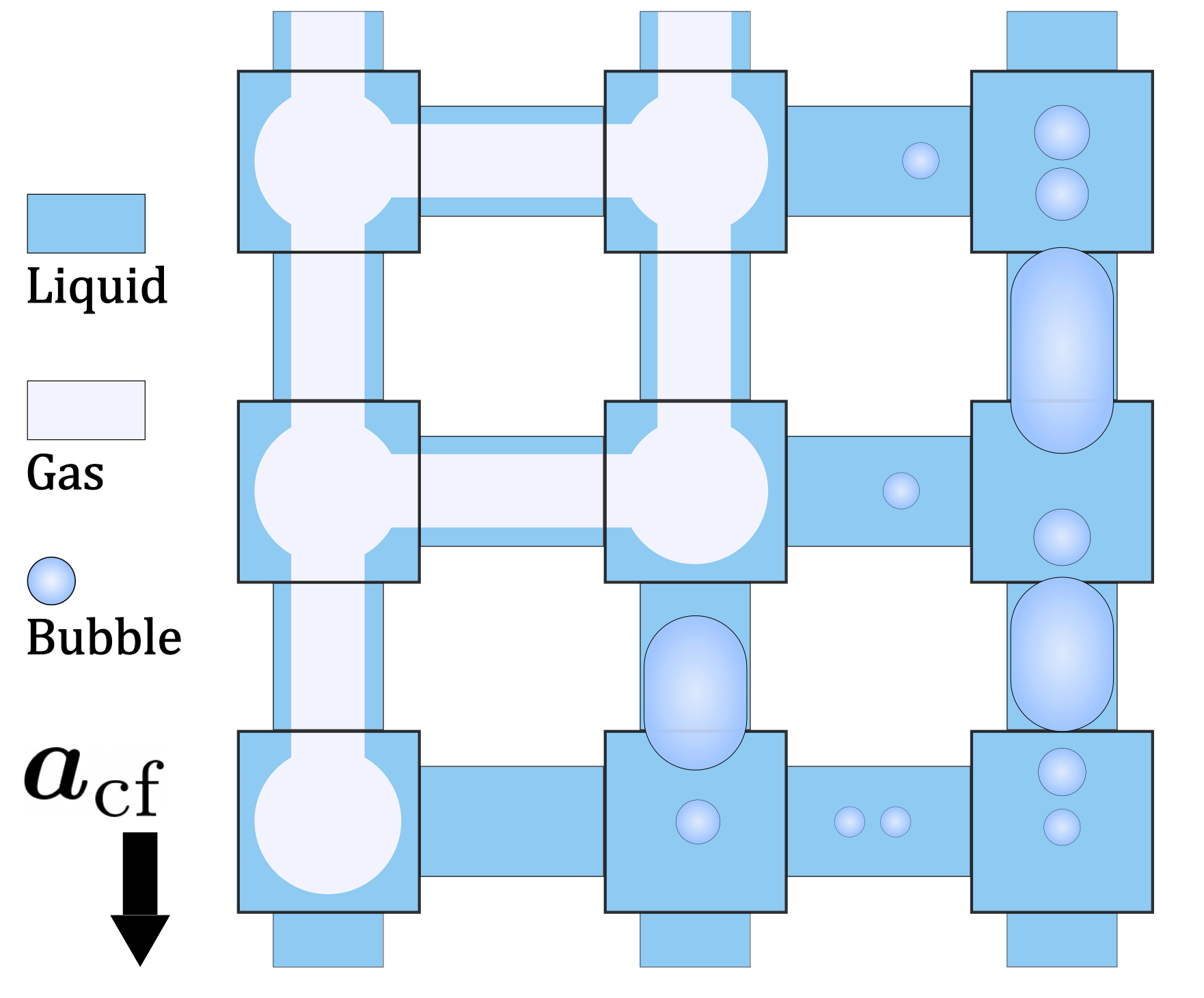}
        \caption{Illustration of the hybrid PNM flow regimes: connected wetting and non-wetting phase (left column) and discrete non-wetting bubbles (right column).}
        \label{fig:pnm-example}
    \end{minipage}
\end{figure}

%The existence of bubbles moving through the porous medium by buoyancy and advection will now be motivated.
While classical nucleation theory limits the size of bubbles formed in the bulk liquid by thermodynamic arguments \citep{hodgsonHomogeneousNucleation1984, taqieddinReviewPhysicochemicalHydrodynamics2017}, bubbles are far more likely to grow from cracks and impurities on the walls, which encourage nucleation \citep{galloHeterogeneousBubbleNucleation2021}. The limiting size of mobile bubbles entering the liquid is therefore given by the detachment radius of the bubbles. Detachment occurs when the acceleration-induced buoyancy force exceeds the surface-tension force retaining the bubble at its nucleation site. Balancing these forces gives \citep{fritzBerechnungMaximalvolumesDampfblasen1935}
\begin{equation}\label{eqn:detachment_radius}
    r_d=\left(\frac{3r_c\sigma_w}{2\, \|\boldsymbol a_{\mathrm{cf}}\|\,\Delta\rho}\right)^{1/3},
\end{equation}
where \(r_c\) is the contact radius of the nucleation site, and $\Delta \rho = \rho_w - \rho_n$ is the density difference between wetting phase density $\rho_w$ and non-wetting phase density $\rho_n$. We use this relation as a simplified detachment model; dynamic coalescence and liquid flow may cause detachment at smaller sizes \citep{lvSelfPropelledDetachmentCoalescence2021,nouri-khorasaniModelOxygenBubbles2017}. The contact radius can span several orders of magnitude. We choose $r_c=0.1~\mathrm{\mu m}$ as a representative pore scale for the catalytic layer \citep{leeDesigningCatalystLayer2023}, which also falls within the range of contact radii inferred for pinned electrolytic bubbles \citep{demirkirLifeFritzDetachment2024}. \autoref{eqn:detachment_radius} is plotted in \autoref{fig:detachment-radius}, showing that bubbles detach with radii in the range $10-100~\mathrm{\mu m}$ in our system.

Large values of $\boldsymbol a_{\mathrm{cf}}$ can therefore lead to detachment of bubbles with radii smaller than the typical pore size, such that they are freely suspended in the liquid and free to move by buoyancy. Furthermore, strong apparent gravitational forces can force bubbles into smaller pore-throats. This motivates introducing buoyancy-driven bubbles as an additional mode of transport in the PNM. \autoref{fig:pnm-example} illustrates the three gas-transport processes through such a pore network.

\subsection{Bubble motion and deformation}
The motion of bubbles within capillaries is characterised by the dimensionless capillary number, Bond number, Reynolds number, and Weber number,

\begin{equation}
    \mathrm{Ca} = \frac{\mu_w u_{b,ij}}{\sigma_w}, \quad 
    \mathrm{Bo} = \frac{\Delta \rho \,\|\boldsymbol a_{\mathrm{cf}}\|\, r_{ij}^2 }{\sigma_w}, \quad
    \mathrm{Re} = \frac{\rho_w u_{b,ij}\,r_{ij}}{\mu_w}, \quad 
    \mathrm{We}= \mathrm{Re}~\mathrm{Ca}\,.
\end{equation}

Here, $\mu_w$ is the wetting-phase dynamic viscosity, $\sigma_w$ is the gas/liquid surface tension, % and \(\theta\) is the wetting-phase contact angle.
and $u_{b,ij}$ is the velocity of a bubble within throat \(ij\), which has inscribed radius $r_{ij}$ and length $l_{ij}$. The parameters used in this study are summarised in \autoref{tab:parameters}.  

The capillary number $\mathrm{Ca}$ quantifies the relative importance of viscous to interfacial forces. For typical values of our system, we can estimate $\mathrm{Ca}(\Gamma=1) \approx 7 \cdot 10^{-6}$ and $\mathrm{Ca}(\Gamma=300) \approx 2\cdot 10^{-3}$. At small capillary numbers, large bubbles take a non-axisymmetric shape when constricted within the channel, leaving a thin wetting film along the centre of the channel walls with larger wetting areas in the corners \citep{magniniMorphologyLongGas2020}. The exact shape of this thin-film region will be discussed in the next section.

The Bond number $\mathrm{Bo}$ gives the ratio of gravitational forces to interfacial forces. For typical values of our system, we can estimate $\mathrm{Bo}(\Gamma=1) \approx 7 \cdot 10^{-4}$ and $\mathrm{Bo}(\Gamma=300) \approx 2\cdot 10^{-1}$. The apparent gravitational forces are therefore negligible under stationary conditions, but not when the system is rotating. The Bond number is not large enough for gravity to dominate the shape. We discuss the exact entry conditions and bubble deformation in throats in \autoref{sec:numerical-algorithm}.

The Reynolds number $\mathrm{Re}$ is the ratio of inertial forces to viscous forces. From the parameters in \autoref{tab:parameters}, we can estimate $\mathrm{Re}(\Gamma=1) \approx 0.16$ and $\mathrm{Re}(\Gamma=300) \approx 40$. At higher Reynolds numbers, inertia influences bubble drag. We expect that inertia slows bubbles relative to the Stokes assumption. For simplicity, we will derive closure relations from the Stokes equations and address the potential effect of inertia in the discussion. 

The Weber number $\mathrm{We}$ compares the inertial forces to the interfacial forces. For typical values of our system, we can estimate $\mathrm{We}(\Gamma=1) \approx 1\cdot 10^{-6}$ and $\mathrm{We}(\Gamma=300) \approx 8\cdot 10^{-2}$. \citet{liBuoyancydrivenMotionBubbles2008} found that inertia-induced bubble deformation is only relevant at large Weber numbers. We neglect such deformation in our system.

In summary, this regime allows us to derive approximate pore-throat-local closures for bubble motion and its impact on flow, based on fixed bubble shapes determined solely by surface tension and independent of flow.

\subsection{Bubble shapes in square capillaries}\label{sec:bubble-shape}
To derive local rules for bubble movement in our pore network, we conduct pore-scale Stokes simulations in capillaries with square cross-sections (described in \autoref{sec:Stokes_models}). In those simulations, each bubble is treated as a rigid, non-deforming inclusion with an immobile no-slip interface. We represent its volume with an equivalent radius \(r_b\), defined as the radius of a sphere with the same volume. A bubble is spherical when it fits within the square throat while maintaining the prescribed wall clearance; larger bubbles adopt an elongated, pill-shaped geometry. In equilibrium, the exact shape can be determined by minimising the surface energy of a volume confined within a square channel with given dimensions.

We characterise the square capillaries (pore throats) by their inscribed radius \(r_{ij}\) and length \(l_{ij}\). For wall-confined bubble geometries, we prescribe a minimum wall clearance \(d_{ij}\). Bubbles are spherical for \(r_b \le r_{ij}-d_{ij}\) and elongated for \(r_b > r_{ij}-d_{ij}\). For elongated bubbles, this clearance represents the imposed thin-film thickness at the centre of the throat walls; for spherical bubbles, \(r_{ij}-r_b\) is instead the minimum geometric clearance.
At small capillary numbers, the thin film region is expected to be non-uniform \citep{ratulowskiTransportGasBubbles1989, wongMotionLongBubbles1995}. \citet{magniniMorphologyLongGas2020} give a scaling law for the minimum film thickness: \(d_{\text{min}}/r_{ij} \propto \mathrm{Ca}\) as \(\mathrm{Ca}\rightarrow 0\). A study by \citet{hazelSteadyPropagationSemiinfinite2002} suggests that increasing the Bond number at a fixed capillary number increases the wetting film thickness. To obtain a tractable family of closure geometries, we prescribe a uniform dimensionless clearance of \(d_{ij}/r_{ij}=0.01\), matching the upper end of the $\Gamma$ range. All fitted closure coefficients reported below are therefore calibrated for this value. The present study does not establish independence from \(d_{ij}/r_{ij}\), and the resistance near the spherical-to-elongated transition may be sensitive to this parameter. This value is on the scale suggested by \citet{hazelSteadyPropagationSemiinfinite2002} and \citet{magniniMorphologyLongGas2020} for our range of capillary numbers. We expect a negligible wetting-phase flux in the film, and for large, elongated bubbles, flow is dominated by flow through the angular corners of the square capillaries.

\subsection{Gas-liquid interface conditions for bubbles}\label{sec:bubble-interface}
A perfectly clean gas-liquid interface is mobile and approaches the free-slip Hadamard-Rybczynski limit, whereas surfactants and impurities can substantially reduce its mobility \citep{stoneEffectsSurfactantsDrop1990,parkinsonTerminalRiseVelocity2008,ervikTransitionSettlingVelocity2017,danovMotionLongBubbles2021}. Because contamination is expected in an electrolyser PTL, we adopt the limiting case of an immobile no-slip interface. The resulting closures should therefore be interpreted as applying to strongly immobilised bubble interfaces rather than to clean bubble interfaces.

\section{Hybrid pore-network model with discrete bubble tracking}\label{sec:pnm_model}
The hybrid PNM consists of a non-compositional, isothermal two-phase PNM and a discrete bubble tracker. Water is the wetting phase and connected oxygen gas is the non-wetting phase, denoted by \(\alpha\in\{w,n\}\). The tracker represents discrete oxygen bubbles, denoted by \(b\). The model is implemented in the numerical software framework DuMu\textsuperscript{x}~\cite{kochDuMux3Opensource2021} using a fully implicit discretisation of the underlying dynamic PNM \citep{weishauptDynamicFullyImplicit2021}.

The primary variables of the hybrid PNM are the wetting phase pressure \(p_w\), non-wetting phase saturation \(S_n\), bubble position \(\boldsymbol x_{b,ij}\), and bubble volume \(V_b\). The mass and momentum balance can be written, for each pore body \(i\) and pore throat \(ij\) as,
\begin{subequations}\label{eqn:pnm-hybrid}
\begin{align}
\label{eqn:pnm-hybrid-mass}
    &\frac{\partial (\rho_\alpha S_\alpha{(V-V_{b})})_i }{\partial t} + \sum_j (\rho_\alpha Q_\alpha)_{ij} = (V q_\alpha)_i \; ,\\
    \label{eqn:pnm-hybrid-momentum}
    &Q_{w,ij} = c_{b,ij}\, g_{w,ij}\, \left(p_{w,i}-p_{w,j}
    -\rho_w \boldsymbol a_{\mathrm{cf}} \cdot (\boldsymbol x_i - \boldsymbol x_j) + p_{b,ij} \right) - \left(\frac{\partial V_{b}}{\partial t}\right)_{ij} \; ,\\
    &Q_{n,ij} = g_{n,ij}\, \left(p_{n,i}-p_{n,j} - \rho_n \boldsymbol a_{\mathrm{cf}} \cdot (\boldsymbol x_i - \boldsymbol x_j) \right)\; ,
\end{align}
\end{subequations}
where \(S_{\alpha}\) is the saturation, \(V_i\) is the pore body volume, \(V_{b,i}\) is volume occupied by bubbles in the pore, \((\rho_\alpha Q_\alpha)_{ij}\) is the advective mass flux through throat \(ij\), and \(q_\alpha\) is a mass source/sink density (in $\text{kg}~\text{m}^{-3}~\text{s}^{-1}$). The vector between the pore body centres is \(\boldsymbol x_i - \boldsymbol x_j\), and \(g_{w,ij}\) and \(g_{n,ij}\) are the wetting and non-wetting throat conductances. The term \(\left(\partial V_{b} / \partial t\right)_{ij}\) represents the volumetric flux in the throat, opposing the bubble movement, when a bubble volume is moved between the pore bodies.

The mass and momentum balance can be written for each bubble \(b\) as,
\begin{subequations}\label{eqn:bubble-tracker}
\begin{align}
    &\rho_{n,i} \frac{\partial V_b}{\partial t} = \left(V q_{b}\right)_i \, , \label{eqn:bubble-tracker-mass} \\
    &\frac{\partial x_{b,ij}}{\partial t} = u_{b,ij} \,, \quad \text{ with } u_{b,ij} = u_{\text{buoy},ij} + u_{\text{adv},ij} \,, \label{eqn:bubble-tracker-momentum}
\end{align}
\end{subequations}
where \(q_{b,i}\) is a bubble mass source (in $\mathrm{kg}~\mathrm{m^{-3}}~\mathrm{s}^{-1}$) with $i$ referring to the pore body the bubble is currently in. The bubble velocity \(u_{b,ij}\) can be decomposed in the linear Stokes flow regime into two additive components: the velocity due to buoyancy \(u_{\text{buoy},ij}\), and the velocity due to advection \(u_{\text{adv},ij}\). (A demonstration of the superposition principle for the numerical results is given in Section S1 of the Supplementary Material.) Closure relations for the local capillary pressure saturation relation and a constraint on the saturations are given by
\begin{equation}
    p_{c,i} \left(S_{w,i}\right) = \frac{2\sigma_w}{r_i \, \left(1- \exp(-6.83\,S_{w,i})\right)}, \quad
    S_{w,i} + S_{n,i} = 1 \,,
\end{equation}
with $p_{c,i} \left(S_{w,i}\right)$ is derived for a cubic pore by \citet{joekar-niasarNonequilibriumEffectsCapillarity2010}.

The discrete bubbles couple to the continuum PNM in four ways. First, their volume reduces the pore volume available to the continuous phases, via $V_b$. Second, detached bubbles modify the hydraulic conductance of liquid-filled throats, affecting $g_{w,ij}$. Third, buoyancy-driven bubble motion contributes a pressure term to the liquid flux law, $p_{b,ij}$. Fourth, changes in discrete bubble pore occupancy (a bubble is always associated with exactly one pore and one pore throat; pores carry the volumes in our pore network formulation) displace liquid between neighbouring pore bodies, modelled by $(\partial V_b / \partial t)_{ij}$. When a discrete bubble is transferred to the connected gas phase, its mass is conserved, and the continuum saturations are updated accordingly.

The conductance correction and buoyancy-induced pressure term are components of the same linear pressure-flow response of a bubble-occupied throat. \autoref{sec:Stokes-double} uses the reciprocal theorem and linearity of the Stokes equations to show that the two terms superpose.

The throat conductance \(g_{\alpha,ij}\) incorporates the geometry (e.g., length, shape, diameter) and the fluid viscosity, as well as the capillary-pressure-dependent distribution \citep{weishauptModelConceptsCoupling2020}. The throat conductance is thus different for single-phase and two-phase flow. The throat conductances implemented for square throats are given below.

\paragraph{Single-phase conductance.}
Throats which are not invaded by gas are fully water-saturated. The conductance for a throat with square cross-section between two pores $i$ and $j$ is, as proposed by \citet{patzekShapeFactorHydraulic2001},
\begin{equation}\label{eqn:single_phase_conductance}
    g_{w,ij}= \frac{K_G A_{ij}^2}{\mu_{w} l_{ij}} \, ,
\end{equation}
where \(K_G=0.5623/16\) is the hydraulic conductance of a square throat, \(\mu_{w}\) is the wetting phase dynamic viscosity, \(A_{ij}\) is the throat's cross-sectional area, and \(l_{ij}\) is the throat length.

\paragraph{Wetting-phase conductance after invasion.}
After gas invasion of a water-wet throat, water remains in the corners. To describe flow in the corner films, we use the wetting phase throat conductance model proposed by~\citet{ransohoffLaminarFlowWetting1988},
\begin{equation}\label{eqn:wetting-phase-conductance}
    g_{w,ij} = \frac{r_a^2}{\mu_{w} l_{ij}}\frac{A_{w,ij} }{\beta} \, ,
\end{equation}
where \(r_a\) is the radius of curvature of the arc menisci, \(A_{w,ij}\) is the wetting-phase cross-section, and \(\beta\) is a dimensionless flow resistance which depends on the corner geometry and contact angle. We use the closed-form expression for \(\beta\) proposed by \citet{zhouHydrocarbonDrainageCorners1997} based on the tabulated values of \citet{ransohoffLaminarFlowWetting1988}, given for completeness in \autoref{app:beta}.

\paragraph{Non-wetting phase conductance.}
When a water-wet throat is invaded by a non-wetting gas phase, the gas is confined in the middle of the throat. The non-wetting phase throat conductance is given by \citet{bakke3DPoreScaleModelling1997} and \citet{toraDynamicNetworkModel2012} as:
\begin{equation}
    g_{n,ij}= \frac{r_{d}^2 A_{n,ij}}{8\, \mu_n l_{ij}} \; , \quad \text{where} \quad r_{d}=\tfrac{1}{2}\left(\sqrt{A_{n,ij}/\pi} + r_{ij}\right) \quad \text{and} \quad A_{n,ij} = \frac{r_{ij}^2}{4 G} - A_{w,ij}
\end{equation}
are the effective throat inscribed radius and the cross-sectional area of the non-wetting phase, respectively.
\(G=1/16\) is the dimensionless shape factor of the square throat.

\section{Bubbles in pore throats: closure relations from Stokes simulations}\label{sec:Stokes_models}
Next, we derive closure relations for the coupling terms in \autoref{eqn:pnm-hybrid} and \autoref{eqn:bubble-tracker}. The closure relations are parametrised by the dimensionless relative bubble length

\[l_{b,ij}^*=\frac{l_{b,ij}}{2r_{\mathrm{eff}}},\qquad r_{\mathrm{eff}}=r_{ij}-d_{ij}, \qquad r^*=\frac{r_{\mathrm{eff}}}{r_{ij}}=0.99\]

where \(r_{\mathrm{eff}}\) is the dimensional effective throat radius, \(l_{b,ij}\) is the length of the bubble, \(r_{ij}\) is the throat inscribed radius, \(d_{ij}\) is the wetting phase thin-film thickness for large bubbles, and \(r^*\) is the dimensionless effective throat radius. The bubble is spherical for \(l_{b,ij}^*\le1\) and elongated for \(l_{b,ij}^*>1\). The bubble length can be calculated from the equivalent bubble radius \(r_b\) with \autoref{eqn:bubble_length}, presented in \autoref{app:Stokes}.

Closing the hybrid model requires three independent geometry-dependent relations. The first is the buoyancy velocity, written as \(u_{\mathrm{buoy},ij}\). The second is the force-free hydraulic conductance of a bubble-occupied throat, represented by \(c_{b,ij}\). The third is a reciprocal coupling coefficient \(\phi\), which determines both the buoyancy-induced pressure contribution and the flow-induced bubble velocity,

\[p_{b,ij}=\phi(l_{b,ij}^*)\frac{F_{b,ij}}{A_{ij}},\qquad u_{\mathrm{adv},ij}=\phi(l_{b,ij}^*)\frac{Q_{w,ij}}{A_{ij}},\]
with the signed throat-parallel buoyancy force \(F_{b,ij}=-\Delta\rho\,(\boldsymbol a_{\mathrm{cf}}\cdot\boldsymbol e_{ij})V_b \). We derive these relations with a combination of numerical data and analytical results in the limit cases of small bubbles and long bubbles. The fitted closures lie within $5\%$ relative error of the simulation results. We perform a mesh refinement and throat-length study in \autoref{app:Stokes}.

\subsection{Simulation setup}
We conduct Stokes simulations of flow around a rigid, non-deforming, no-slip obstacle in a square throat. We vary throat length, throat radius, bubble volume, effective acceleration, and the imposed pressure difference. We solve in the bubble frame: the bubble appears stationary, and the throat walls move with velocity \(-u_{b,ij}\). The laboratory-frame velocity is recovered by adding \(u_{b,ij}\) to the computed liquid velocity field.

We obtain the bubble geometries using Surface Evolver (v2.7) \citep{brakkeSurfaceEvolver1992}, which minimises the surface energy of a volume confined within a square channel. Depending on the bubble size and channel width, bubbles either remain spherical or elongate within the channel. Surface Evolver provides a bubble surface mesh, which we then use to create the fluid-domain mesh for the Stokes simulation. Additional details are presented in \autoref{app:Stokes}.

We solve two types of Stokes problems with different boundary conditions. In the buoyancy case ($\Gamma \neq 0$), we adjust the bubble-frame wall velocity and the non-hydrostatic boundary pressure so that the total force on the bubble and the net liquid flux both vanish. This yields the buoyancy velocity at zero net liquid flux and the associated viscous pressure drop across the bubble. 
In the advection case, a pressure difference drives the flow at $\Gamma = 0$, and we adjust the bubble-frame wall velocity until the hydrodynamic force on the bubble vanishes. This allows us to compute the liquid-phase throat conductance for a force-free bubble and the advective bubble velocity.

All four closure relations can be obtained from the two sets of simulation cases. \autoref{app:bubble-velocity} provides additional details on how we set up the Stokes simulations and compute the bubble velocities; an example of the resulting flow field appears in \autoref{fig:mesh} and Section S2 of the Supplementary Material. We present the numerical algorithms for obtaining net liquid flux and a force-free bubble in \autoref{app:Stokes}.

\subsection{Buoyancy velocity model} \label{sec:Stokes-buoyancy-vel}
We compute the bubble buoyancy velocity from the buoyancy scenarios. When the bubble surface is modelled with a no-slip condition, we expect the velocity of small bubbles to approach Stokes' law, obtained by balancing the drag and buoyancy forces acting on a rigid sphere suspended in an infinite liquid as
\begin{equation}\label{eqn:Stokes-velocity}
    u_S = \frac{2}{9} \frac{\Delta \rho \, \boldsymbol a_{\mathrm{cf}} \, r_b^2}{\mu_w} \; .
\end{equation}

Within a channel, the first-order wall correction is $k \,r_b/r_{ij}$, where $k$ is a coefficient determined by \citet{faxenWiderstandGegenBewegung1922} as $k_c=2.10443$ for a cylindrical tube, and by \citet{happelSettlingSphereAxis1974} as $k_s=1.903266$ for a square channel \citep{happelLowReynoldsNumber1983}. A higher-order expansion by \citet{bohlinDragRigidSphere1960} for the cylinder can be re-scaled such that the leading-order correction matches the results for the square channel,
\begin{equation}\label{eqn:gvelocity_model}
    \frac{u_{\text{buoy},ij}}{u_S} = f_{\text{buoy}}(l_{b,ij}^*) = \left[1+k_c\,\lambda+k_c^2\,\lambda^2 + a\left(\frac{A_b}{A_w}\right)^n(2\, l_{b,ij}^*)^m \right]^{-1} ,
\end{equation}  
where the quantities \(A_b\) and \(A_w\) are the bubble and wetting-phase cross-sectional areas, \(l_{b,ij}^*\) is the relative bubble length, and

\begin{equation}
    \lambda=\dfrac{r^*\,l_{b,ij}^*}{k_c/k_s} \, .
\end{equation}

For a spherical bubble, \(r^*\,l_{b,ij}^* = r_b/r_{ij}\) such that the first-order term recovers the Happel-Bart square channel correction. We fit the parameters \(a=8.106\), \(n=1.333\), and \(m=1.864\) using nonlinear least squares. The first three terms of the denominator are a Bohlin-inspired wall resistance, while the final term accounts for the additional resistance from liquid flow through the corners and films as the bubble length increases.

\subsection{Throat conductance model}\label{sec:throat-conductance-model}
We compute the bubble impact on throat conductance from the advection cases. It is expressed as an excess length: the extra length of a bubble-free throat that would give the same reduction in conductance. The force-free hydraulic conductance of a bubble-occupied throat can then be written as
\begin{equation}\label{eqn:conductance_model}
    c_{b,ij} = \frac{g_{b,ij}}{g_{w,ij}} = \frac{1}{1+ \left(r_{\mathrm{eff}}/l_{ij} \right) \hat E}\, , \qquad \text{where} \quad \hat E = \dfrac{E}{r_{\mathrm{eff}}} \,.
\end{equation}

Here, \(g_{w,ij}\) is the analytical single-phase throat conductance from \autoref{eqn:single_phase_conductance}, \(g_{b,ij}\) is the throat conductance in the presence of a bubble, and \(r_{\mathrm{eff}}/l_{ij}\) is the aspect ratio of the throat. $r_{\mathrm{eff}}=r^*\,r_{ij}$ is the effective throat inscribed radius when there is a thin liquid film of thickness \(d_{ij}=0.01\,r_{ij}\). We obtain the dimensionless excess length \(\hat E\) from the Stokes simulations. When there are multiple bubbles within a throat, we assume that the excess lengths $\hat E$ add in series. The proposed model for $\hat E$ is given by
\begin{equation}\label{eqn:excess_length_model}
    \hat E =
    \begin{cases}
    C_1 \dfrac{\pi}{3} \left(r^*\right)^2 \left(l_{b,ij}^*\right)^3 \left[1 + C_2\dfrac{r^*\,l_{b,ij}^*}{1-r^*\,l_{b,ij}^*}\right], & l_{b,ij}^* \leq 1, \\
    2 l_{b,ij}^* \left[\dfrac{w(l_{b,ij}^*)}{\kappa_c} + \dfrac{1-w(l_{b,ij}^*)}{C_3} - 1 \right], & l_{b,ij}^* > 1.
    \end{cases}
\end{equation}
where \(w(l_{b,ij}^*)\) is a weighting function describing how much of the bubble-occupied region behaves like the two rounded end caps rather than the elongated bubble body,
\begin{equation}
    w(l_{b,ij}^*) = \operatorname{min} \left\lbrace 1, \operatorname{max} \left\lbrace 0, \tilde w(l_{b,ij}^*) \right\rbrace \right\rbrace, \quad \tilde w(l_{b,ij}^*) = \frac{1}{l_{b,ij}^*}\left[ l_\infty + (1-l_\infty) \exp\left( -\frac{l_{b,ij}^*-1}{l_c}\right) \right]
\end{equation}
where \(l_\infty\) and \(l_c\) are fitted parameters representing the asymptotic dimensionless cap length and the rate at which the cap contribution relaxes from the transition value to this asymptotic value, respectively. \(\kappa\) is the effective relative conductance of the bubble-occupied part of the throat. Since hydraulic resistances add in series, the model interpolates the inverse conductance, \(1/\kappa\), between the conductance associated with the bubble caps, \(\kappa_c\), and that of the elongated bubble body (\(C_3\)),
\begin{equation}
\kappa_c = \frac{1}{ 1 + \hat E_{\mathrm{sph}}(1)/2},
\end{equation}
where \(\hat E_{\mathrm{sph}}(1)\) is the spherical-branch value evaluated at \(l_{b,ij}^*=1\):
\begin{equation}
    \hat E_{\mathrm{sph}}(1) = C_1 \dfrac{\pi}{3} \left(r^*\right)^2 \left[1 + C_2 \dfrac{r^*}{1-r^*} \right] \,.
\end{equation}

\begin{figure}[pos=ht!,width=0.49\linewidth]
    \centering
    \begin{minipage}[t]{0.49\linewidth}
        \vspace{0pt}
        \centering
        \includegraphics[width=\linewidth]{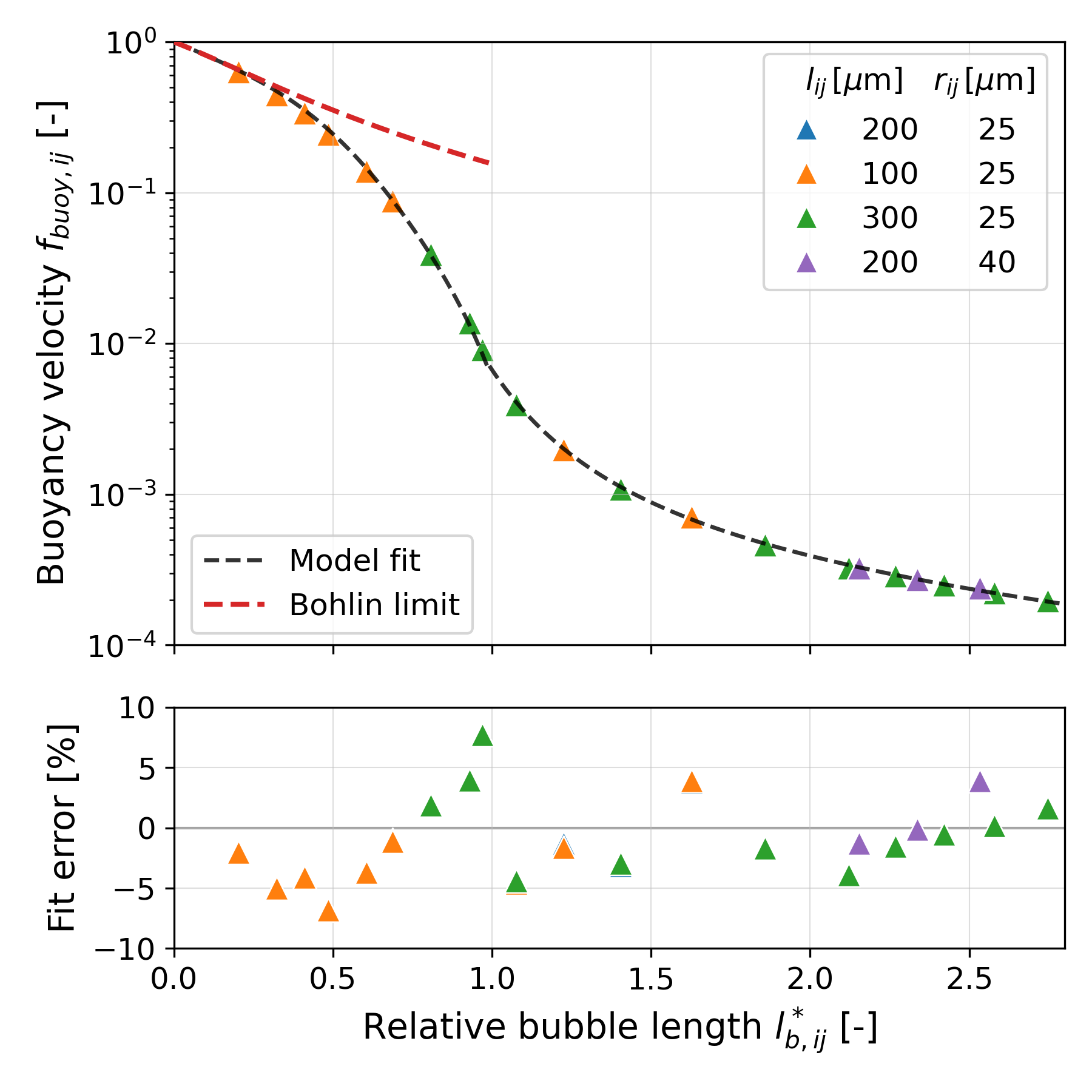}
        \caption{Bubble buoyancy velocity model, fitted to Stokes simulation data where the bubble is moving due to buoyancy. The top plot shows the fitted data, and the bottom plot shows the relative error of the fit to each data point. The full model is given by \autoref{eqn:gvelocity_model}.}
        \label{fig:model_gvelocity}
    \end{minipage}
    \hfill
    \begin{minipage}[t]{0.49\linewidth}
        \vspace{0pt}
        \centering
        \includegraphics[width=\linewidth]{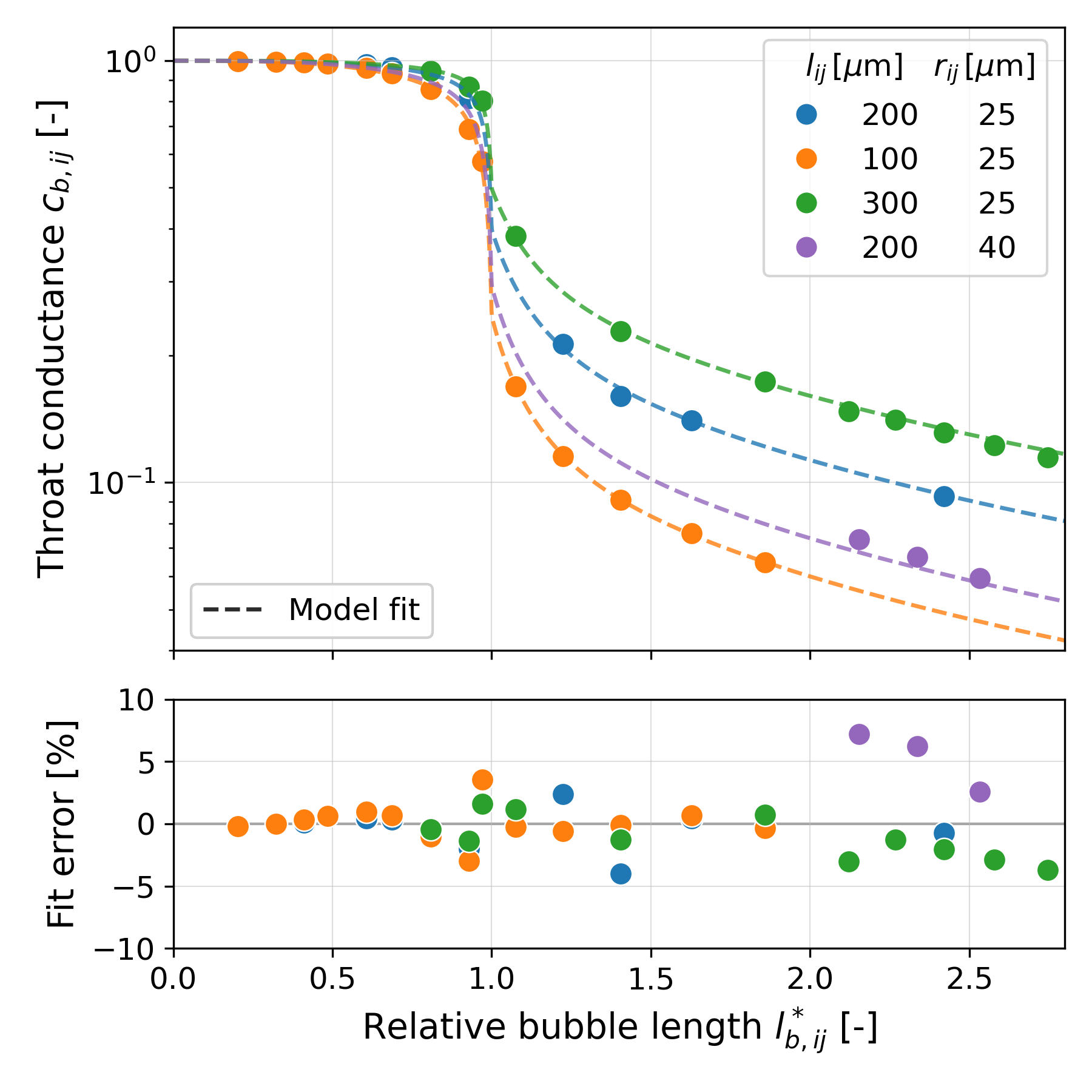}
        \caption{Throat conductance model, fitted to Stokes simulation data where a potential difference is driving a liquid flux in the throat, with the bubble moving by advection. The top plot shows the fitted data and the bottom plot shows the relative difference of the fit to each data point. The model is given by \autoref{eqn:conductance_model} and the excess length by \autoref{eqn:excess_length_model}.}
        \label{fig:model-conductance}
    \end{minipage}
\end{figure}

The excess length is split into three parts: the resistance in the small bubble limit ($C_1$) when $l_{b,ij}^*\ll 1$, the near-confinement regime ($C_2$) when $l_{b,ij}^*\approx 1$, and the corner-flow-dominated regime ($C_3$) when $l_{b,ij}^*\gg 1$.
The fitted parameters are \(C_1 = 0.708\), \(C_2= 0.160\), \(C_3= 0.054\), \(l_\infty = 0.293\), and \(l_c = 0.209\). The fitted model is shown in \autoref{fig:model-conductance}.

\subsection{Advection velocity and viscous pressure model}\label{sec:Stokes-double}
We compute the advection velocity from the advection case, combined with the corresponding viscous pressure drop from the buoyancy case. Lorentz' reciprocal theorem can be used to determine both states with a single coefficient \citep{lorentzGeneralTheoremMotion1996} as we briefy explain. In two settings with identical fluid domain and fluid viscosity but potentially different boundary conditions, it holds

\begin{equation}\label{eqn:reciprocal-theorem}
    \oint_{S} \boldsymbol u'\cdot \sigma \cdot \boldsymbol n ~dS = \oint_{S} \boldsymbol u \cdot \sigma' \cdot \boldsymbol n ~dS \,,
\end{equation}
where \(S\) is the surface bounding the fluid domain. Let us define \(\left(\boldsymbol u,\sigma\right)\) as the liquid velocity and stress in the advection run, and \(\left(\boldsymbol u',\sigma'\right)\) as the liquid velocity and stress in the buoyancy run. The stress tensors are given by

\begin{equation}
    \sigma = -pI + \mu \left(\nabla \boldsymbol u + (\nabla \boldsymbol u)^T\right) \qquad \text{and} \quad \sigma' = -p'I + \mu \left(\nabla \boldsymbol u' + (\nabla \boldsymbol u')^T\right) \,.
\end{equation}

With a no-slip condition on the outer walls, the bubble is moving, and the flow is fully developed at the inlet and outlet, we may write \citep{brennerPressureDropDue1970}

\begin{align}
    \oint_{S} \boldsymbol u'\cdot \sigma \cdot \boldsymbol n ~dS = \Delta p_{\text{adv}}\,Q_{\text{buoy}} - u_{\text{buoy}} \,F_{\text{adv}} \,, \quad
    \oint_{S} \boldsymbol u\cdot \sigma' \cdot \boldsymbol n ~dS = \Delta p_{\text{buoy}}\,Q_{\text{adv}} - u_{\text{adv}} \,F_{\text{buoy}} \,,
\end{align}
where $\Delta p$ is the respective non-hydrostatic pressure drop between inlet and outlet, $Q$ is the flux, \(u\) is the bubble velocity, and $F$ is the force acting on the bubble, each for the advection and buoyancy cases. Since the advection system has $F_{\text{adv}}=0$, and the buoyancy system has $Q_{\text{buoy}}=0$, then the reciprocal theorem gives
\begin{equation}
    \frac{\Delta p_{\text{buoy}}}{F_{\text{buoy}}} = \frac{u_{\text{adv}}}{Q_{\text{adv}}}\,.
\end{equation}

Multiplying by the throat area $A_{ij}$, writing out the buoyancy force $F_{\text{buoy}}=\Delta \rho \, \|\boldsymbol a_{\mathrm{cf}}\| \,V_b$, and writing the mean liquid velocity as \(u_w = Q_{\text{adv}}/A_{ij}\),

\begin{equation}\label{eqn:model-double}
    \phi = \frac{p_b \, A_{ij}}{\Delta \rho \,\|\boldsymbol a_{\mathrm{cf}}\|\,V_b} = \frac{u_{\text{adv}}}{u_w} \,,
\end{equation}
where $\phi$ has the meaning of a non-dimensional coupling function for both the viscous pressure drop $p_b$ in the buoyancy case and the bubble velocity $u_{\text{adv}}$ in the advection case. It is therefore only necessary to run one of the simulation cases and fit the function \(\phi\) to the data, in order to obtain \(p_b\) and \(u_{\text{adv}}\).

The linearity of the Stokes problem, independent of \autoref{eqn:reciprocal-theorem}, guarantees that the two closures superpose without double counting. A force balance on the bubble splits its velocity exactly into the buoyancy and advection contributions of \autoref{eqn:bubble-tracker-momentum}, and the pressure difference across the throat correspondingly into a bubble term \(p_{b,ij}\) and a resistance term. The resistance appearing in that split is the one measured with a force-free bubble, i.e.\ the advection scenario of \autoref{sec:throat-conductance-model}, so \(c_{b,ij}\) and \(p_{b,ij}\) may be applied together in \autoref{eqn:pnm-hybrid-momentum}.

Based on two analytical limits a function $\phi(l_{b,ij}^*)$ can be constructed, where \(l_{b,ij}^*\) is the relative bubble length,

\begin{equation}\label{eqn:phi}
    \phi(l_{b,ij}^*) = \phi_\infty + \left( \phi_0 - \phi_\infty \right)\exp \left(- \dfrac{(r^*\,l_{b,ij}^*)^2}{24\,K_G\left(\phi_0 - \phi_\infty\right)} - c_4 (l_{b,ij}^*)^4 - c_p (l_{b,ij}^*)^p \right) \,.
\end{equation}

For a spherical bubble, \(r^*\,l_{b,ij}^*=r_b/r_{ij}\), such that the leading-order correction of \citet{faxenWiderstandGegenBewegung1922} is recovered; \(\phi_0=2.09626\) is the ratio of the centreline velocity to the cross-sectional mean velocity for fully developed Stokes flow in a square duct. The quadratic coefficient \(1/(24K_G)\) follows from the small-sphere limit calculated by \citet{faxenWiderstandGegenBewegung1922}. The limit \(\phi_\infty\) is a geometric expression for the corner flow resistance,

\begin{equation}\label{eqn:phi_inf}
    \phi_\infty = \frac{A_{ij}}{A_b+kA_w}
\end{equation}

where the cross sectional areas are for the throat $A_{ij}$, bubble $A_{b}$, and wetting area (corner flow) $A_w$. The ratio of the mean liquid velocity in the corners and the bubble velocity, when corner flow dominates, is calculated in \autoref{app:corner-k} as \(k=0.428\), with a wetting film thickness \(d=0.01\) and corner curvature radius approaching \(\frac{r_a}{r_{ij}}\rightarrow0.499\) for large bubbles. We get \(\phi_\infty = 1.04378\). 

\begin{figure}[pos=ht!,width=0.50\linewidth]
    \centering
    \begin{minipage}[t]{0.50\linewidth}
        \vspace{0pt}
        \centering
        \includegraphics[width=\linewidth]{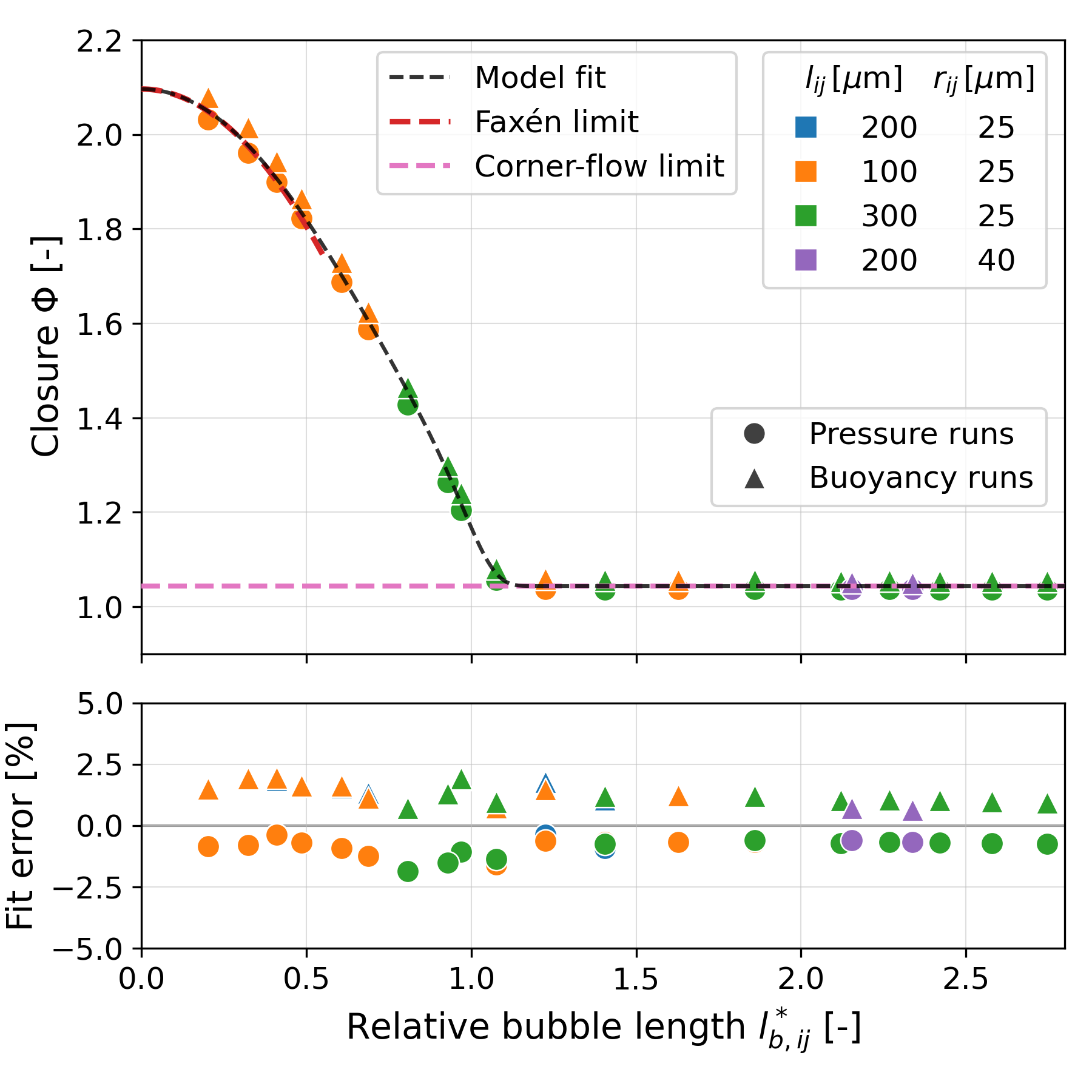}
        \caption{Fitted $\phi$ for the bubble advection velocity and buoyancy viscous pressure. Data from both the advection and buoyancy runs, with numerical limits highlighted. The top plot shows the fitted data, and the bottom plot shows the relative error of the fit to each data point. The model is given by \autoref{eqn:phi}.}
        \label{fig:model-double}
    \end{minipage}
\end{figure}

We fit $\phi$ to both the advection and buoyancy data using nonlinear least squares, yielding the free parameters \(c_4=0.466\), \(c_p=0.569\), and \(p=15.55\). \autoref{fig:model-double} shows the closure fit and its residuals.

\section{Numerical algorithm}\label{sec:numerical-algorithm}
\begin{figure}[pos=ht!]
    \centering
    \includegraphics[width=0.97\linewidth]{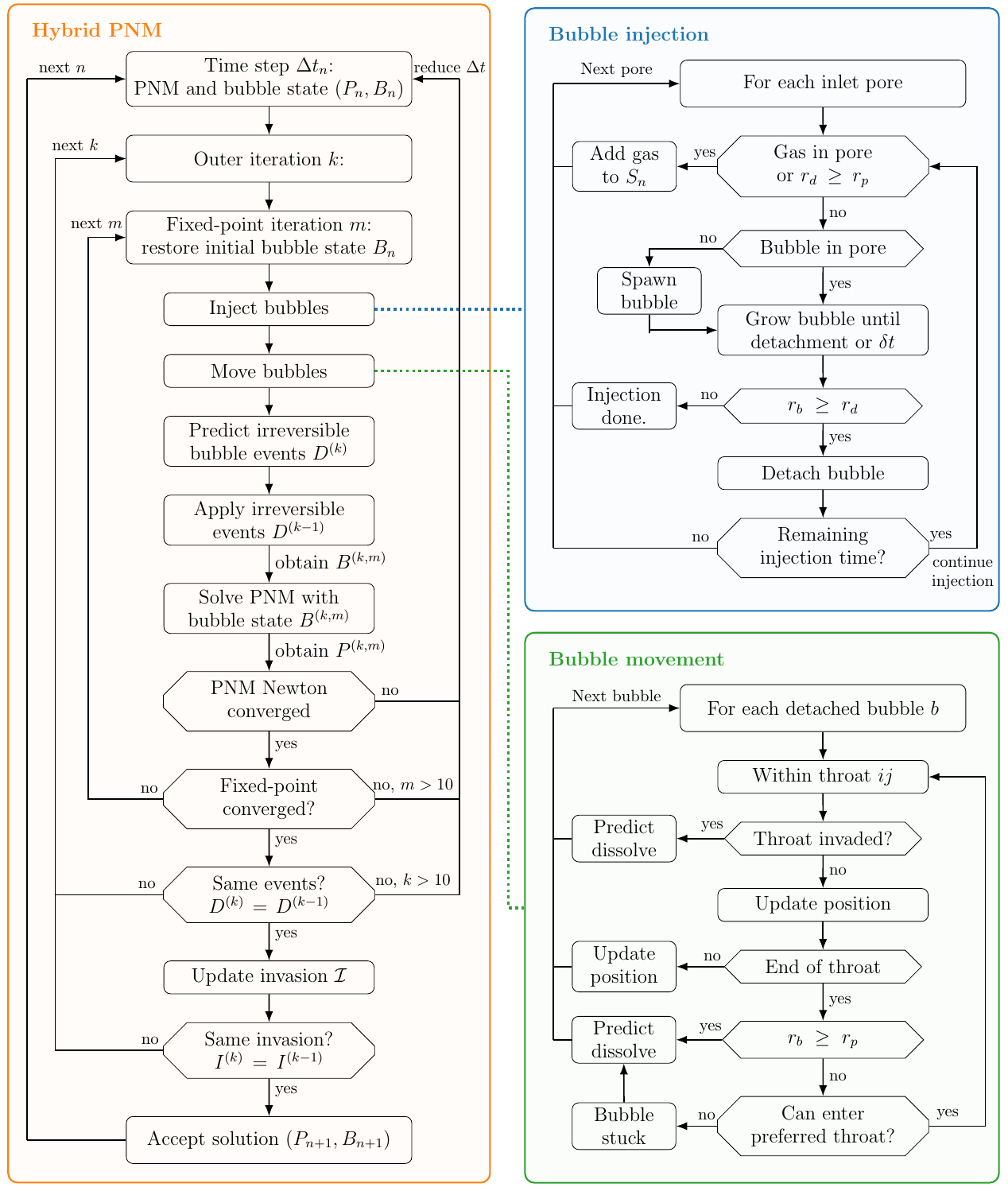}
    \caption{Main algorithm of the hybrid PNM (left), and sub-algorithms for injecting (top right) and moving bubbles (bottom right).}
    \label{fig:full-algorithm}
\end{figure}

This section describes how the bubble tracker algorithm couples with the pore-network model algorithm, outlines the bubble tracker steps, and is complemented by a flowchart in \autoref{fig:full-algorithm}. 

Within one time step \(\Delta t_n\), the coupled PNM-bubble system is advanced from \((P_n, B_n, I_n)\) at time \(t_n\) to \((P_{n+1}, B_{n+1}, I_{n+1})\) at time \(t_{n+1} = t_n + \Delta t_n\). Here \(P\) stands for the primary variables of the pore network, \(p_w\) and \(S_n\) in every pore body, and \(B\) stands for the primary variables of the bubble tracker, bubble positions \(x_{b,ij}\) and bubble volumes \(V_b\). The set of gas-invaded throats is denoted by \(I\). Moreover, we introduce \(D\), the set of irreversible bubble events, as an auxiliary state during time-step integration.

A fully implicit time step of the PNM can be expressed by
\begin{equation}\label{eqn:op-pnm}
    P = \mathcal{P}(B;\, I;\, P_n, B_n, \Delta t)
\end{equation}
such that the operator $\mathcal{P}$ returns the solution of \autoref{eqn:pnm-hybrid}, together with the capillary pressure relation and the saturation constraint, at \(t_{n+1}\) for a given bubble state \(B\) and invasion set \(I\). The state at \(t_n\) enters through the storage term of \autoref{eqn:pnm-hybrid-mass}, with the saturations \(S_\alpha\) and bubble volume \(V_b\) obtained from the primary variables of \(P_n\) and \(B_n\), respectively. In addition, the previous state at \(t_n\) enters through the change of bubble volume between neighbouring pores; \((\partial V_b/\partial t)_{ij}\) in \autoref{eqn:pnm-hybrid-momentum}. The bubble state at \(t_{n+1}\) enters through \(V_b\), \(c_{b,ij}\) and \(p_{b,ij}\). One time step of the bubble tracker can be expressed by
\begin{equation}\label{eqn:op-tracker}
    B = \mathcal{B}(P;\, D, I;\, B_n, \Delta t)
\end{equation}
where the operator $\mathcal{B}$ starts from \(B_n\), injects and grows bubbles in the source pores, integrates \autoref{eqn:bubble-tracker-momentum} over \(\Delta t\) with \(u_{\mathrm{buoy},ij}\) and \(u_{\mathrm{adv},ij}\) evaluated from \(P\), and applies the events contained in \(D\).

The event detector $\mathcal{D}$ is defined such that
\begin{equation}\label{eqn:op-events}
    D = \mathcal{D}(B, P;\, I)
\end{equation}
returns the set of irreversible events $D$ (exiting, merging, and transfer to the connected gas phase) whose trigger conditions are met along the trajectory that produced the bubble state \(B\). The invasion update $\mathcal{I}$ is defined such that
\begin{equation}\label{eqn:op-invasion}
    I = \mathcal{I}(P;\, I_n)
\end{equation}
updates \(I_n\) with every throat whose entry pressure is exceeded in the PNM state \(P\).

Bubble tracker and PNM are coupled. The solution of the time-step integration can be written as a fixed point of the composed operators,
\begin{equation}\label{eqn:fp-condition}
    B_{n+1} = \mathcal{B}\big(\mathcal{P}(B_{n+1};\,I_{n+1});\,D,I_{n+1}\big), \qquad P_{n+1} = \mathcal{P}(B_{n+1};\,I_{n+1}) \, ,
\end{equation}
and the consistent events and invasions of the step,
\begin{equation}\label{eqn:fp-consistency}
    D = \mathcal{D}(B_{n+1}, P_{n+1};\,I_{n+1}), \qquad I_{n+1} = \mathcal{I}(P_{n+1}) \, .
\end{equation}
\autoref{eqn:fp-condition} and \autoref{eqn:fp-consistency} are enforced by two nested iterations: an inner fixed-point iteration on \(B\) with \((D,I)\) frozen, and an outer iteration on \((D,I)\), cf.~\autoref{fig:full-algorithm}.

\paragraph{Inner iteration.}
With \(D^{(k-1)}\) and \(I^{(k-1)}\) fixed, the inner iteration \(m = 1, 2, \dots\) (of outer iteration \(k\), see below) reads
\begin{equation}\label{eqn:fp-inner}
    B^{(k,m)} = \mathcal{B}\big(\mathcal{P}(B^{(k,m-1)};\,I^{(k-1)});\,D^{(k-1)},I^{(k-1)}\big), \qquad B^{(k,0)} = B_n \, ,
\end{equation}
so that every sweep restarts the tracker from \(B_n\) and moves the bubbles with the network state of the previous sweep. The iteration is stopped when

\begin{subequations}\label{eqn:fp-tol}
\begin{align}
    &\big\| B^{(k,m)} - B^{(k,m-1)} \big\| < \epsilon_B, \qquad
    \| B - B' \| := \max_{B} \,|\boldsymbol x_B - \boldsymbol x_B'| \, , \qquad \text{and} \label{eqn:fp-tol-a}\\
    &\big\| P^{(k,m)} - P^{(k,m-1)} \big\| < \epsilon_P, \qquad
    \| P - P' \| := \max_{ij} \,|Q_{ij} - Q_{ij}'| \, ,
\end{align}
\end{subequations}

where \(\boldsymbol x_b\) is the position of bubble \(b\) in the network and \(\epsilon_B = 10^{-9}~\mathrm{m}\), and \(Q_{ij}\) is the flux in throat $ij$ and \(\epsilon_P = 10^{-15}~\mathrm{m^3/s}\). Only the positions and fluxes enter the norm; within the inner iteration, the bubble volumes are fixed by the injection rate and by \(D^{(k-1)}\). 

Because a bubble either does or does not cross into the next throat within \(\Delta t\), the right-hand side of \autoref{eqn:fp-inner} is discontinuous in \(B\), and the iterations can oscillate between two states in which the crossing alternately occurs and does not occur. The two states are accepted as equivalent: the alternating pattern is detected with the same tolerance, \(\| B^{(k,m)} - B^{(k,m-2)} \| < \epsilon_B\) and \(\| B^{(k,m-1)} - B^{(k,m-3)} \| < \epsilon_B\), and the last iterate is then accepted as the solution.

The converged inner state is denoted \(B^{(k,\star)}\), with \(P^{(k,\star)} = \mathcal{P}(B^{(k,\star)};\,I^{(k-1)})\). The pore network state $P$ is therefore determined by the position of the bubbles at the end of the time step, regardless of how they moved during the time step.

\paragraph{Outer iteration.}
The event-predictor and invasion update of iteration $k$ is given as

\begin{equation}
    D^{(k)} = \mathcal{D}\left(P^{(k,\star)}, B^{(k,\star)};\, D^{(k-1)},\, I^{(k-1)}\right) , \qquad I^{(k)} = \mathcal{I}\left(P^{(k,\star)}\right)\,,
\end{equation}

and the outer iteration $k=1,2,\dots$ is accepted when the events and invasion of iteration $k$ is the same as in iteration $k-1$, i.e. \(D^{(k)} = D^{(k-1)}\) and \(I^{(k)} = I^{(k-1)}\).

\paragraph{Adaptive time-step control.}
The time-step will be restarted from the initial states $P_n$ and $B_n$, with reduced $\Delta t_n$, if either the inner or outer fixed-point iteration does not converge within $10$ iterations. A Newton failure can also reduce $\Delta t_n$. A minimum time-step size of $\Delta t_{\min}=10^{-9}~\mathrm{s}$ is set such that the simulation is cancelled if the solver has not yet converged.

\subsection{Spawning and bubble growth}
A new bubble forms in the source pore if no bubbles are currently present. If, instead, a non-detached bubble exists in a source pore, the gas source is added to the existing bubble, allowing it to grow until it meets the detachment criterion, \autoref{eqn:detachment_radius}.
%If there is still time left for injection within the current time step, then a new bubble will be formed. 
If there is a continuous non-wetting phase in the source pore, then the gas source is added directly to it. We assume the gas within the bubbles has a fixed density, consistent with our treatment of the non-wetting phase.

\subsection{Bubble movement}
Bubbles move along the pore throats. Our algorithm assigns bubbles to one specific throat at a given time. The bubble velocity is the sum of the velocity contributions due to buoyancy and advection. The buoyancy velocity \(u_{\text{buoy},ij}\) is determined by the effective gravitational strength $\Gamma$, and the advection velocity \(u_{\text{adv},ij}\) is determined by the liquid velocity in the throat $u_{w,ij}$. These two are discussed in \autoref{sec:Stokes_models}.

When a bubble reaches the end of a pore throat and enters the pore body, the next throat is selected according to the minimum resistance to the bubble. This resistance is determined according to the gravitational field and the advective flow field. If buoyancy is the only driving force in the network, meaning advective fluxes are negligible, then the selected throat is the one most closely aligned with the gravitational field. 
If the bubble is larger than the throat's inscribed radius, it must deform to enter the throat. Entry occurs when the pressure forcing on the bubble exceeds the increase in capillary pressure associated with the bubble deformation, $\Delta p_c$, 
\begin{equation}\label{eqn:entry_condition}
    \Delta \rho \, \|\boldsymbol a_{\mathrm{cf}}\| \, l_b + \Delta p_{b,ij} \geq \Delta p_c \,, \quad \text{where}\quad \Delta p_c = 2\,\sigma_w \left(\frac{1}{r_{ij}} - \frac{1}{r_b}\right)\,,
\end{equation}
in which \(\Delta \rho \, \|\boldsymbol a_{\mathrm{cf}}\| \, l_b\) is the pressure forcing due to buoyancy effects with bubble length $l_b$ in the direction of $\boldsymbol a_{\mathrm{cf}}$, and \(\Delta p_{b}\) is the pressure difference across the bubble. In practice, we use the throat length $l_{ij}$ as a surrogate for $l_b$ and the pressure difference in the throat $\Delta p_{ij}$ as a surrogate for $\Delta p_{b,ij}$.
If the entry condition is not met, the bubble cannot enter the throat. In our model, we transfer such an immobilised bubble into the continuous gas phase, assuming remobilisation is unlikely. The non-wetting saturation added to the pore can then invade the neighbouring pores as usual. 

\paragraph{Bubble merging} Bubble merging occurs at the end of each time step if two bubbles are in direct contact, even if the bubbles are in different throats. The merged bubble is then placed at the volume-weighted centre of the two merging bubbles in the appropriate throat. Bubbles that overtake one another within a throat during a time step also merge, with overtaking
being detected by comparing the order of the bubbles at the start and end of the step.

\paragraph{Transfer to continuum phase} A bubble is deleted and its mass transferred into the continuous gas phase if the bubble no longer fits within the pore without deforming, i.e., the bubble radius is bigger than the pore inscribed radius \(r_b \geq r_p\), or if the bubble ends up in a pore body or throat that is invaded by a continuous gas phase. Moving the mass from a bubble to the non-wetting phase is currently non-reversible.

\section{Model verification}\label{sec:verification}
This section describes the simulation setup, the boundary conditions, convergence criteria, and numerical methods, as well as eight verification tests.

As discussed in \autoref{sec:gravity}, we assume in the following that $\boldsymbol a_{\text{cf}}$ is uniform throughout the network and perpendicular to the bottom layer of pores.
%$\|\boldsymbol a_{\text{cf}}\|$ is then quantified by the effective gravitational strength $\Gamma$, caused by rotation, as a multiple of \(g_0\).
%
The pores at the bottom of the pore network represent the catalytic layer of the PEM electrolyser, where the reaction \(2~\mathrm{H_2O} \rightarrow \mathrm{O_2} + 4 ~\mathrm{H}^+ + 4 ~\mathrm{e}^-\) occurs. As protons leave through the proton-exchange membrane and electrons through an external circuit, this results in an effective oxygen gas source at the bottom boundary of our network that models the diffusion layer between the catalytic layer and the flow channel at the top of the domain. The total gas source \(\dot m\) is specified in units $\mathrm{kg/s}$, while a volumetric source term $q_{b}=q_{n}=\dot m/V_{\text{src}}$ is calculated based on the total volume of the source pores $V_{\text{src}}$. $q_{b}$ is then added to each source pore, such that each pore source is proportional to its size. For each $\mathrm{kg}$ of created oxygen in the source pores, a proportional amount of water must be consumed, given by the specific water consumption \(c_{w}\) of the system, such that $q_w=-q_{b}/c_w$. 
The pores at the top of the network border the electrolyser flow channel, a free-flow domain that supplies the PTL with water.

\subsection{Initial and boundary conditions}
\begin{figure}[pos=ht!,width=\linewidth]
    \centering
    \begin{subfigure}[t]{0.48\linewidth}
        \centering
        \includegraphics[width=\linewidth]{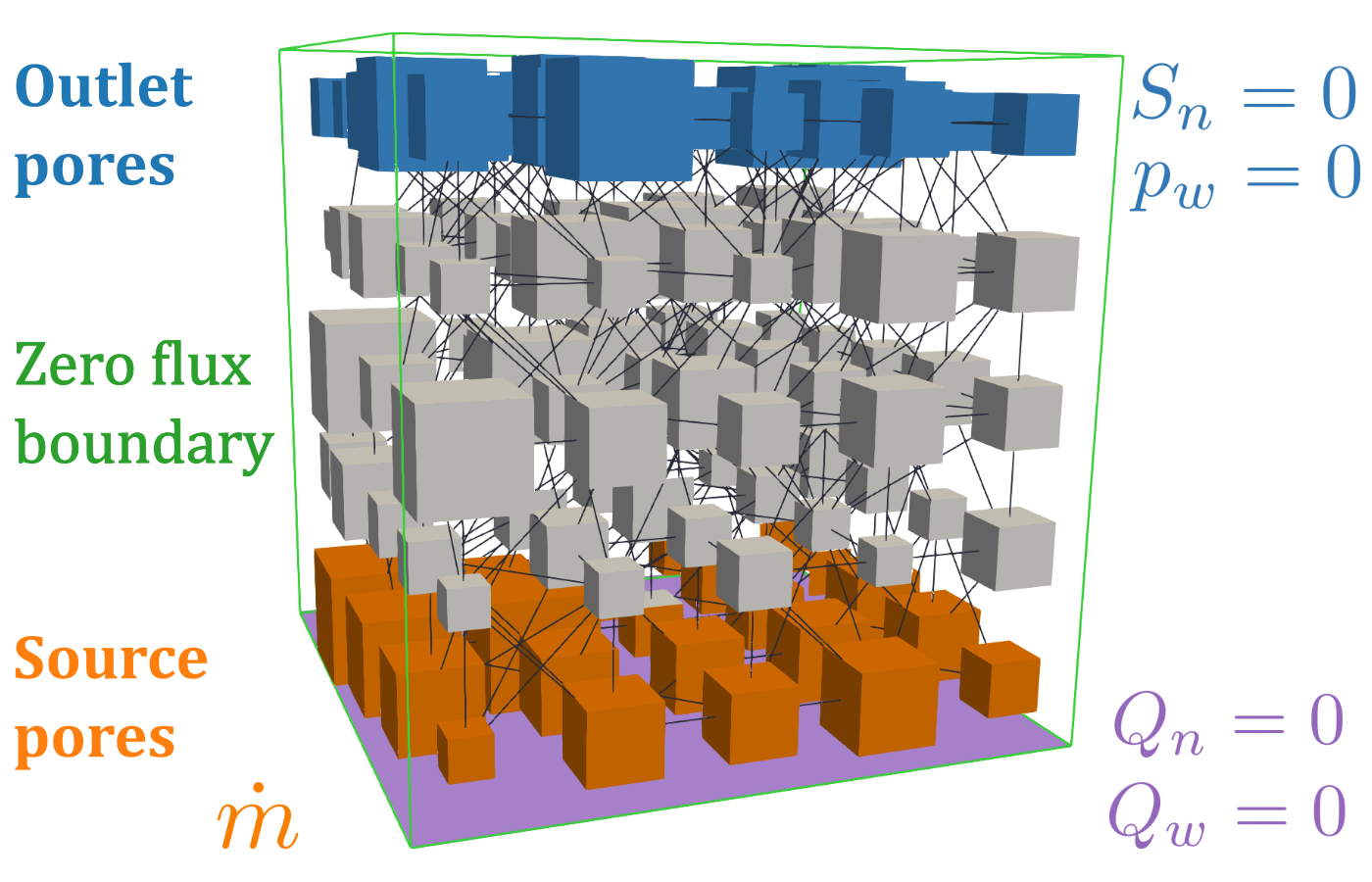}
        \caption{Closed system.}
    \end{subfigure}
    \hfill
    \begin{subfigure}[t]{0.48\linewidth}
        \centering
        \includegraphics[width=\linewidth]{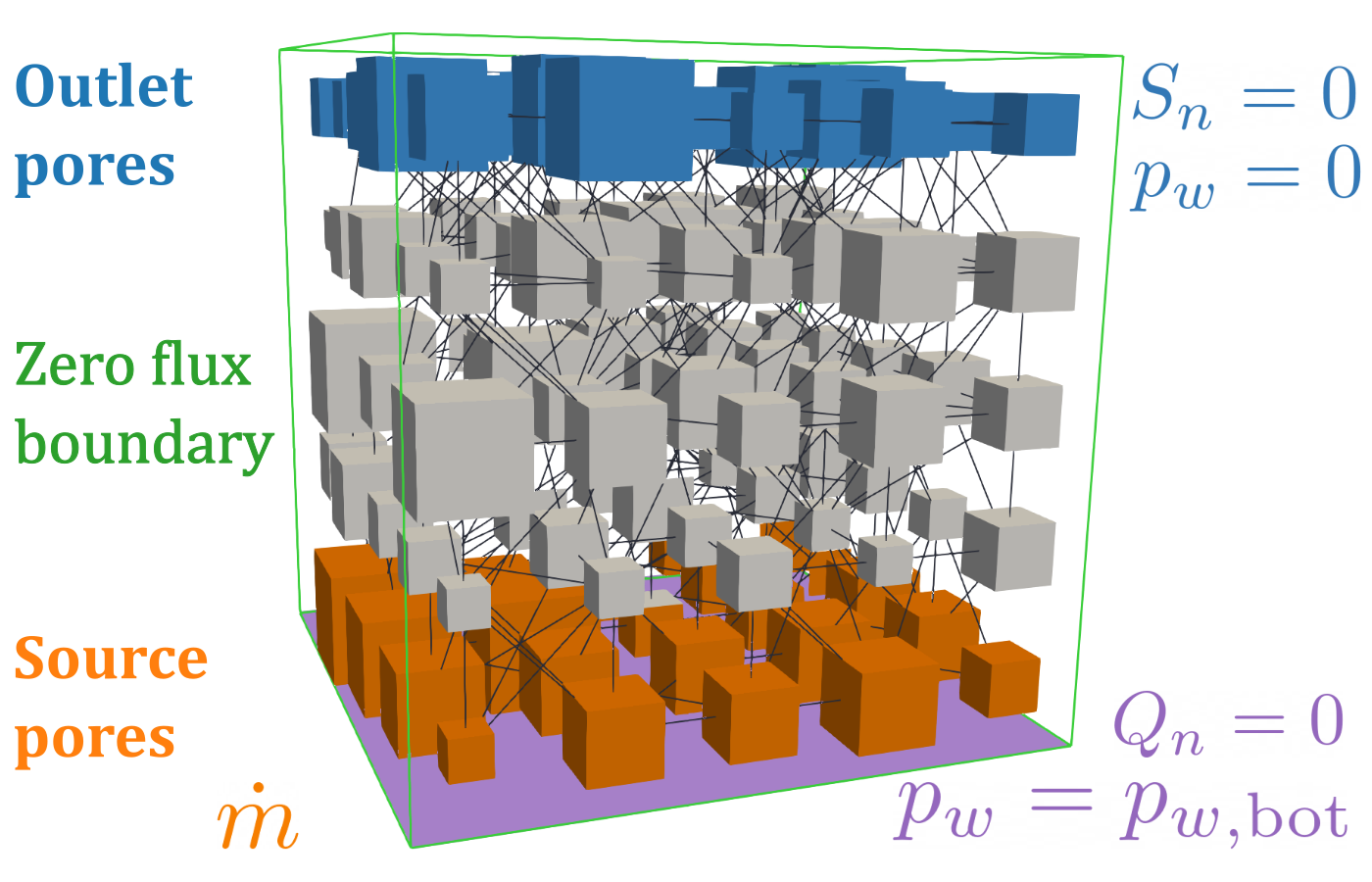}
        \caption{Open system.}
    \end{subfigure}
    \caption{Boundary conditions of the hybrid PNM, illustrated on a $5\times5\times5$ pore network. The source pores (bottom, orange) contain a gas mass source $\dot m$ and a corresponding liquid mass sink, and can be modelled either zero wetting phase flux (left) or fixed wetting phase pressure (right) boundary. The outlet pores (top, blue) has a fixed non-wetting saturation and wetting pressure, such that gas is removed when it reaches these pores. The remaining boundaries (green) are no-flow boundaries. Pore bodies are represented by boxes and throats by black lines.}
    \label{fig:boundary-conditions}
\end{figure}

The system is initially fully water-saturated in every pore, \(S_{w,i}=1\), with water the wetting phase and oxygen the non-wetting phase. \autoref{fig:boundary-conditions} illustrates the two sets of boundary conditions used subsequently. %The boundary conditions are as follows.

\paragraph{Top boundary (outlet pores).}
At the top boundary we strongly enforce the non-wetting saturation \(S_{n,top}=0\) and wetting-phase pressure \(p_{w,\mathrm{top}}=0~\mathrm{Pa}\), relative to the reference pressure $p_{\text{ref}}=10^{5}~\mathrm{Pa}$. Any gas entering the pores at the top boundary is therefore removed from the system and immediately replaced by water, mimicking the free-flow channel of the electrolyser that borders the top layer of pores.

\paragraph{Bottom boundary (source pores).}
We set the bottom boundary conditions in two ways, depending on the system we want to model. If we model an open system, driven by a prescribed pressure difference between the top and bottom of the domain, we enforce strongly \(p_{w,\mathrm{bot}}=\Delta p + \rho_w \boldsymbol a_{\mathrm{cf}}\, h\), where \(\rho_w \|\boldsymbol a_{\mathrm{cf}}\| h\) is the hydrostatic pressure, and no-flux for the non-wetting phase. For a closed system with a reactive boundary layer, we enforce no-flux for both phases. In both cases, the saturation is therefore free to evolve. We implement the non-wetting phase source terms and the wetting phase sink terms due to the reaction as volumetric sources and sinks in the source pores.

\paragraph{Side boundaries.}
The remaining boundaries are no-flow boundaries modelling symmetry.

\subsection{Verification}
\begin{figure*}[pos=ht!]
    \centering
    \begin{subfigure}[t]{0.18\linewidth}
        \centering
        \includegraphics[height=0.25\textheight]{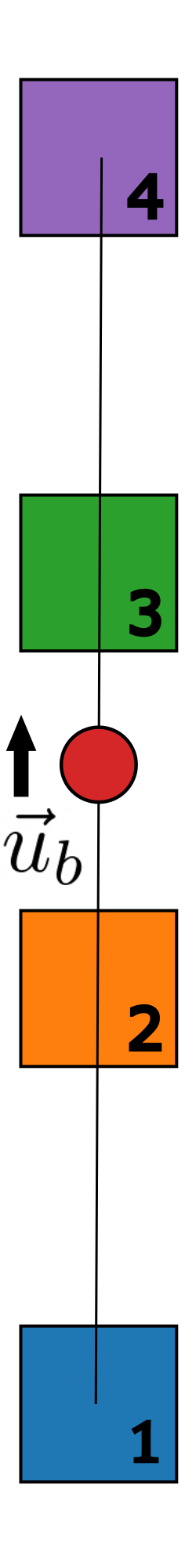}
        \caption{Chain network.}
        \label{fig:pnm-column}
    \end{subfigure}
    \hfill
    \begin{subfigure}[t]{0.25\linewidth}
        \centering
        \includegraphics[height=0.25\textheight]{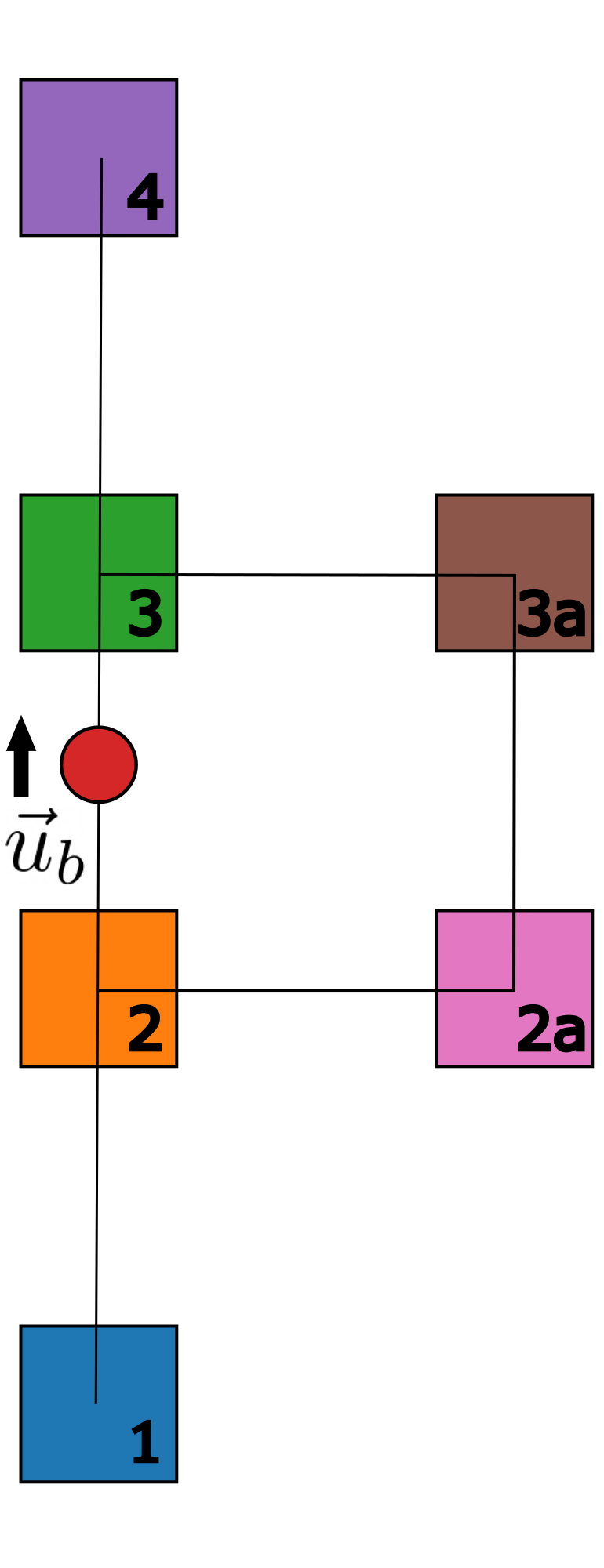}
        \caption{Loop network.}
        \label{fig:pnm-loop}
    \end{subfigure}
    \hfill
    \begin{subfigure}[t]{0.25\linewidth}
        \centering
        \includegraphics[trim=0 50mm 0 10mm,clip,height=0.25\textheight]{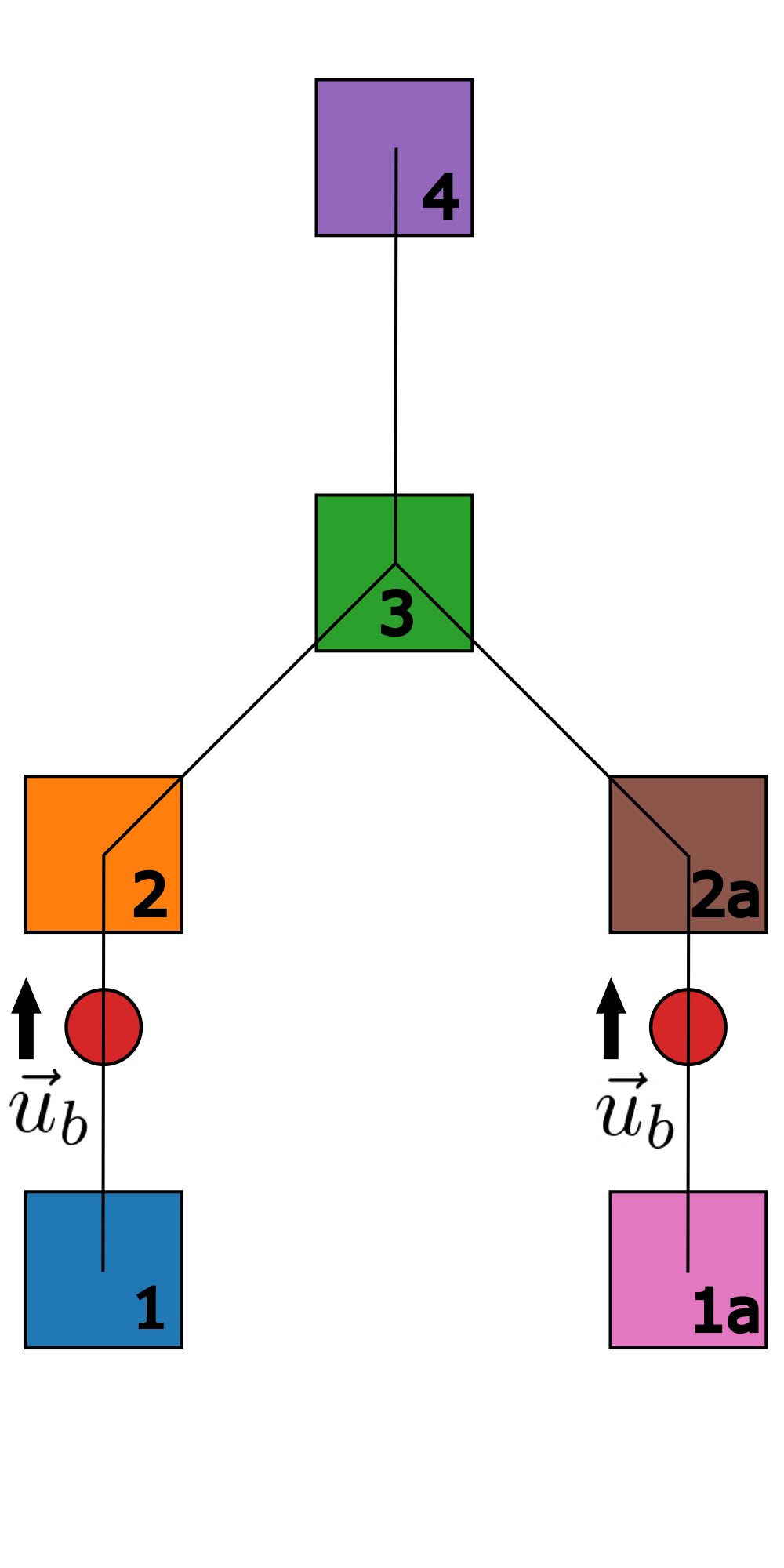}
        \caption{Fork network.}
        \label{fig:pnm-fork}
    \end{subfigure}
    \hfill
    \begin{subfigure}[t]{0.25\linewidth}
        \centering
        \includegraphics[height=0.25\textheight]{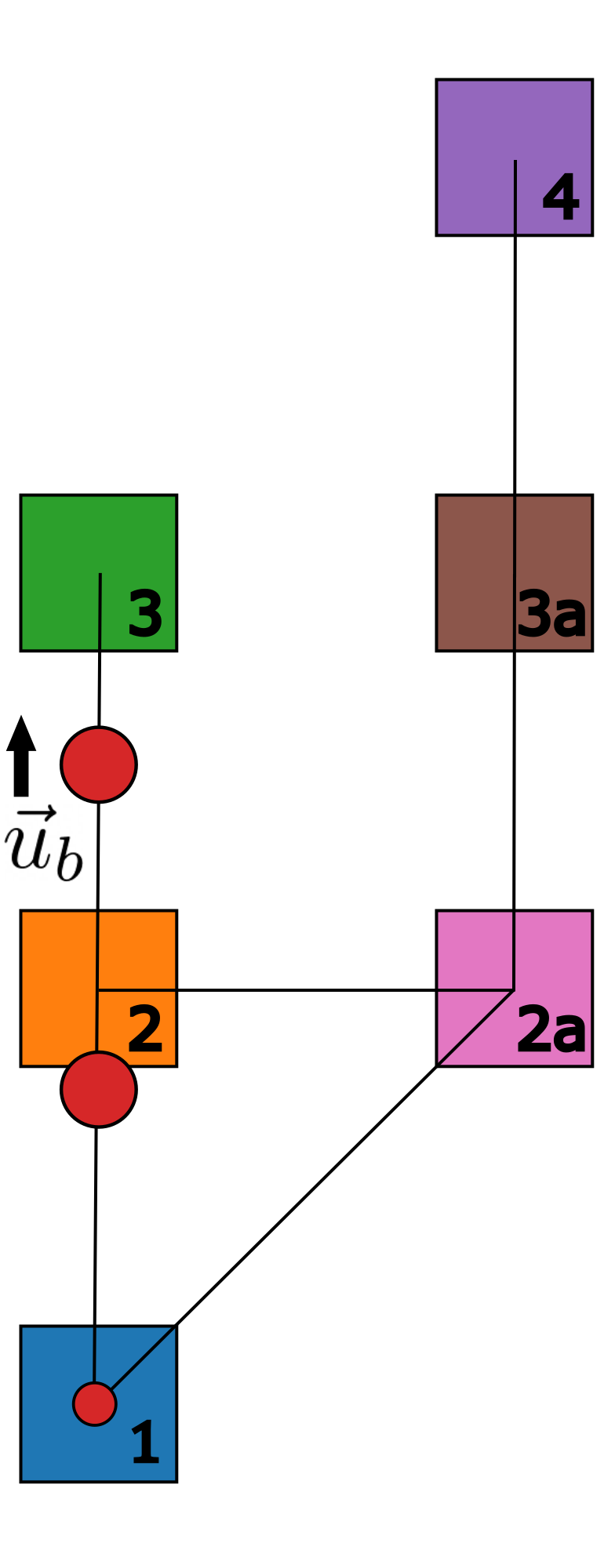}
        \caption{Trap network.}
        \label{fig:pnm-trap}
    \end{subfigure}
    \caption{
        Selection of pore networks used for verification of the model.
        Network (a) is used for verifying bubble volume tracking,
        bubble-induced wetting-phase flux, bubble growth and detachment,
        and bubble rising due to buoyancy and advection.
        Network (b) is used to verify the effects of the viscous pressure drop.
        Network (c) is used for testing merging of bubbles.
        Network (d) is used for testing bubble trapping in the network.
    }
    \label{fig:test-illustrations}
\end{figure*}

We construct eight tests to verify the implementation of the local rules for bubble--PNM interactions. We use small networks in which the expected balances and state transitions can be computed analytically. Tests 1-4 provide quantitative comparisons using analytical relations for pore volume, flux, pressure, and velocity. Tests 5-8 verify discrete event handling, including merging, repeated creation, trapping, phase transfer, and motion through throats of different radii. The main model and network parameters are summarised in \autoref{tab:verification_setup}; the corresponding network configurations are given in \autoref{fig:test-illustrations}. An additional discussion of the cases and simulation results in each case is presented in Section S4 of the Supplementary Material.

\paragraph{Test 1 - Local flux conservation.}
A single bubble, with a radius of \(50~\mathrm{\mu m}\), is spawned in a chain of pores (\autoref{fig:pnm-column}), and given a fixed prescribed velocity of \(0.05~\mathrm{m/s}\).
This case verifies the correctness of the flux and storage modifications induced by a bubble changing pores. We also check bubble volume conservation. The flux occurs within the single time step in which the bubble changes pores. Between these steps, the network state and bubble velocity remain constant.
We find that the measured fluxes induced by bubble movement are exact to machine precision.

\paragraph{Test 2 - Bubble growth and detachment.}
Next, we verify bubble growth and detachment in the source pores. We add a mass source of oxygen to the bottom pore, with a proportional water mass sink. During the growth stage, we expect a volumetric source term away from the source pores due to the density differences of the liquid and gas. We conduct the test on a simple network, as depicted in \autoref{fig:pnm-column}. At \(\Gamma=50\), we fix the detachment radius to \(50 ~\mathrm{\mu m}\) to get an artificially large bubble.
We find that the measured fluxes induced by bubble movement are exact to machine precision.

\paragraph{Test 3 - Bubble advection.}
The third test verifies the impact of throat conductance due to a bubble moving by advection. The same setup is used as in Test 2, but now there is an additional pressure delta of \(10 ~\mathrm{Pa}\) between the bottom pore (1) and top pore (4) such that a steady flux of water is flowing through the system, while gravity is disabled. The detachment radius is set to \(65 ~\mathrm{\mu m}\) to detach an artificially large bubble.
We find that the measured fluxes induced by bubble movement and in the pressure are exact to machine precision.

\paragraph{Test 4 - Viscous pressure drop.} We add a bypass branch to form a loop network, \autoref{fig:pnm-loop}. A buoyancy-driven bubble leads to an additional pressure drop in its current pore throat. This pressure redistributes the liquid through the bypass branch, slowing the bubble as it moves through the bypassed throat. Test 4 verifies the correct coupled pressure-flow response in a loop network. The agreement with the analytical loop solution tests the non-trivial reciprocal coupling used by the network implementation. The effective gravitational strength is \(\Gamma =50\) and the bubble is allowed to grow until it reaches the detachment radius given by \autoref{eqn:detachment_radius}.
We find that the induced pressure-drop magnitudes and the bubble-velocity changes are exact to machine precision under the closure laws.

\paragraph{Test 5 - Bubble merging.}
The fifth test verifies the correctness of the implementation of the bubble merging rule and tests bubble velocity in tilted channels. A forked network, as depicted in \autoref{fig:pnm-fork}, is used. Bubbles are forced to move through diagonal throats and merge upon contact. The bubbles are driven by an effective gravitational strength \(\Gamma=50\) and the bubbles are allowed to grow until the detachment radius given by \autoref{eqn:detachment_radius}.
We find that the bubble volumes are conserved and the bubble velocities match the expected solution to machine precision.
%The deviations of the bubble volumes in the pores are smaller than \(2\cdot10^{-14}~\%\), and the bubble velocity deviations are smaller than \(3 \cdot 10^{-13}~\%\).

\paragraph{Test 6 - Continuous bubble creation.}
The sixth test verifies the detaching and merging behaviour when bubbles are created continuously. We perform the test on a simple network, as depicted in \autoref{fig:pnm-column}. The bubbles are driven by an effective gravitational strength \(\Gamma = 20\) and bubbles are allowed to grow to the detachment size specified by \autoref{eqn:detachment_radius}.
The bubble volumes are conserved to machine precision.
%The deviations in bubble volumes within the pores are less than \(7\cdot10^{-13}~\%\). 

\paragraph{Test 7 - Bubble trapping.}
The seventh test verifies that bubbles get trapped as intended and then get transferred to the non-wetting phase, conserving mass. We use a network with a dead end that traps bubbles, \autoref{fig:pnm-trap}. The bubbles are buoyancy-driven at \(\Gamma = 50\) and grow until they reach the detachment radius given by \autoref{eqn:detachment_radius}.
We find that the bubble volumes are conserved to machine precision.
%The deviations in bubble volumes in the pores are less than \(4\cdot10^{-13}~\%\).

\paragraph{Test 8 - Throat narrowing.}
We test that bubble velocity changes with the throat inscribed radius. In a simple chain network (\autoref{fig:pnm-column}), throat radii are set to consecutively decrease from bottom to top: \(70~\mathrm{\mu m},\) to \( 60~\mathrm{\mu m}\) to \( 50~\mathrm{\mu m}\). At \(\Gamma = 20\), we impose a potential difference of \(10~\mathrm{Pa}\) across the network. The bubble detachment radius is set to \(60~\mathrm{\mu m}\) to test large bubbles.
We find that the bubble volumes are conserved and that fluxes match to machine precision. Note that due to the discrete nature of the network, the velocities in each individual throat are constant rather than varying over the length.
%The deviations of the bubble volumes in the pores are smaller than \(3\cdot10^{-14}~\%\), and the flux deviation is smaller than \(2\cdot 10^{-12}~\%\).

\begin{table}[pos=ht!]
    \centering
    \caption{Simulation parameters used for the verification tests.}
    \label{tab:verification_setup}
    \small
    \begin{tabular}{lcccccccc}
        \toprule
        Quantity& Test 1& Test 2& Test 3& Test 4& Test 5& Test 6& Test 7& Test 8 \\
        \midrule
        Network& Chain& Chain& Chain& Loop& Fork& Chain& Trap& Chain \\
        Pore radius [$\mathrm{\mu m}$]& 100& 100& 100& 100& 70& 100& 70& 100 \\
        Throat inscribed radius [$\mathrm{\mu m}$]& 70& 70& 70& 70& 70& 70& 70& $70,\ 60,\ 50$ \\
        Gas source [$\mathrm{kg/s}$]& --& $10^{-10}$& $10^{-10}$& $10^{-10}$& $10^{-10}$& $10^{-10}$& $10^{-10}$& $10^{-10}$ \\
        $c_w$ [-] & -- &0.888 &0.888 &0.888 &0.888 &0.888 &0.888 &0.888 \\
        G-level [$\Gamma$]& --& 50& --& 50& 50& 20& 50& 20 \\
        Pressure delta [$\mathrm{Pa}$]& --& --& 10& --& --& --& --& 10 \\
        Detachment radius [$\mathrm{\mu m}$]& --& 50& 65& Eq.\ref{eqn:detachment_radius}& Eq.\ref{eqn:detachment_radius}& Eq.\ref{eqn:detachment_radius}& Eq.\ref{eqn:detachment_radius}& 60 \\
    \bottomrule
    \end{tabular}
\end{table}

\section{Gas transport at high apparent gravitational forces}\label{sec:application}
\subsection{Simulation setup}
We use the developed hybrid PNM to investigate how higher apparent gravitational forces ($\Gamma \gg 1$), e.g. in a centrifuge, affect bubble flow through the porous medium. We use a network of \(5\times5\times5\) pores, where gas is generated at a constant rate \(q_{b,i}\) in each pore at the bottom boundary (here called source pores). Although the network is small, its height is comparable to that of a real electrolyser PTL. We fix the contact radius such that the detachment radius is given by \autoref{eqn:detachment_radius} and is determined entirely by the effective acceleration $\|\boldsymbol a_{\mathrm{cf}}\|$.

The hybrid PNM runs until a steady state is reached and maintained for one second of simulation time. Here, steady state means no invasion events in the preceding $0.5$ seconds and an outlet exit rate within \(2\%\) of the gas source rate. (The margin of error is needed as bubble gas flux fluctuates even in steady state since it's computed from discrete particles crossing the domain boundary.) Of the $341$ simulations run for the results in the following section, $17$ were discarded for not reaching a steady state within $10~\mathrm{s}$ of simulated time, and an additional $9$ failed to converge in either the outer coupling or the Newton solver.

\begin{table}[pos=ht!]
    \centering
    \caption{Parameters used to generate the $5\times5\times5$ and $10\times10\times10$ pore networks, as well as the simulation setups.}
    \label{tab:5x5_PNM}
    \begin{tabular}{lll}
        \toprule
        Quantity & Symbol & Value \\
        \midrule
        Domain size $5 \times 5 \times 5 $ & & $1.2 ~\mathrm{mm} \times 1.2 ~\mathrm{mm} \times 1.2 ~\mathrm{mm}$ \\
        \hphantom{Domain size} $10 \times 10 \times 10 $ & & $2.4 ~\mathrm{mm} \times 2.4 ~\mathrm{mm} \times 2.4 ~\mathrm{mm}$ \\        
        Pore geometry & & Cube \\
        Throat cross-section & & Square \\
        Mean pore inscribed radius & $\bar r_p$ & $10-100~\mathrm{\mu m}$ \\
        Standard deviation of $\bar r_p$ & $\sigma _p$ & $20~\mathrm{\mu m}$ \\
        $\bar r_p$ distribution & & Log-normal \\
        Pore/throat deletion probability & & $0.60$ \\
        Maximum time step range & $\Delta t_{\max}$ & $10^{-5}~\mathrm{s}\,-\,8\cdot 10^{-4}~\mathrm{s}$\\ 
        Specific water consumption & $c_{w}$ & $0.888~\mathrm{kg ~{O_2}/kg~{H_2O}}$ \\
        Oxygen mass source & $\dot m$ & $5\cdot 10^{-11}~\mathrm{kg/s} $ \\
        Contact radius & $r_c$ & $0.1~\mathrm{\mu m}$ \\
        G-level & $\Gamma$ & $1-300$ \\
        Pressure delta & $\Delta p$ & $0~\mathrm{Pa}$\\
        Detachment radius & $r_d$ & $10-100~\mathrm{\mu m}$ (\autoref{eqn:detachment_radius}) \\
        \bottomrule
    \end{tabular}
\end{table}

\subsection{Pore network geometry}
We generate the pore network by creating a structured grid from the specified domain size and number of pores. The vertices represent pore bodies, while connections form pore throats. These throats are randomly removed according to the specified deletion probability. If all throats connected to a pore are deleted, the pore is removed. We sample each pore inscribed radius from a log-normal distribution defined by the mean pore inscribed radius $\bar r_p$ and its standard deviation $\sigma_p$, capping excessively large radii to prevent neighbouring pores from overlapping. We then calculate each throat inscribed radius from the radii of its two adjacent pores and their centre-to-centre distance using the averaging relation proposed by \citet{joekar-niasarInsightsRelationshipsCapillary2008}. The parameters used to generate the networks are summarised in \autoref{tab:5x5_PNM}.

\subsection{Time-step independence}
The pore-network model can be sensitive to time-step size, as discrete events (invasion or bubble events) are not explicitly resolved within a time step \citep{wuNewFullyImplicit2024, schneiderImprovementFullyimplicitTwophase2025}. The Newton solver proposes the time step size based on convergence history up to an upper limit $\Delta t_{\max}$, and halves the time step when either the inner iteration, the outer iteration, or the Newton algorithm fails to converge. In this section, we propose an estimate $\Delta t_{\max}$, such that further refinement does not significantly impact the network.

Assuming that a typical bubble has a size set by the detachment radius and a velocity equal to the Stokes velocity of a rising spherical bubble in an infinite medium, we estimate the maximum time-step size as

\begin{equation}\label{eqn:max-dt}
    \Delta t_{\max}^\star = \frac{l_{t,min}}{2\,u_S(r_d)} \,,
\end{equation}

where \(l_{t,min}\) is the length of the shortest throat in the system, and $u_S(r_d)$ is the Stokes velocity of a bubble with the same size as the detachment radius $r_d$. A time step size smaller than $\Delta t_{\max}^\star$ ensures that a typical bubble does not cross more than half a throat length in a single time step, provided buoyancy is the dominant velocity component. (If advection plays a larger role in the overall bubble transport, then a different estimate might be needed for the maximum time step size. This is not the case in our simulations.)

Time-step independence is tested by varying the maximum time-step size \(\Delta t_{\max}\) between $10^{-5}~\mathrm{s}$ and $10^{-3}~\mathrm{s}$, for a single network realisation.
We set the mean pore radius to $\bar r_p=70~\mathrm{\mu m}$, and the bubble detachment radius $r_d$ is given by \autoref{eqn:detachment_radius} with $r_c=0.1~\mathrm{\mu m}$. We estimate the maximum time-step size for this network to be $\Delta t_{\max}^\star=1.2\cdot10^{-4}~\mathrm{s}$.

The test shows that refining $\Delta t_{\max}$ beyond the upper limit $\Delta t_{\max}^\star$ has little impact on the evolution and steady-state configuration. The network invasion patterns remain the same, and the gas exit ratio is within $1\%$ of the value obtained for the smallest tested value of $\Delta t_{\max}$. The mean number of time-step reductions is also less than $0.02~\%$, so nearly all of the time-step sizes equal the prescribed maximum size. Solver statistics for the time-step independence study are presented in Section S6, with the resulting invasion patterns given in Section S5, of the Supplementary Material.

\subsection{Model predictions for varying $\Gamma$}
How do invasion patterns and gas transport vary with $\Gamma$? In this section, the maximum time step size for the adaptive time stepping algorithm is set to \(\Delta t_{\max}=5\cdot 10^{-5}~\mathrm{s}\) such that \autoref{eqn:max-dt} is fulfilled for all runs. We set the mean pore radius to $\bar r_p=70~\mathrm{\mu m}$, and the bubble detachment radius $r_d$ is given by \autoref{eqn:detachment_radius} with $r_c=0.1~\mathrm{\mu m}$. We then test $\Gamma \in \lbrace 1,5,10,20,40,60,80,100,150,200,300 \rbrace$, each for $11$ random realisations of the $5\times5\times5$ pore network.
The remaining parameters for the simulation are given in \autoref{tab:5x5_PNM}. Each simulation is run until a steady state is reached.

\begin{figure}[pos=ht!]
    \centering
    \begin{subfigure}[t]{0.5\linewidth}
        \centering
        \includegraphics[width=\linewidth]{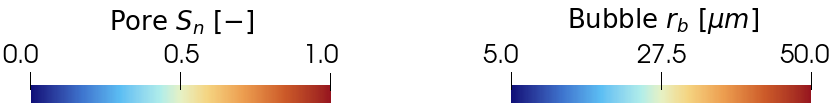}
    \end{subfigure}
    \par\vspace{0.1em}
    \begin{subfigure}[t]{0.32\linewidth}
        \centering
        \includegraphics[width=\linewidth]{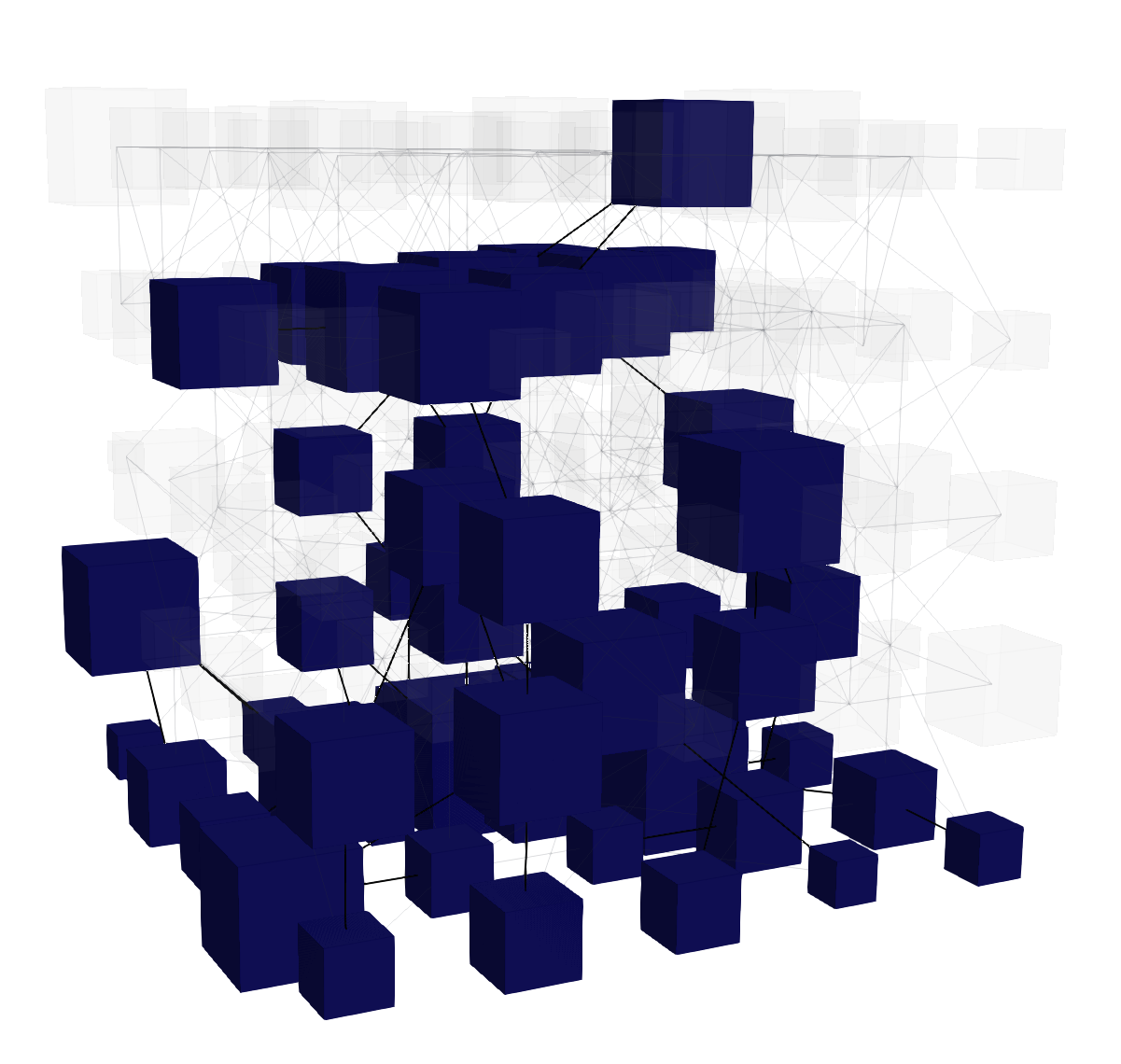}
        \caption{$\Gamma=1$}
    \end{subfigure}\hfill
    \begin{subfigure}[t]{0.32\linewidth}
        \centering
        \includegraphics[width=\linewidth]{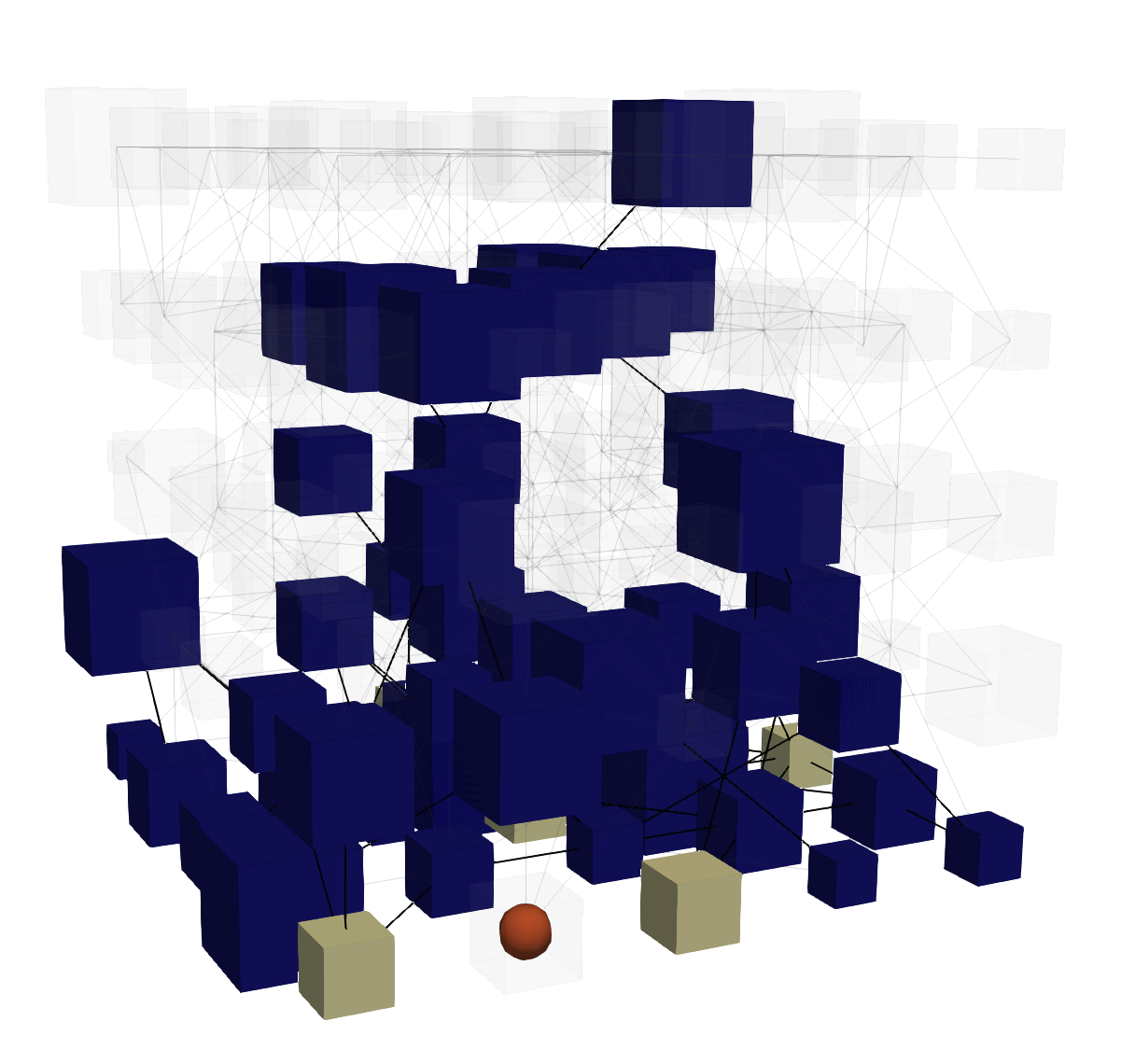}
        \caption{$\Gamma=10$}
    \end{subfigure}\hfill
    \begin{subfigure}[t]{0.32\linewidth}
        \centering
        \includegraphics[width=\linewidth]{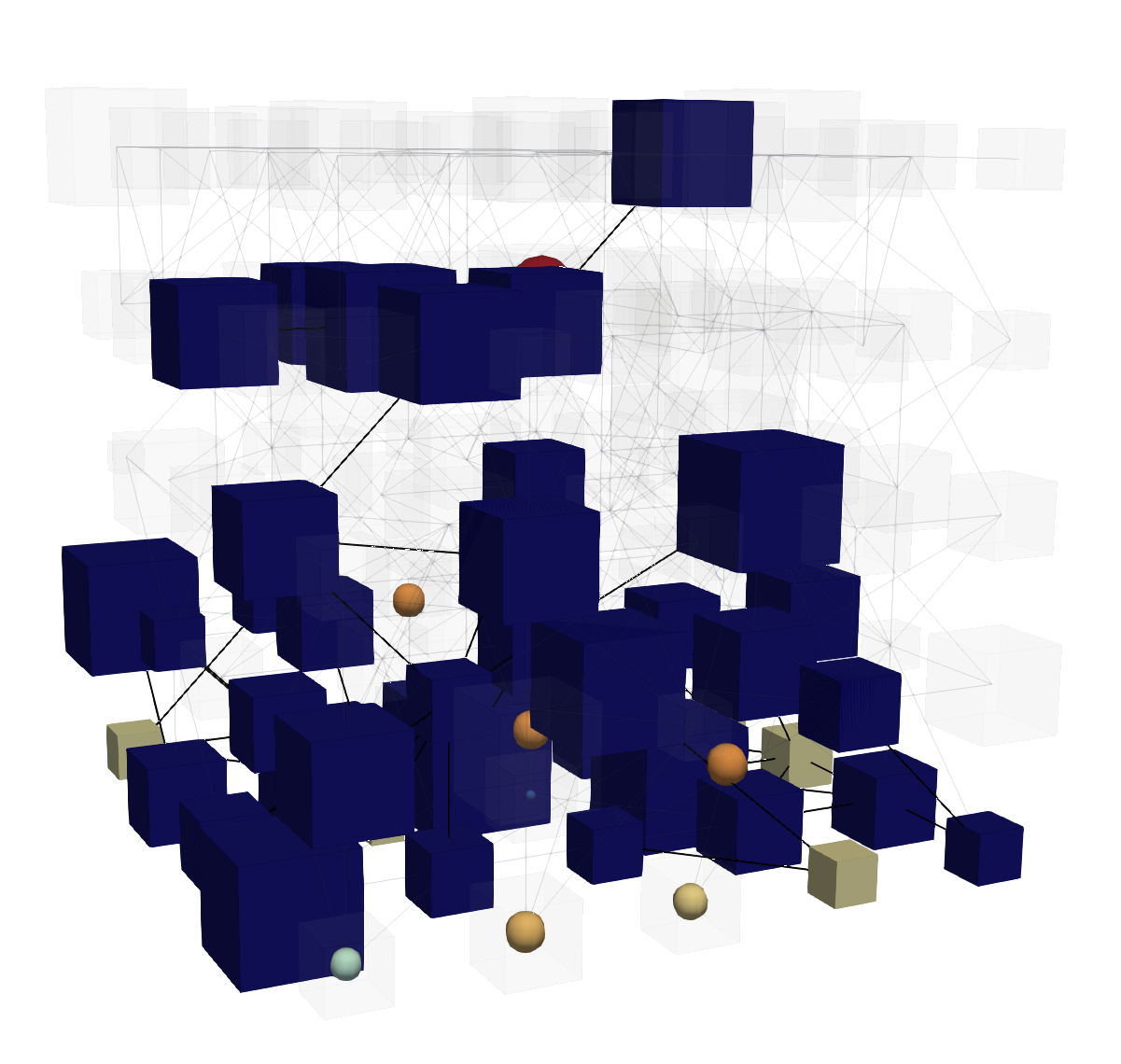}
        \caption{$\Gamma = 20$}
    \end{subfigure}
    \par\vspace{1em}
    \begin{subfigure}[t]{0.32\linewidth}
        \centering
        \includegraphics[width=\linewidth]{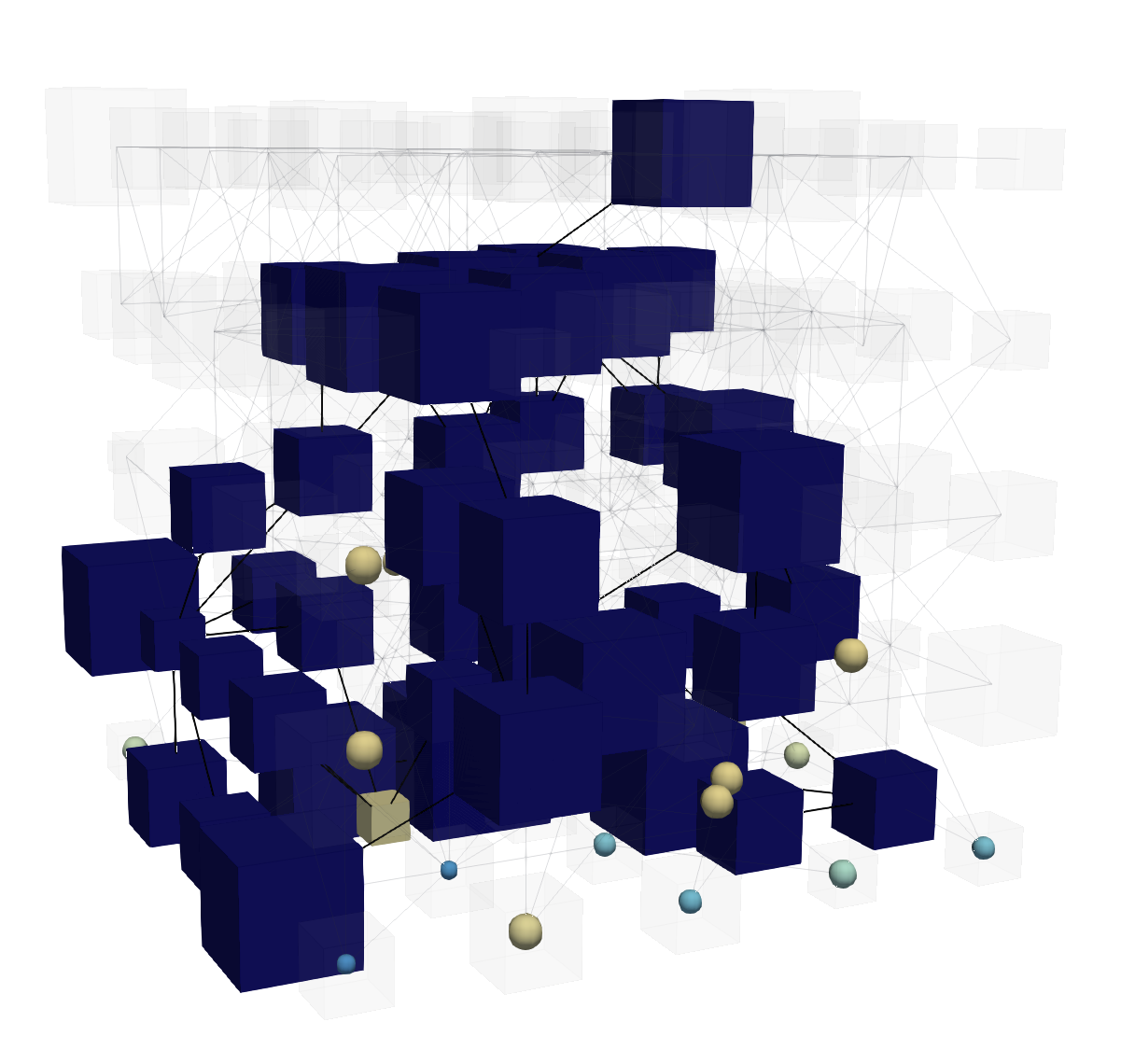}
        \caption{$\Gamma=40$}
    \end{subfigure}\hfill
    \begin{subfigure}[t]{0.32\linewidth}
        \centering
        \includegraphics[width=\linewidth]{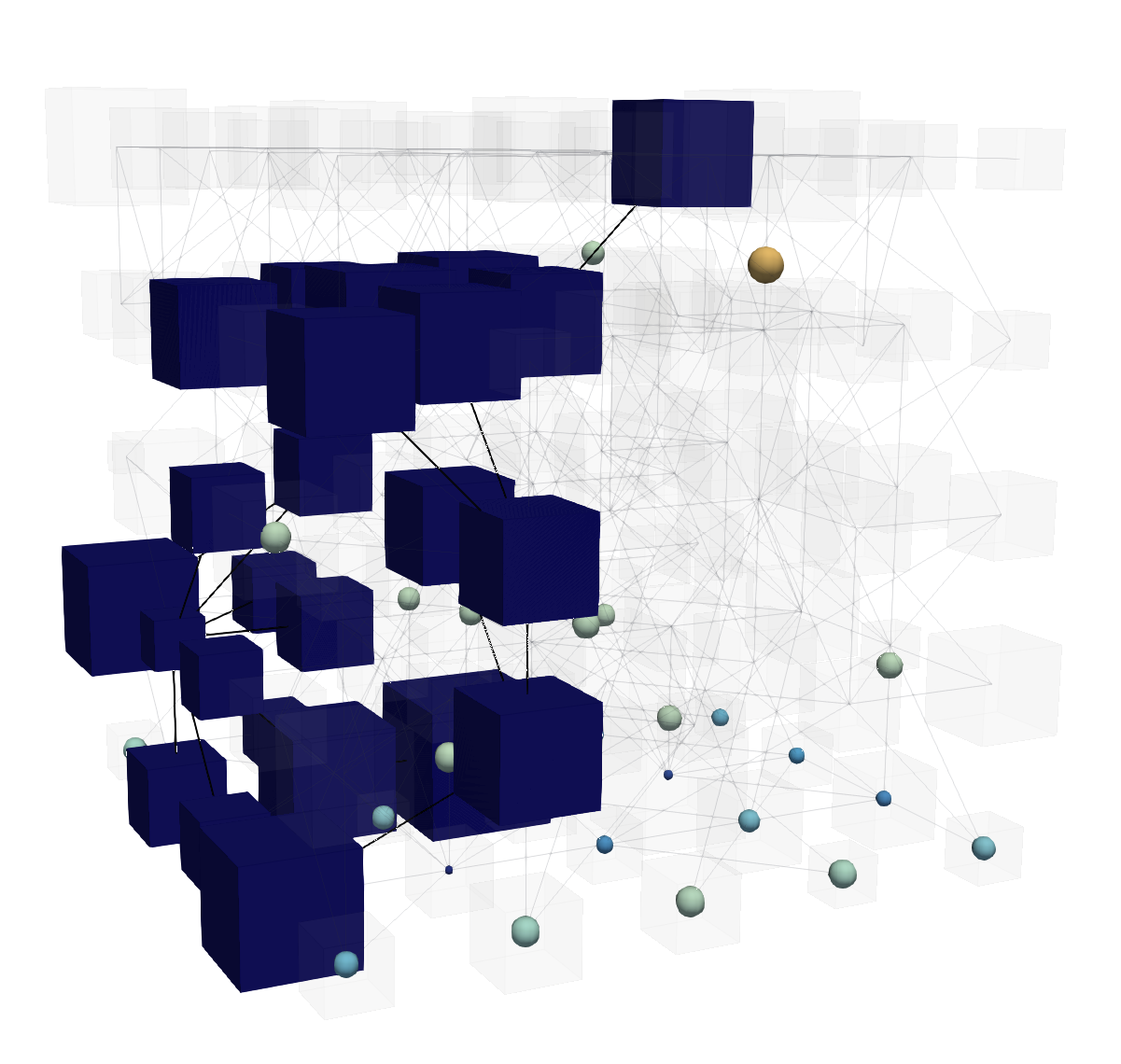}
        \caption{$\Gamma = 60$}
    \end{subfigure}\hfill
    \begin{subfigure}[t]{0.32\linewidth}
        \centering
        \includegraphics[width=\linewidth]{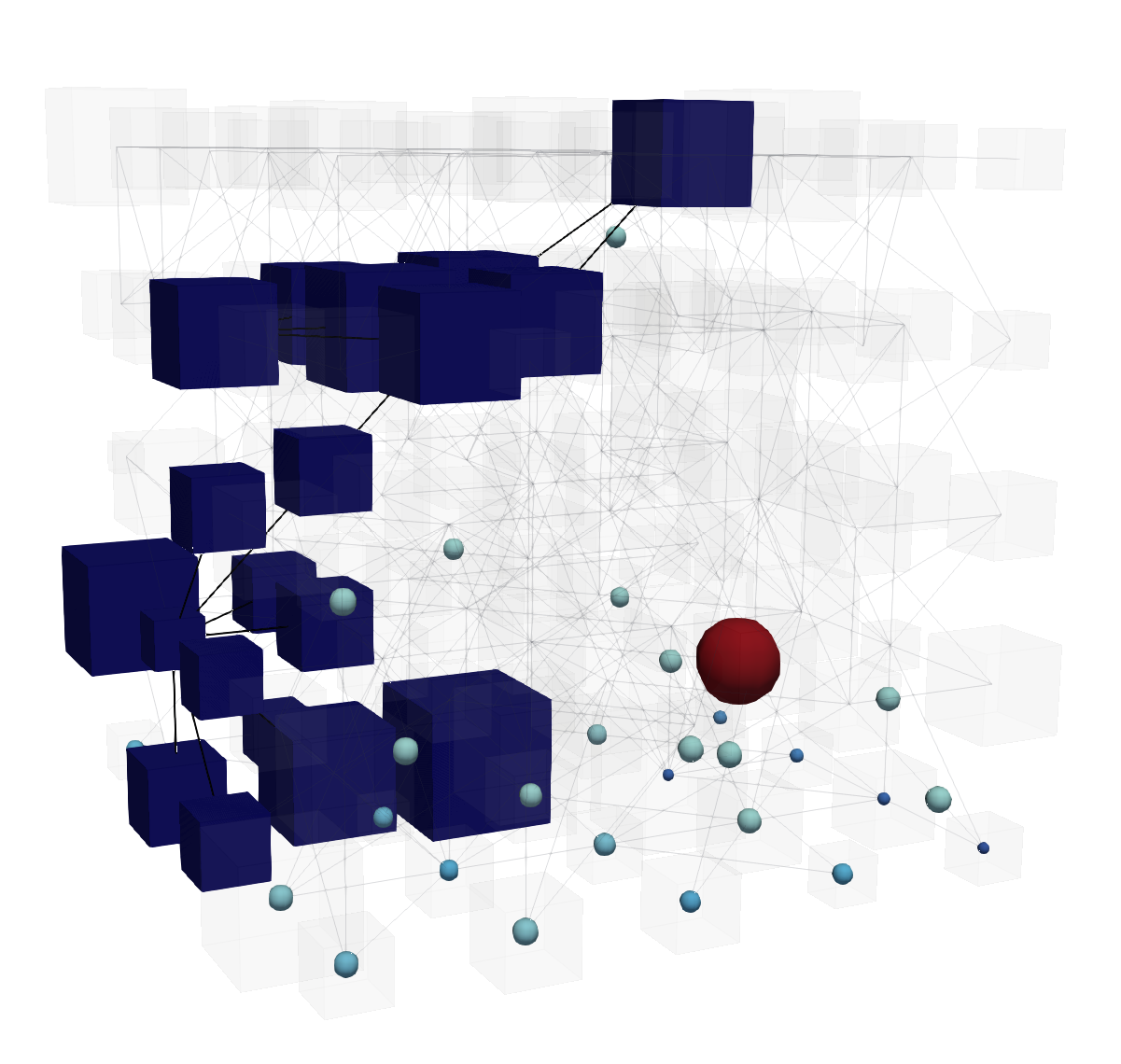}
        \caption{$\Gamma = 80$}
    \end{subfigure}
    \par\vspace{1em}
    \begin{subfigure}[t]{0.32\linewidth}
        \centering
        \includegraphics[width=\linewidth]{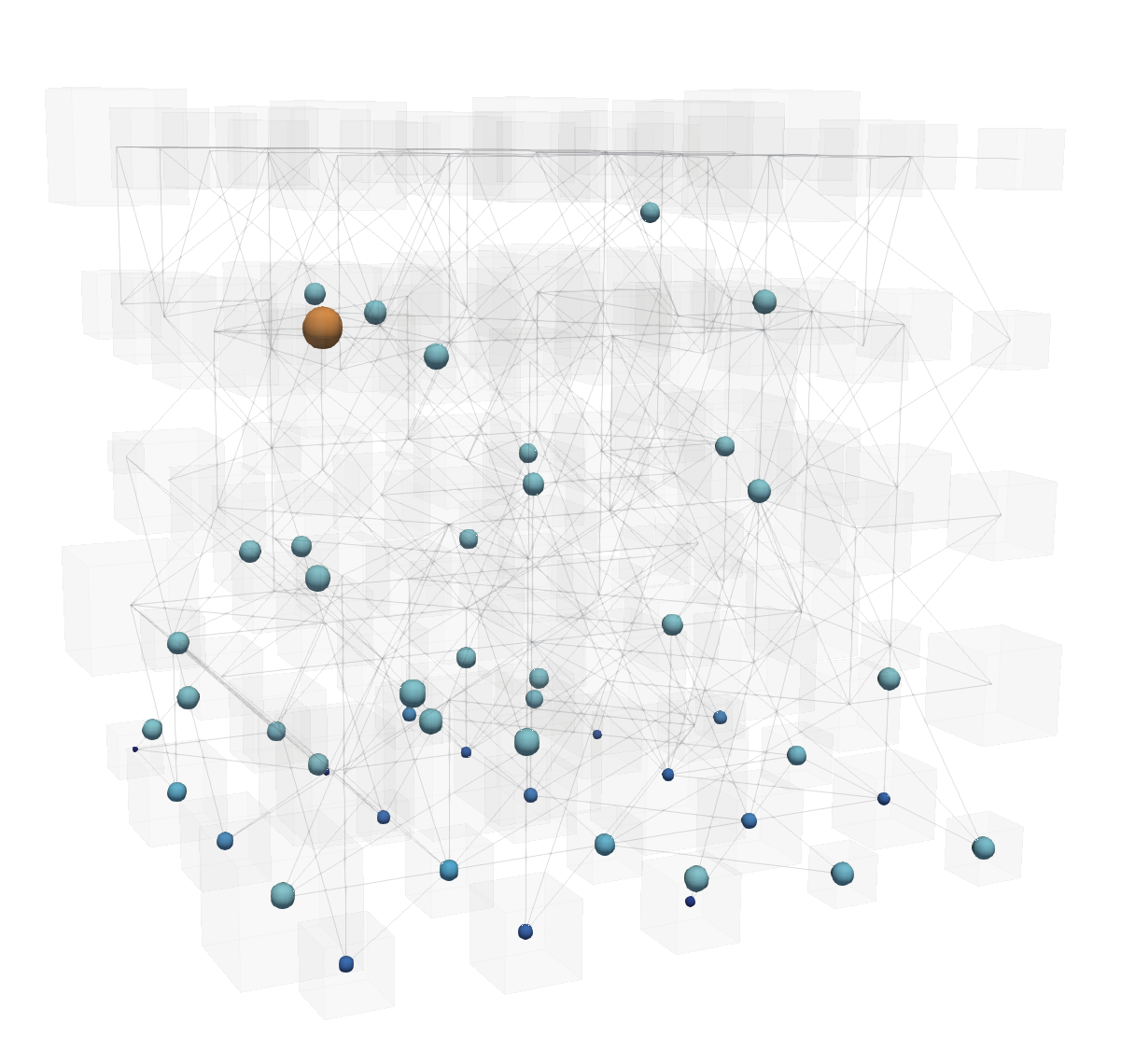}
        \caption{$\Gamma=100$}
    \end{subfigure}\hfill
    \begin{subfigure}[t]{0.32\linewidth}
        \centering
        \includegraphics[width=\linewidth]{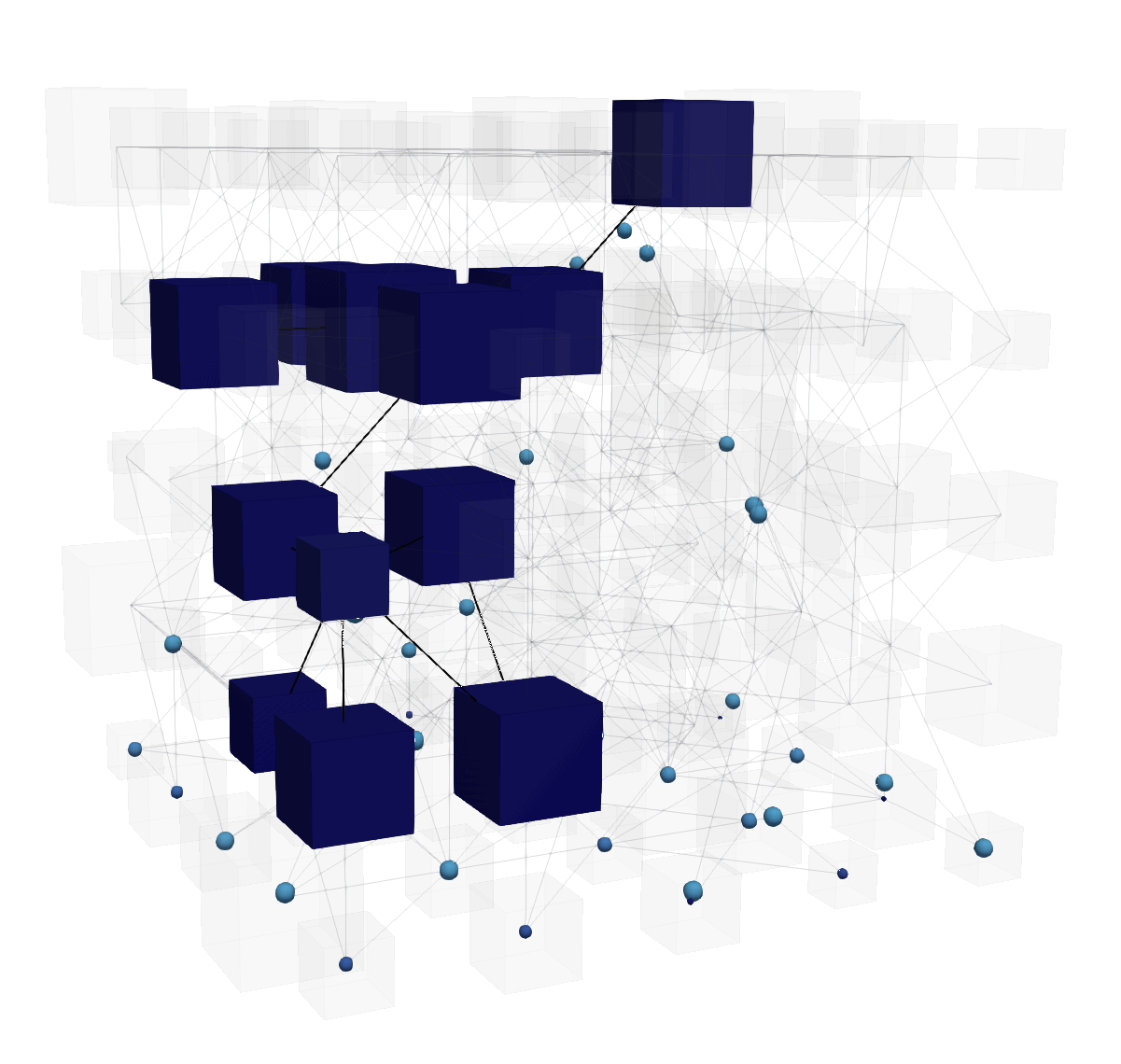}
        \caption{$\Gamma = 200$}
    \end{subfigure}\hfill
    \begin{subfigure}[t]{0.32\linewidth}
        \centering
        \includegraphics[width=\linewidth]{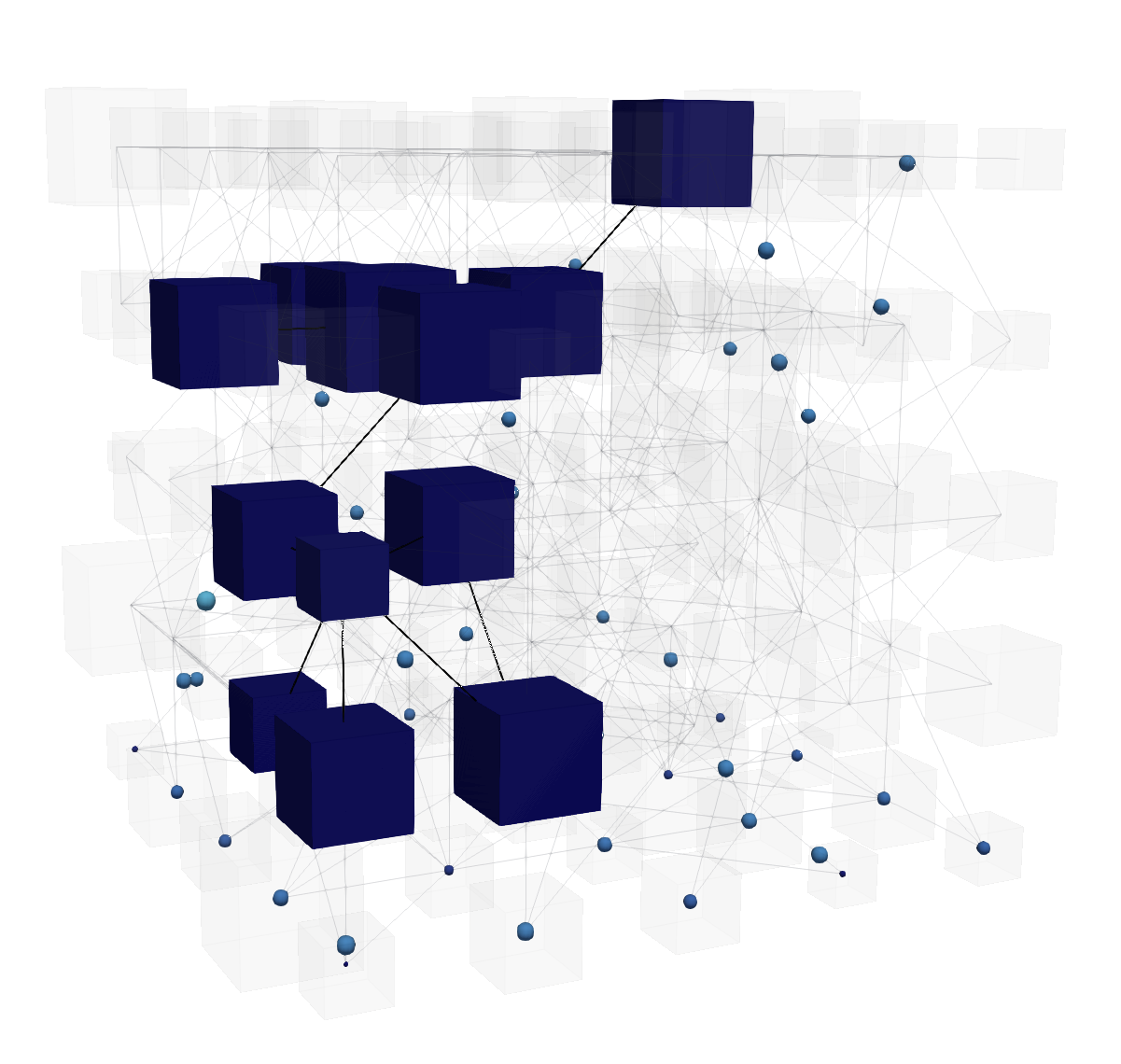}
        \caption{$\Gamma = 300$}
    \end{subfigure}
    \caption{Steady-state invasion patters for nine values of $\Gamma$, using \(\Delta t_{\max}=5\cdot 10^{-5}~\mathrm{s}\) and $\bar r_p=70~\mathrm{\mu m}$. %The results show that there are fewer invaded pores at stronger apparent gravitational forces ($\Gamma \gg 1$), and that pores closer to the top of the network are invaded because smaller bubbles can travel farther before becoming trapped. 
    The coloured pore-bodies are invaded, while the transparent ones are liquid-saturated.}
    \label{fig:gravity-independence}
\end{figure}

\autoref{fig:gravity-independence} compares the steady-state invasion patterns for nine $\Gamma$ values using single realizations of the network. Stronger apparent gravitational forces ($\Gamma \gg 1$) correlate with fewer invaded pores in the network. Pores higher up in the network also tend to be invaded at higher $\Gamma$ but remain water-saturated at low $\Gamma$. We explain this with the observation that small bubbles at high $\Gamma$ travel farther before becoming trapped.

\begin{figure}[pos=ht!]
    \centering
    \begin{subfigure}[t]{0.49\linewidth}
        \centering
        \includegraphics[width=\linewidth]{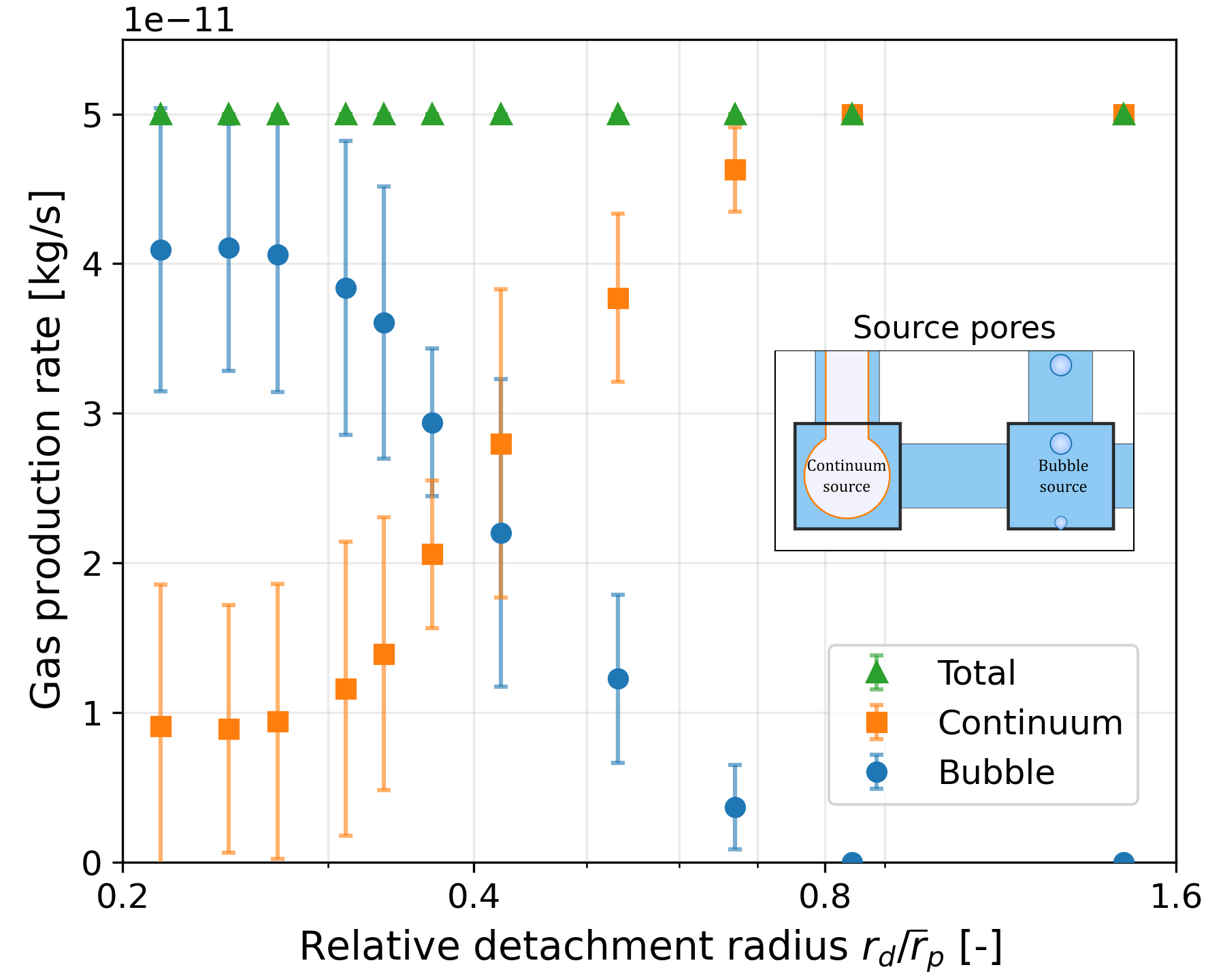}
        \caption{Gas production rate, fixed $\bar r_p=70~\mathrm{\mu m}$.}
        \label{fig:steady-gas-production-rates-varyingG}
    \end{subfigure}\hfill
    \begin{subfigure}[t]{0.49\linewidth}
        \centering
        \includegraphics[width=\linewidth]{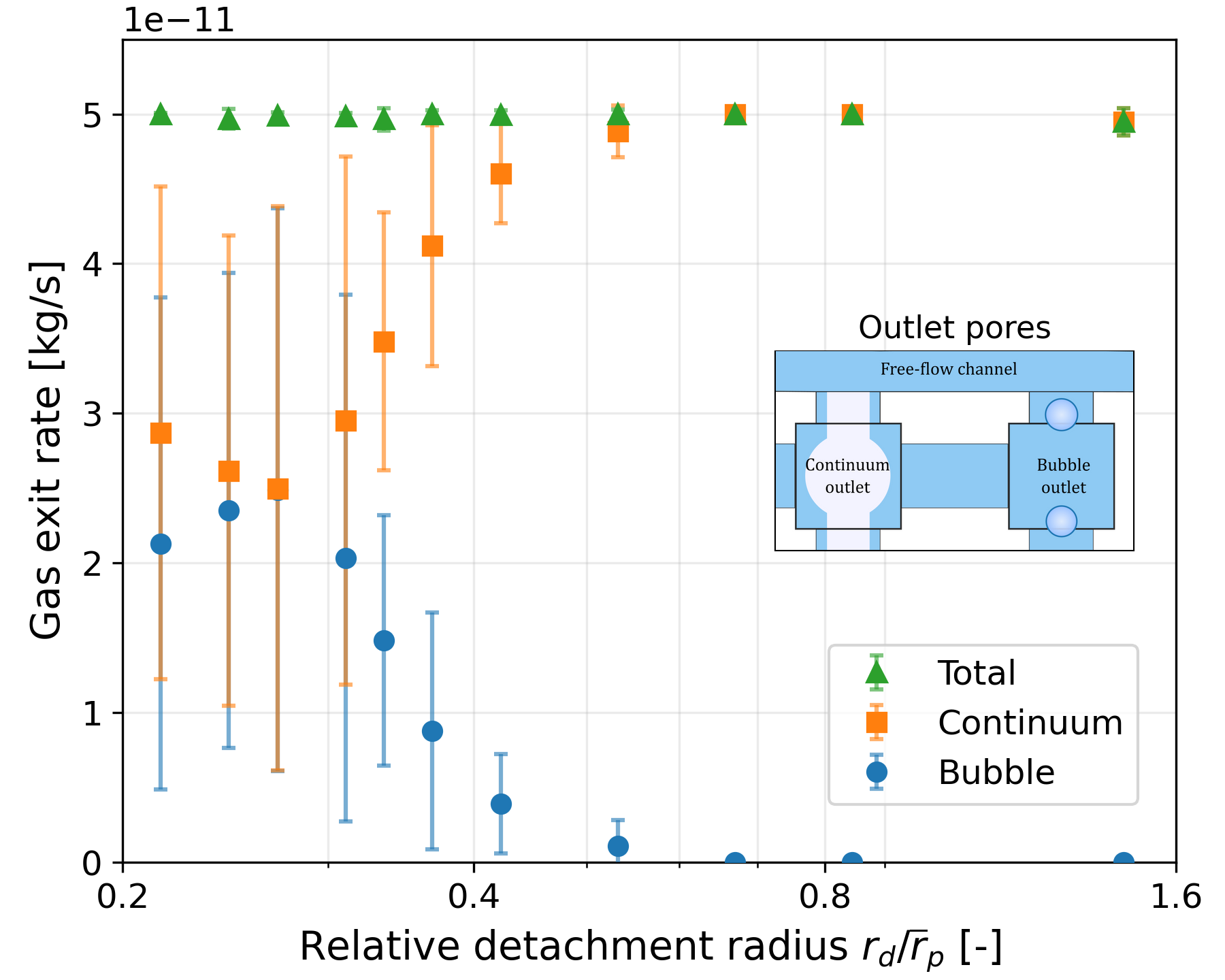}
        \caption{Gas exit rate, fixed $\bar r_p=70~\mathrm{\mu m}$.}
        \label{fig:steady-gas-exit-rates-varyingG}
    \end{subfigure}
    \caption{Gas transport rates vs the relative detachment radius $r_d/\bar r_p$. In these runs $\bar r_p=70~\mathrm{\mu m}$ and the contact radius is fixed such that $r_d$ is uniquely determined from $\|\boldsymbol a_{\mathrm{cf}}\|$ by \autoref{eqn:detachment_radius}. The error bars indicate the corresponding standard deviation.}
    \label{fig:steady-gas-rates-varyingG}
\end{figure}

\autoref{fig:steady-gas-rates-varyingG} shows the ensemble-mean gas production and exit rates as a function of the relative detachment radius $r_d/\bar r_p$. The plot shows that the amount of gas transported as bubbles increases as the detached bubble size decreases, i.e., as $\Gamma$ increases. The importance of the mean pore radius comes from the fact that bubbles are moved into the continuous gas phase if the bubble radius becomes larger than the inscribed radius of a pore it tries to enter or if the entry condition \autoref{eqn:entry_condition} is not met. Since the inscribed throat radius scales with the pore inscribed radius, the ratio $r_d/\bar r_p$ represents how easily bubbles can move through the network. Invasion dominates the network, as $r_d/\bar r_p\rightarrow 1$, since the bubbles can no longer move freely. Some bubbles can move through parts of the network before becoming stuck, causing the discrepancy between the bubble transport at the inlet and the outlet. This is consistent with the observations made from \autoref{fig:gravity-independence}.

\subsection{Model predictions for varying mean pore radii $\bar r_p$}
We next investigate the impact of adjusting the mean pore inscribed radius $\bar r_p$ while keeping $\Gamma=100$ fixed. The other parameters are kept the same as in the previous section, i.e. \(\Delta t_{\max}=5\cdot 10^{-5}~\mathrm{s}\) and the bubble detachment radius $r_d$ is given by \autoref{eqn:detachment_radius} with $r_c=0.1~\mathrm{\mu m}$ and $\Gamma=100$. We then test $\bar r_p \in \lbrace 10,20,30,40,50,60,70,80,90,100 \rbrace ~\mathrm{\mu m}$, each for $11$ random realisations of the $5\times5\times5$ pore network.

\begin{figure}[pos=ht!]
    \centering
    \begin{subfigure}[t]{0.49\linewidth}
        \centering
        \includegraphics[width=\linewidth]{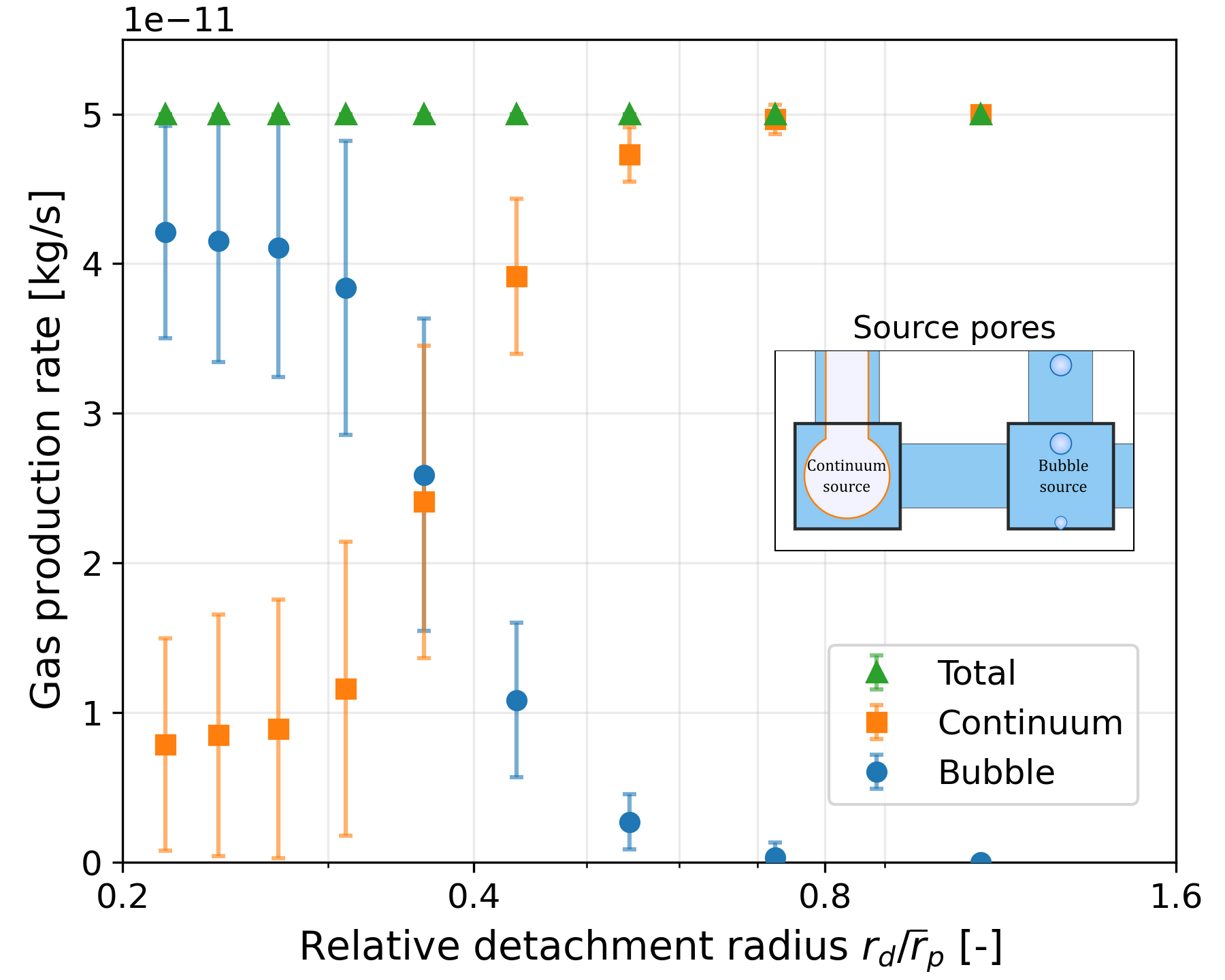}
        \caption{Gas production rate, fixed $\Gamma=100$.}
        \label{fig:steady-gas-production-rates-varyingP}
    \end{subfigure}\hfill
    \begin{subfigure}[t]{0.49\linewidth}
        \centering
        \includegraphics[width=\linewidth]{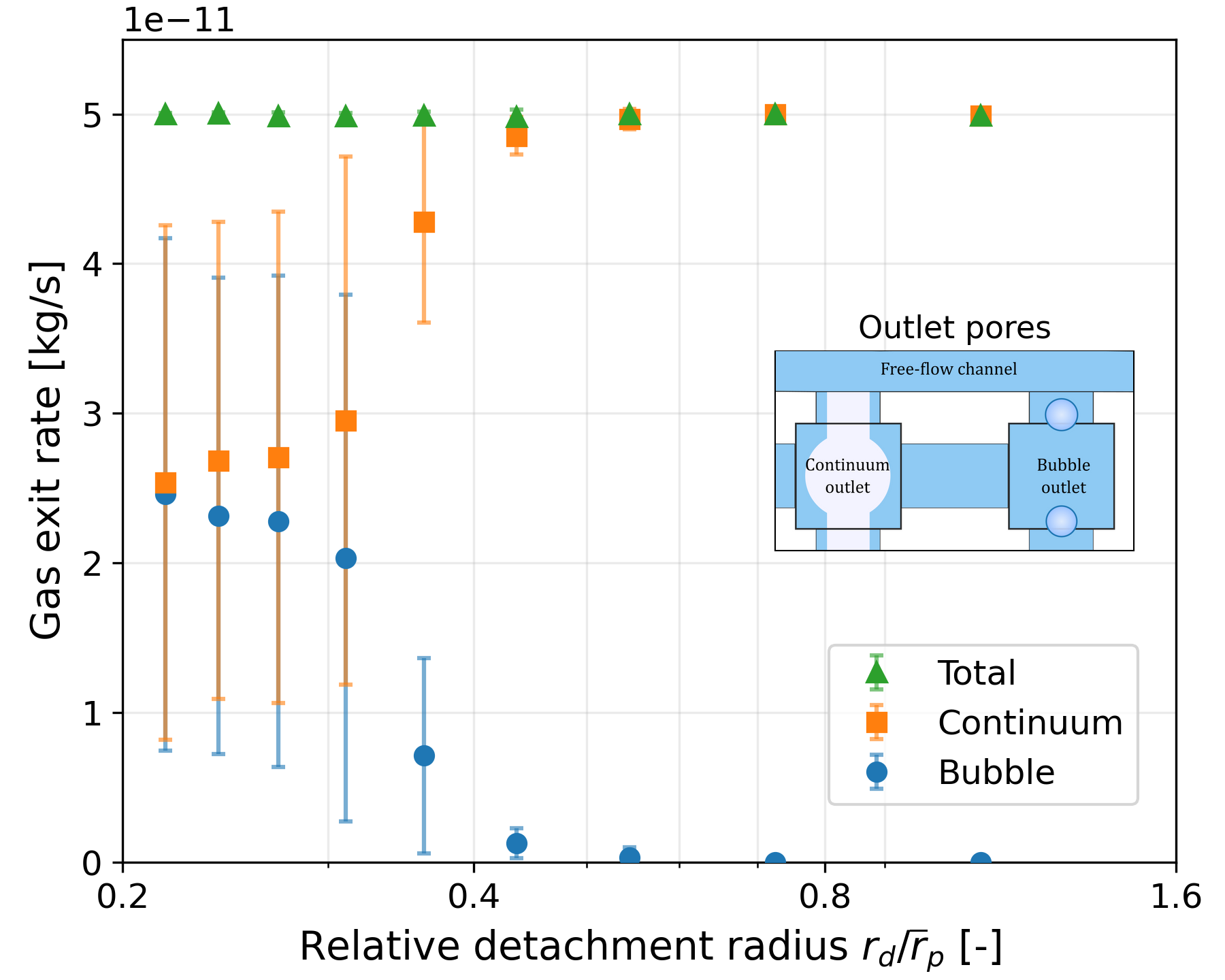}
        \caption{Gas exit rate, fixed $\Gamma=100$.}
        \label{fig:steady-gas-exit-rates-varyingP}
    \end{subfigure}
    \caption{Gas transport rates vs the relative detachment radius $r_d/\bar r_p$. In these runs, the mean pore size $\bar r_p$ is varying, while the detachment radius $r_d$ is constant across all runs (determined by \autoref{eqn:detachment_radius}, for $\Gamma=100$ and $r_c=0.1~\mathrm{\mu m}$). We obtain the mean values from simulations on the $5\times5\times5$ pore network for $11$ different network realisations. The error bars indicate the corresponding standard deviation.}
    \label{fig:steady-gas-rates-varyingP}
\end{figure}

We then hold the detachment radius and gravity forcing constant across all runs, and bubble movement depends on whether the pores can accommodate the bubbles without deforming and whether the throat-entry condition (\autoref{eqn:entry_condition}) is met. The results are plotted in \autoref{fig:steady-gas-rates-varyingP}, showing that varying $\bar r_p$ at fixed $\Gamma$ qualitatively matches varying $\Gamma$ at fixed $\bar r_p$.
The difference between \autoref{fig:steady-gas-rates-varyingG} and \autoref{fig:steady-gas-rates-varyingP} suggests that the bubble fraction in gas transport is not determined solely by $\Gamma$ controlling the detachment radius; bubble dynamics are also influenced by network heterogeneity. 

\subsection{Model predictions for varying $\Gamma$ with fixed $r_d/\bar r_p$}
Next, we investigate the impact of $\Gamma$ while keeping $r_d/\bar r_p$ fixed. It should then be possible to quantify part of the difference between \autoref{fig:steady-gas-rates-varyingG} and \autoref{fig:steady-gas-rates-varyingP} by decoupling the detachment radius from the entry condition \autoref{eqn:entry_condition}. We then artificially fix the detachment radius to $r_d =50~\mathrm{\mu m}$, independent of $\Gamma$. We fix the mean inscribed radius of the pores at $\bar r_p=70~\mathrm{\mu m}$, so the detached bubbles are roughly the same size as the inscribed radii of the pores and throat of the network. The maximum time step size is kept at \(\Delta t_{\max}=5\cdot 10^{-5}~\mathrm{s}\). We then test $\Gamma \in \lbrace 1,5,10,20,40,60,80,100,150,200,300 \rbrace$, each for $11$ random realisations of the $5\times5\times5$ pore network.

\begin{figure}[pos=ht!]
    \centering
    \begin{subfigure}[t]{0.49\linewidth}
        \centering
        \includegraphics[width=\linewidth]{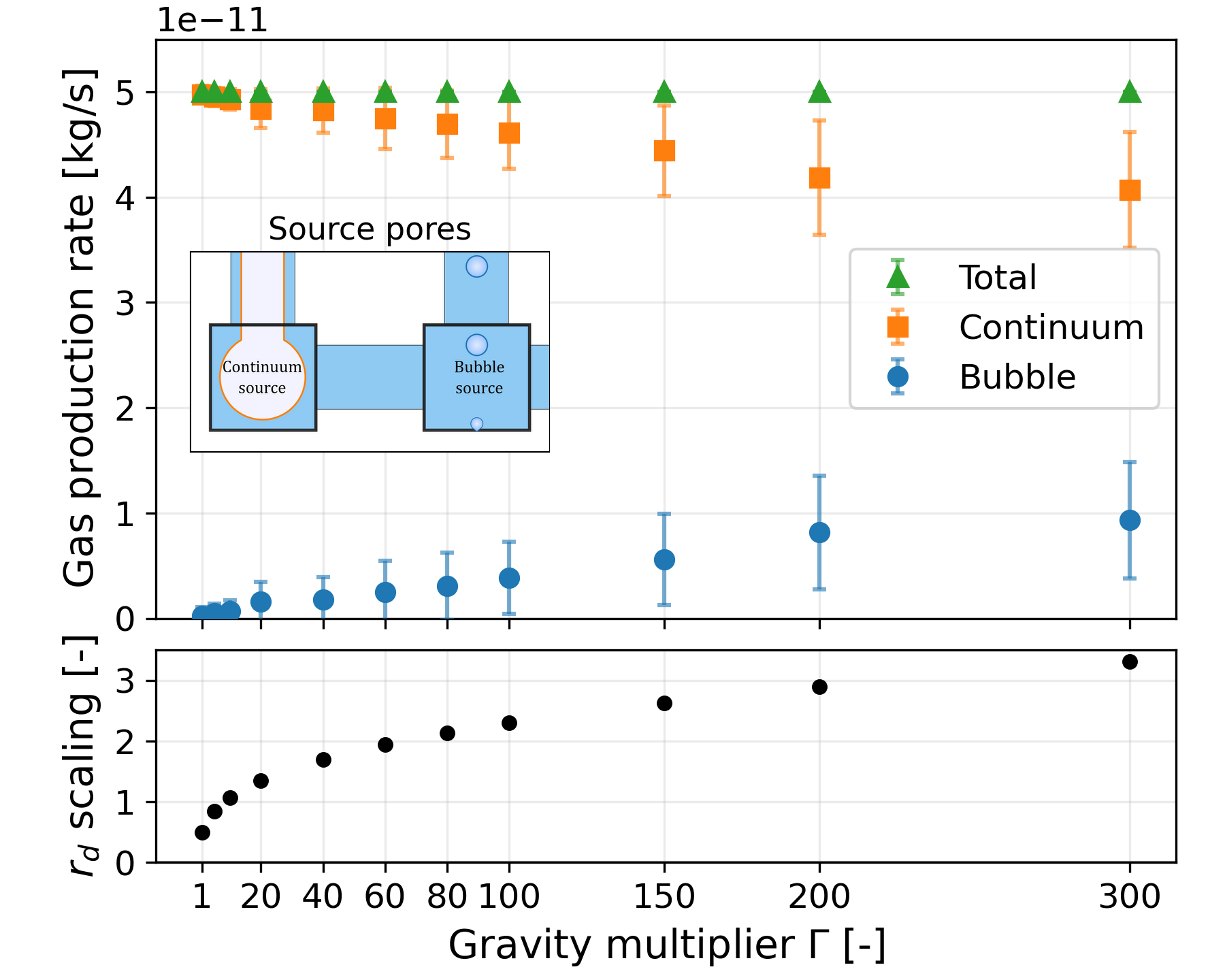}
        \caption{Gas production rate, fixed $r_d/\bar r_p=0.71$.}
        \label{fig:steady-gas-production-rates-fixedRd}
    \end{subfigure}\hfill
    \begin{subfigure}[t]{0.49\linewidth}
        \centering
        \includegraphics[width=\linewidth]{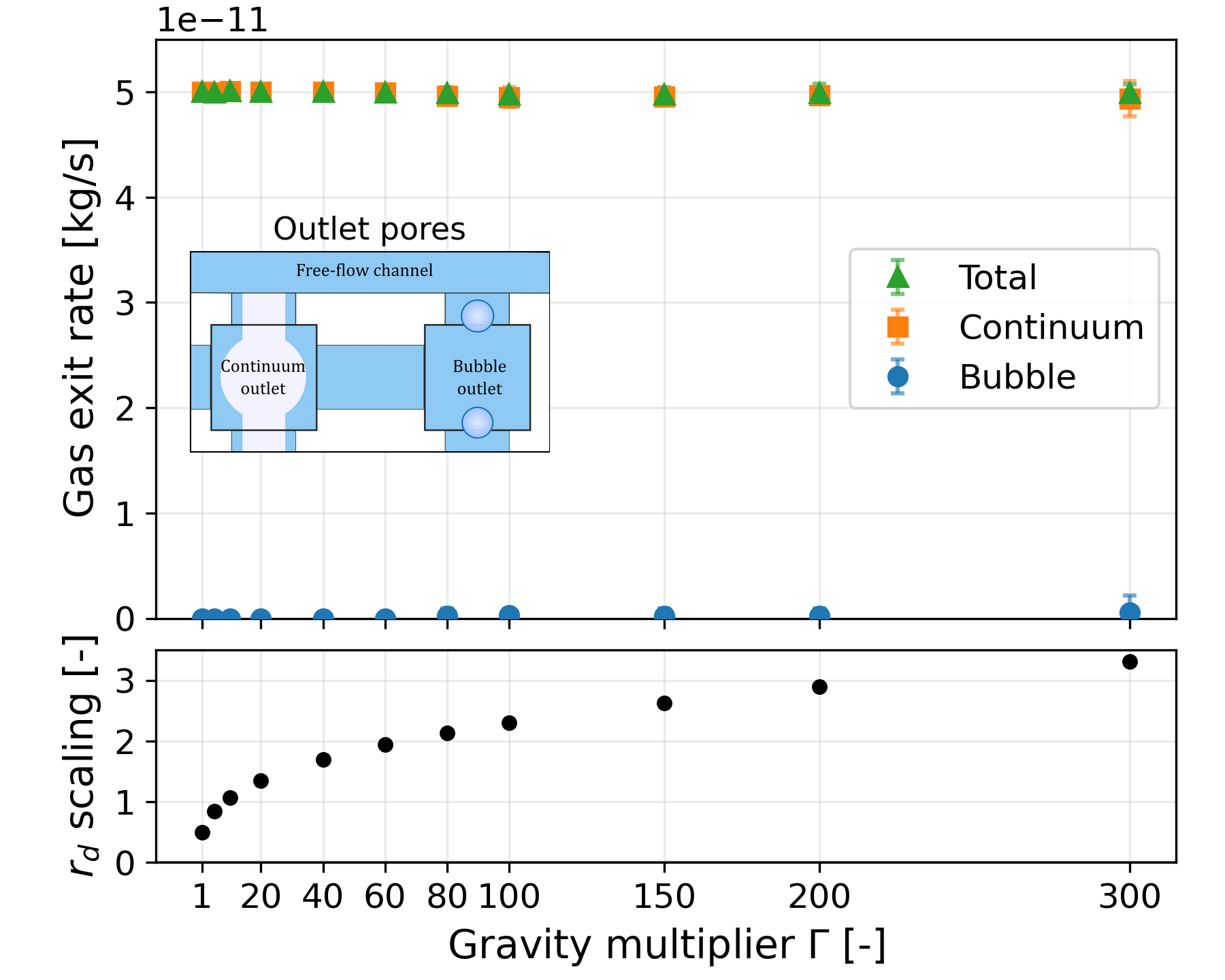}
        \caption{Gas exit rate, fixed $r_d/\bar r_p=0.71$.}
        \label{fig:steady-gas-exit-rates-fixedRd}
    \end{subfigure}
    \caption{Gas transport rates vs the effective gravitational force $\Gamma$, for fixed detachment radius $r_d=50~\mathrm{\mu m}$ and fixed mean pore radius $\bar r_p=70~\mathrm{\mu m}$. The mean values are obtained from simulations on the $5\times5\times5$ pore network for $11$ different network realisations. The error bars indicate the corresponding standard deviation. The bottom plot indicates the scaling of the fixed detachment radius ($50~\mathrm{\mu m}$) relative to the one given by \autoref{eqn:detachment_radius} for each value of $\Gamma$.}
    \label{fig:steady-gas-rates-fixedRd}
\end{figure}

The results are plotted in \autoref{fig:steady-gas-rates-fixedRd}, showing that the fraction of created gas in the bubble phase increases with buoyancy through the entry condition \autoref{eqn:entry_condition}, even though the deformation is only a few $\mathrm{\mu m}$ at $\Gamma=100$. The exit rate is barely affected, meaning these bubbles will eventually immobilise within the network and form a continuous gas phase. We recall that in this verification test, the detachment radius is constant and enlarged to highlight the regime where bubble radii are close to throat radii and the entry condition is thus decisive for whether bubbles are mobile or not.

\subsection{Larger pore networks and simulation times}
We will now investigate how the model scales to a larger network of $10\times10\times10$ pores. Again setting \(\Delta t_{\max}=5\cdot 10^{-5}~\mathrm{s}\) (such that \autoref{eqn:max-dt} is fulfilled), $\bar r_p=70~\mathrm{\mu m}$, and $r_d$ given by \autoref{eqn:detachment_radius} with $r_c=0.1~\mathrm{\mu m}$. The mass source is increased proportionally to the (four-times) larger bottom area of the network, relative to the $5\times5\times5$ cases, such that $\dot m = 2\cdot10^{-10}~\mathrm{kg/s}$. The remaining parameters are given in \autoref{tab:5x5_PNM}. We then test $\Gamma \in \lbrace 10,20,40,60,80,100,150,200,300 \rbrace$, each for $3$ random realisations of the $10\times10\times10$ pore network.

\begin{figure}[pos=ht!]
    \centering
    \begin{subfigure}[t]{0.5\linewidth}
        \centering
        \includegraphics[width=\linewidth]{pictures/large_networks/colorbar_cropped.png}
    \end{subfigure}
    \par\vspace{0.1em}
    \begin{subfigure}[t]{0.46\linewidth}
        \centering
        \includegraphics[width=\linewidth]{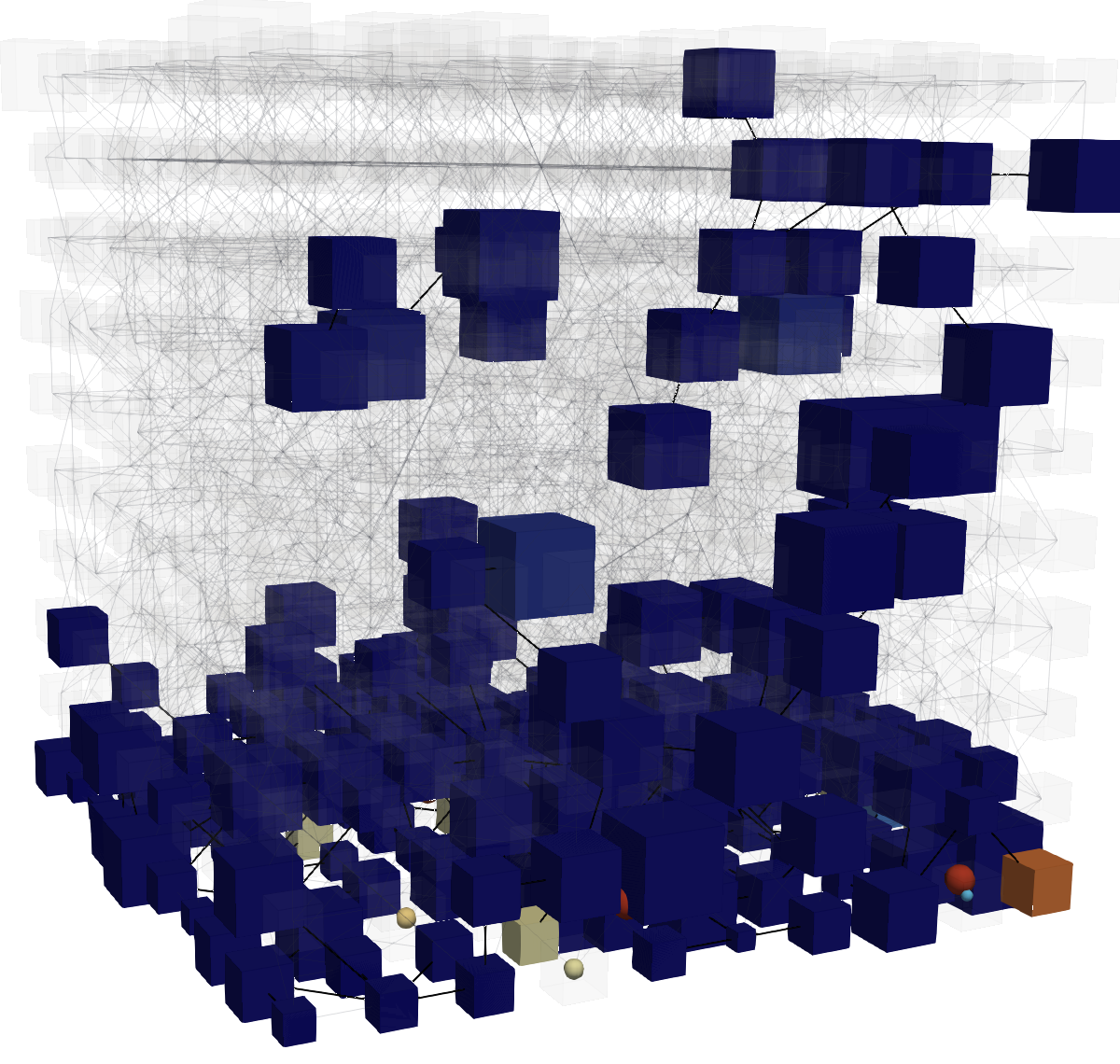}
        \caption{$\Gamma=10$}
    \end{subfigure}\hfill
    \begin{subfigure}[t]{0.46\linewidth}
        \centering
        \includegraphics[width=\linewidth]{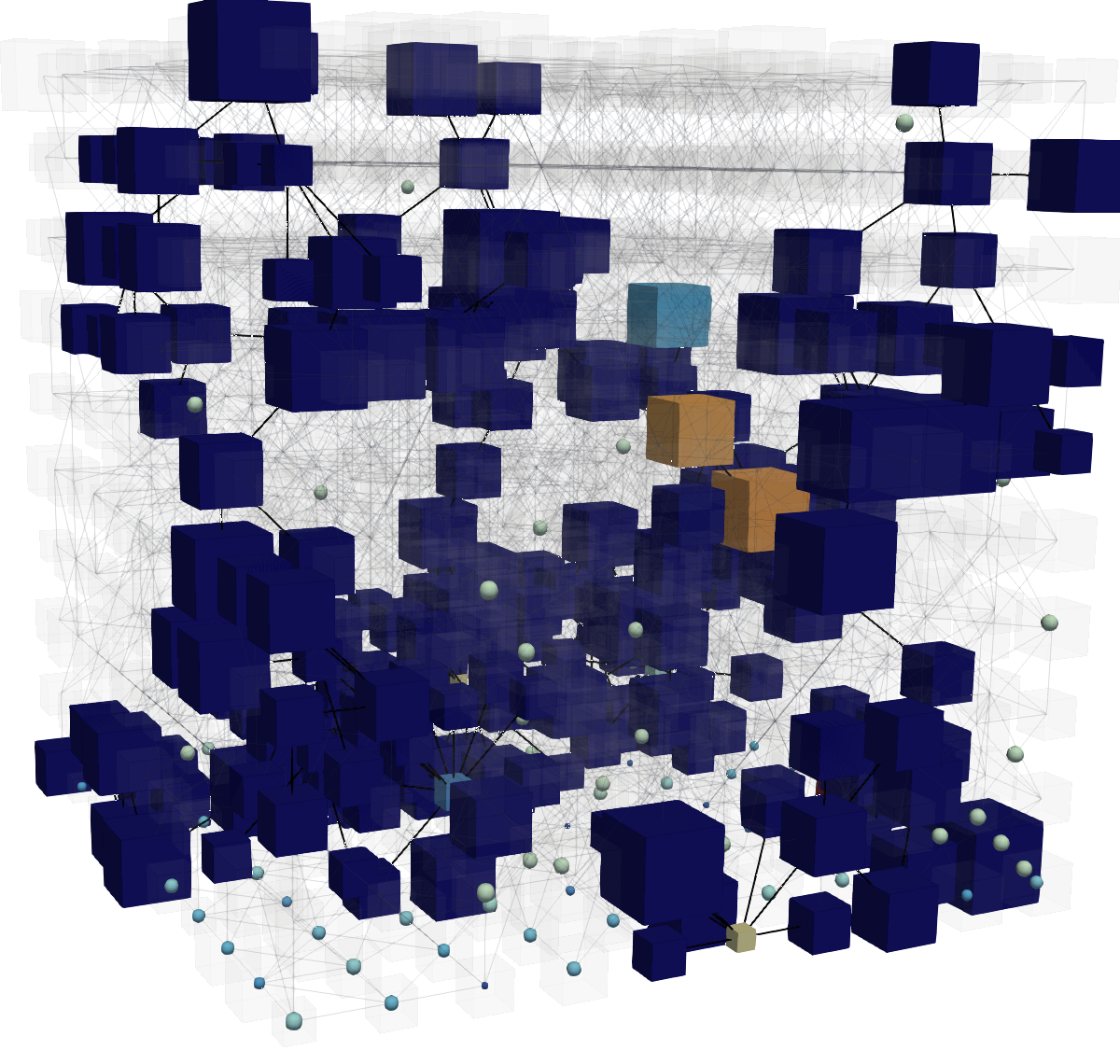}
        \caption{$\Gamma=60$}
    \end{subfigure}
    \par\vspace{1em}
    \begin{subfigure}[t]{0.46\linewidth}
        \centering
        \includegraphics[width=\linewidth]{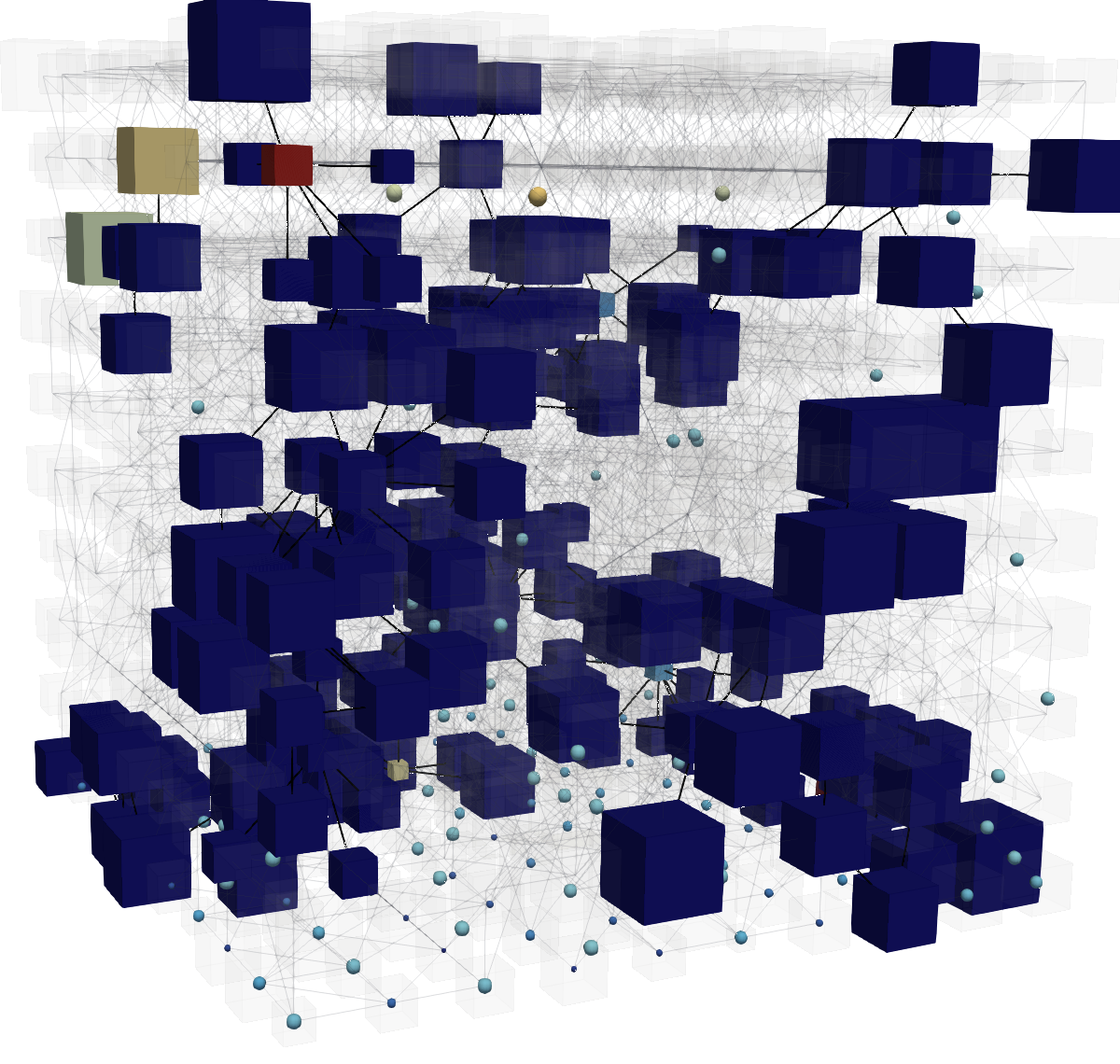}
        \caption{$\Gamma=100$}
    \end{subfigure}\hfill
    \begin{subfigure}[t]{0.46\linewidth}
        \centering
        \includegraphics[width=\linewidth]{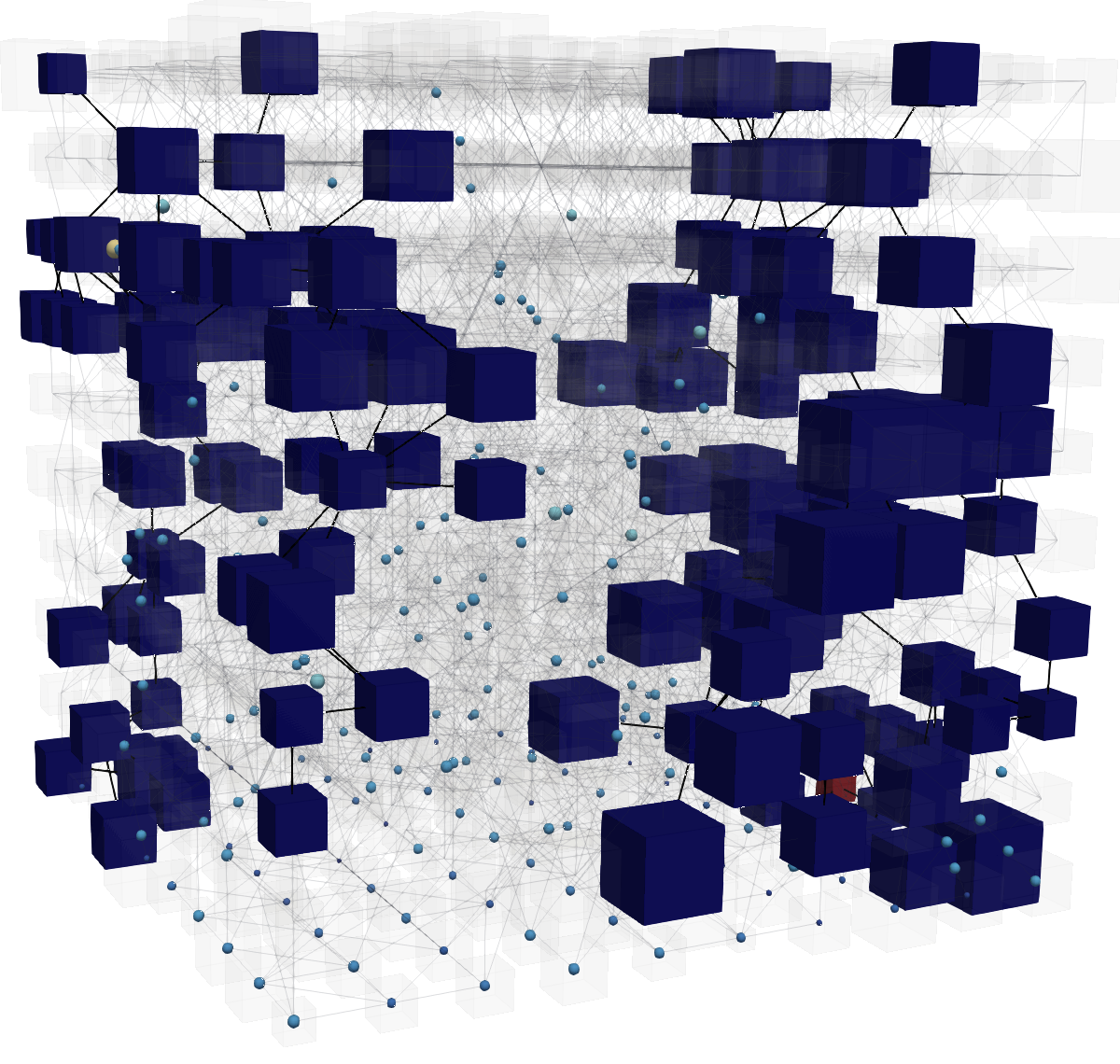}
        \caption{$\Gamma=200$}
    \end{subfigure}
    \caption{Steady-state invasion patters in a $10\times10\times10$ pore network, for four values of $\Gamma$, using \(\Delta t_{\max}=5\cdot 10^{-5}~\mathrm{s}\) and $\bar r_p=70~\mathrm{\mu m}$. The coloured pore bodies are invaded, while the transparent ones are liquid-saturated.}
    \label{fig:10x10x10}
\end{figure}

\begin{figure}[pos=ht!]
    \centering
    \begin{subfigure}[t]{0.49\linewidth}
        \centering
        \includegraphics[width=\linewidth]{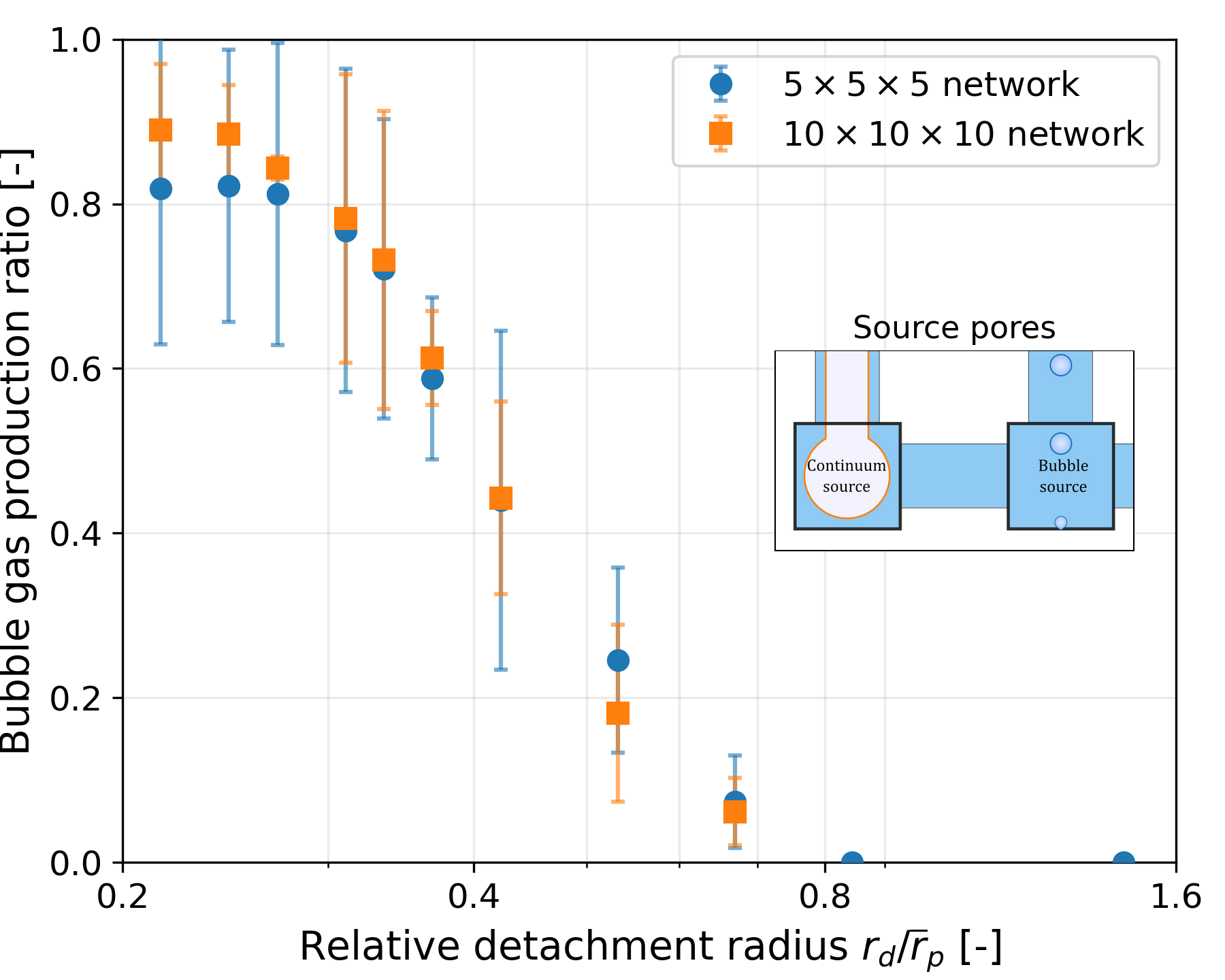}
        \caption{Gas production rate, fixed $\bar r_p=70~\mathrm{\mu m}$.}
        \label{fig:steady-gas-production-rates-comparison}
    \end{subfigure}\hfill
    \begin{subfigure}[t]{0.49\linewidth}
        \centering
        \includegraphics[width=\linewidth]{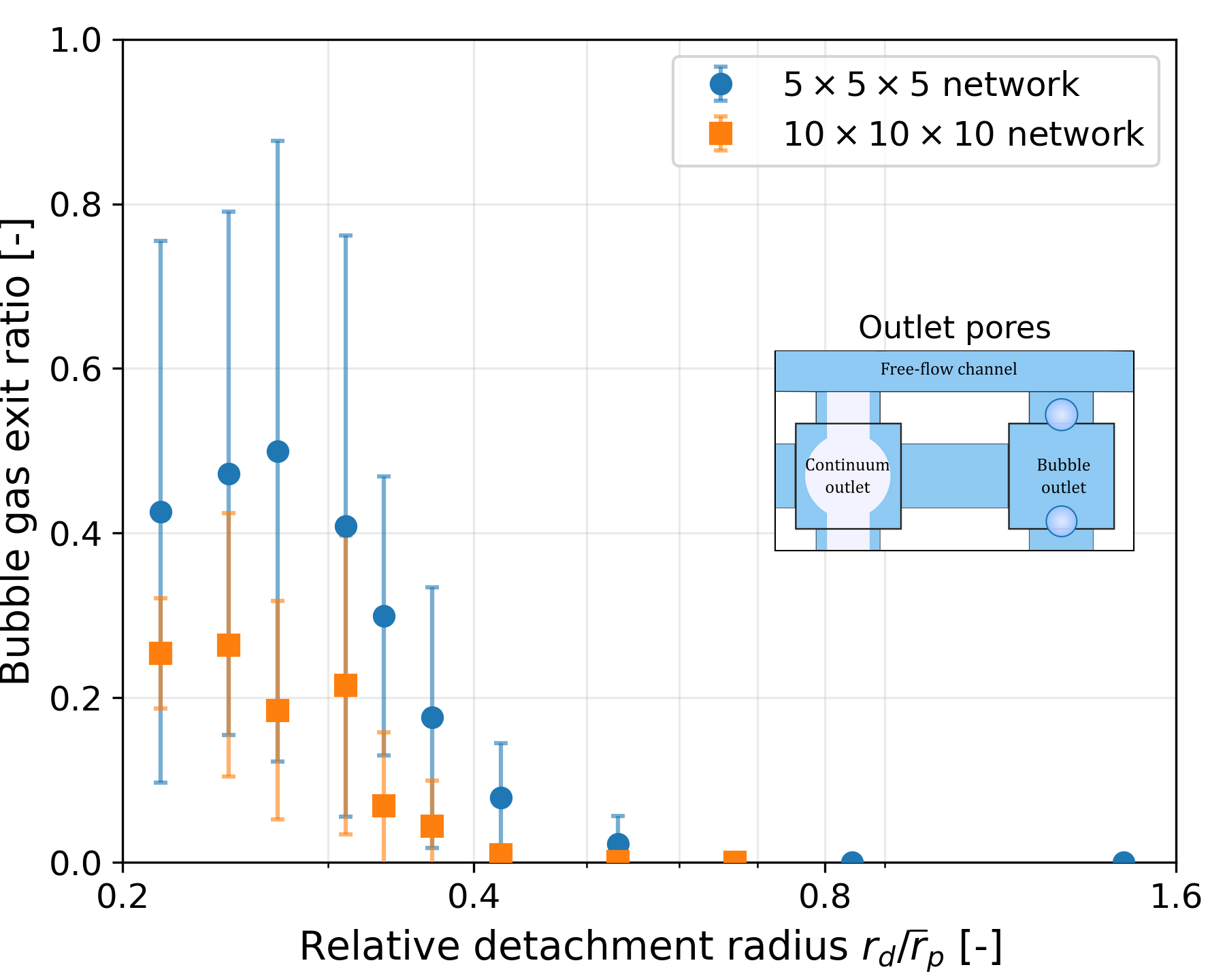}
        \caption{Gas exit rate,  fixed $\bar r_p=70~\mathrm{\mu m}$.}
        \label{fig:steady-gas-exit-rates-comparison}
    \end{subfigure}
    \caption{Ratio of gas transported as bubbles vs the relative detachment radius $r_d/\bar r_p$, for the $5\times5\times5$ and $10\times10\times10$ networks. In these runs $\bar r_p=70~\mathrm{\mu m}$ and the contact radius is fixed such that $r_d$ is uniquely determined from $\|\boldsymbol a_{\mathrm{cf}}\|$ by \autoref{eqn:detachment_radius}. The error bars indicate the corresponding standard deviation.}
    \label{fig:steady-gas-rates-comparison}
\end{figure}

We show a selection of steady-state invasion patterns in \autoref{fig:10x10x10} and the ratio of gas transported as bubbles for different $\Gamma$ in \autoref{fig:steady-gas-rates-comparison}. The observed invasion patterns are similar to those in the smaller networks, where larger $\Gamma$ leads to invasion of pores further up into the network and increased bubble transport. The ratio of bubble transport to continuous gas transport at the inlet is comparable between the small and large networks for all values of $\Gamma$, while the ratio at the outlet is much smaller in the large network. This is expected because the larger network offers more opportunities for bubbles to become trapped.

\subsection{Model limitations}
One limitation of the model is how it treats gas injection. We assume a constant mass source rate. In a reaction-controlled process, the interface area blocked by the bubble might influence the reaction and thus the gas production rate. 

We modelled the detachment process in a simplified manner, as the exact location of bubble growth is not resolved and the effective acceleration $\boldsymbol a_{\mathrm{cf}}$ is constant and assumed perpendicular to the growth surface. This makes the detachment radius constant throughout the network. Real detachment processes are complex and determined by the exact structure of the PEM catalyst surface and the gas/liquid flow around it.

While stuck bubbles could in reality impact the local flow around them, we choose to immediately transfer them to the continuous gas phase. This is only appropriate if re-mobilisation is unlikely. The transferred gas no longer reduces the liquid phase throat conductivity, although it can invade neighbouring pores if the capillary pressure is high enough. 

The transition from bubbles to a continuous non-wetting phase is currently unidirectional, even though small non-wetting-phase saturation could be modelled as individual bubbles, particularly for an isolated gas cluster.

We assume the bubbles, non-wetting phase, and wetting phase have constant properties. The bubbles are assumed to be non-deforming (except elongation within throats), and the wetting-film thickness around elongated bubbles is assumed constant.

The obtained closure relations hold for Stokes flow and are therefore limited to small Reynolds numbers. A detailed histogram of the Reynolds number of all bubbles within a simulation is given in Section S3 of the Supplementary Material for $\Gamma \in\{10, 100, 300\}$, and shows that the vast majority of bubbles have $\mathrm{Re}<10$, with some bubbles in the range $10<\mathrm{Re}<100$, and a few outliers $\mathrm{Re}>100$.

We obtained the closure relations assuming a uniform minimum wetting film thickness, negligible Coriolis effects, and negligible dynamic bubble deformation. Violation of these assumptions introduces modelling errors from the closures. We have not quantified these effects.

\section{Conclusion}\label{sec:conclusion}
In this paper, we presented a novel hybrid PNM with discrete bubbles. The model is based on a dynamic two-phase PNM and coupled with a bubble tracker. Fixed-point iteration and tracking of irreversible bubble events ensure that the PNM and bubble tracker converge to a common solution at each time step. Closure relations for the bubble movement and impact in square throats were obtained numerically for spherical and elongated oxygen bubbles moving through water by buoyancy and advection. We verified the model through a series of tests designed to isolate specific features, ensuring the movement and impact align with expectations. The model is developed for the specific regime and scale where flow has negligible impact on bubble shape. Dynamic bubble shapes influenced by flow are beyond the scope of the current study but play a role in spatially larger systems.

The model predicts that, over an ensemble of random network realisations with parameters characteristic of a PEM electrolyser PTL, the amount of gas carried by bubbles increases as the effective gravitational strength $\Gamma$ increases, with approximately $40~\%$ of the produced gas leaving the network as bubbles when $\Gamma>100$. Bubble transport occurs because the detachment radius decreases, allowing small bubbles to move through the network, and because increasing $\Gamma$ forces bubbles into pore throats they normally could not enter.

\section*{Data and code availability}
The source code and scripts required to reproduce the results in this work are available at
\url{https://git.iws.uni-stuttgart.de/dumux-pub/skjelanger2026a}. The simulation data obtained in this study are provided in the same repository as CSV files. Additional output data are available from the authors upon reasonable request.

\printcredits

\section*{Acknowledgements}
This research is funded by HyValue partners and the Research Council of Norway (grant number 333151). 
We further acknowledge funding by the Deutsche Forschungsgemeinschaft (DFG, German Research Foundation), within the framework of the Collaborative Research Centre “Interface-Driven MultiField Processes in Porous Media” (SFB 1313, Project Number 327154368).
Some computations were performed on resources provided by Sigma2 -- the National Infrastructure for High-Performance Computing and Data Storage in Norway.
Kristoffer Skjelanger would like to thank his supervisors at Western Norway University of Applied Sciences, Jonathan Økland Torstensen and Carina Bringedal, for mentoring, support, and fruitful discussions.
A special thanks goes to Bernd Flemisch for facilitating the research stay, at the University of Stuttgart, which initiated this research.

\section*{Declaration of generative AI use}
During the preparation of this work, the author(s) used OpenAI ChatGPT as a tool for writing and troubleshooting the code. After using this tool, the author(s) reviewed, tested, and edited the code as needed and take full responsibility for the content of the published article.
The authors also used AI tools during the final pre-submission review of the manuscript to check for grammatical errors, conflicting symbols, missing definitions, and other editorial issues.

\bibliographystyle{cas-model1-num-names}
% \bibliographystyle{cas-model2-names}

\bibliography{refs}

\clearpage

\appendix
\section{Bubble rising in a square pore throat}\label{app:Stokes}
We obtain the closure relations, which parameterise bubbles interacting with the liquid-phase flow, from data based on precomputed bubble shapes, Stokes-equation solutions in the bubble frame in a narrow square channel, and auxiliary analytical and numerical computations in the limits of small bubbles or long, elongated bubbles. 
\autoref{app:fluid} describes how bubble shapes were obtained and turned into meshes using in Stokes simulations. \autoref{app:corner-k} describes how we approximate a scaling factor in the limit of long bubbles. \autoref{app:Stokes} describes the Stokes setup and the algorithms used to obtain the relevant dataset for the closures in \autoref{app:bubble-velocity} and \autoref{app:viscous-pressure}. 
The Stokes simulation were conducted in DuMu\textsuperscript{x} \citep{kochDuMux3Opensource2021}.
The results are mesh converged to a tolerance acceptable in our application, as shown in \autoref{app:mesh}.
\autoref{fig:mesh} shows the obtained Stokes flow results, with additional plots in Section S2 of the Supplementary Material.

\subsection{Bubble shape and fluid domain}\label{app:fluid}
We now present the procedure for generating the fluid domain around spherical and elongated bubbles in a square channel. We enforce a minimum wetting film thickness of $0.01~r_{ij}$ between the channel and bubble surface, consistent with the discussion in \autoref{sec:bubble-shape}. The bubble shape is represented by the relative length $l_{b,ij}^*$ defined in \autoref{eqn:bubble_length}, such that the bubble is spherical when $l_{b,ij}^*\leq1.0$ and elongated when $l_{b,ij}^*>1.0$.

We obtain the bubble surfaces using Surface Evolver (v2.7) \citep{brakkeSurfaceEvolver1992}, which minimises the surface energy of a volume confined within a square channel. The convergence criteria of the Surface Evolver are: The relative surface energy does not change by more than \(10^{-5}\) on successive iterations, the bubble volume deviation is smaller than \(10^{-15}~\mathrm{m^3}\), and the bubble is constricted by the channel within an error margin of \(15\%\) of the thin film thickness. 

We construct the fluid domain by subtracting the triangulated bubble surface from the square channel using a CGAL boolean difference, and an unstructured tetrahedral mesh with CGAL's 3D mesh generator \citep{cgal:eb-26a}. The mesh criteria impose a minimum facet angle of ($12^\circ$), a maximum surface-approximation distance of ($0.02~\mathrm{\mu m}$), and a maximum tetrahedral radius-edge ratio of ($4.0$). The feature edges, surface facets, and cell sizes scale with the characteristic mesh size ($h$). We optimise computational resources by meshing and simulating only one quarter of the fluid domain, since the problem is axisymmetric along the throat. 

\begin{figure}[pos=ht!]
    \centering
    \begin{minipage}{\linewidth}
    \begin{subfigure}[t]{0.19\textwidth}
        \centering
        \includegraphics[width=\textwidth,trim=0 1 0 7,clip]{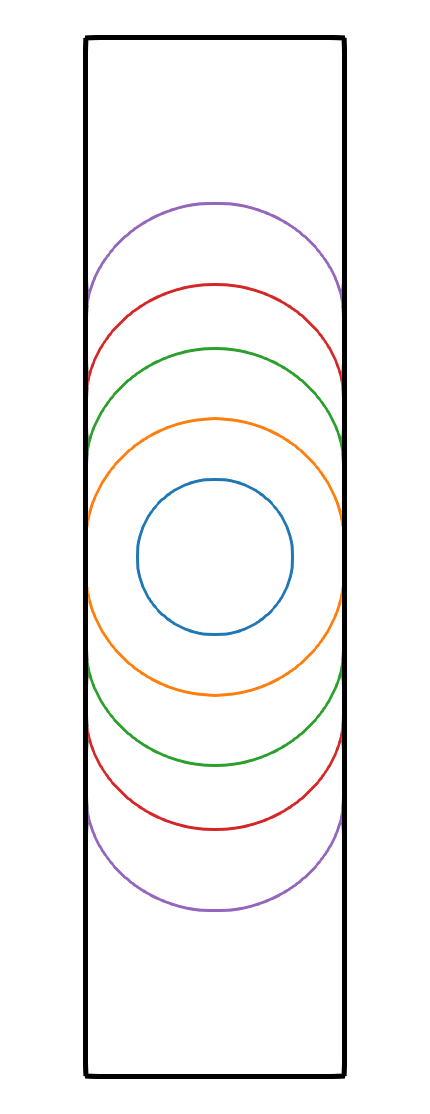}
        \caption{}
        \label{fig:mesh-outlines}
        \end{subfigure}
    \hfill
    \begin{subfigure}[t]{0.192\textwidth}
        \centering
        \includegraphics[width=\textwidth,trim=0 30 5 0,clip]{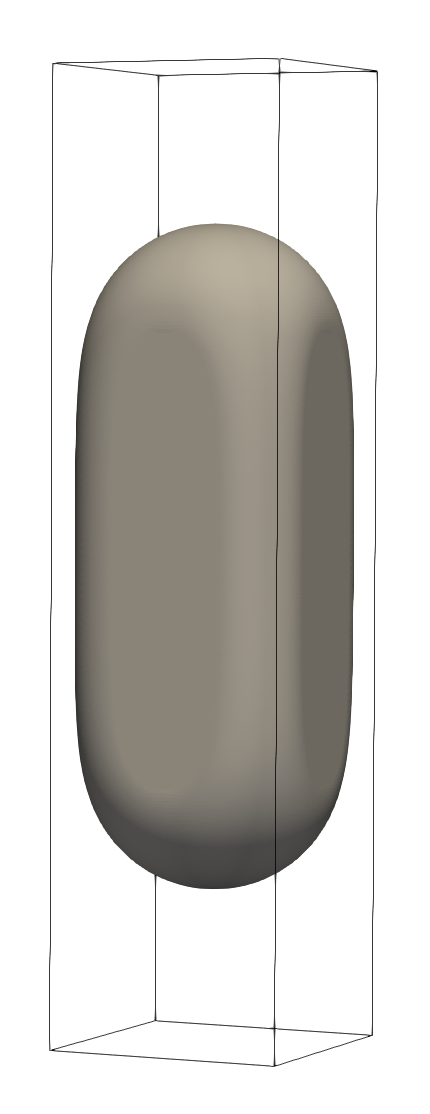}
        \caption{}
        \label{fig:mesh-bubble}
        \end{subfigure}
    \hfill
    \begin{subfigure}[t]{0.192\textwidth}
        \centering
        \includegraphics[width=\textwidth,trim=0 30 5 0,clip]{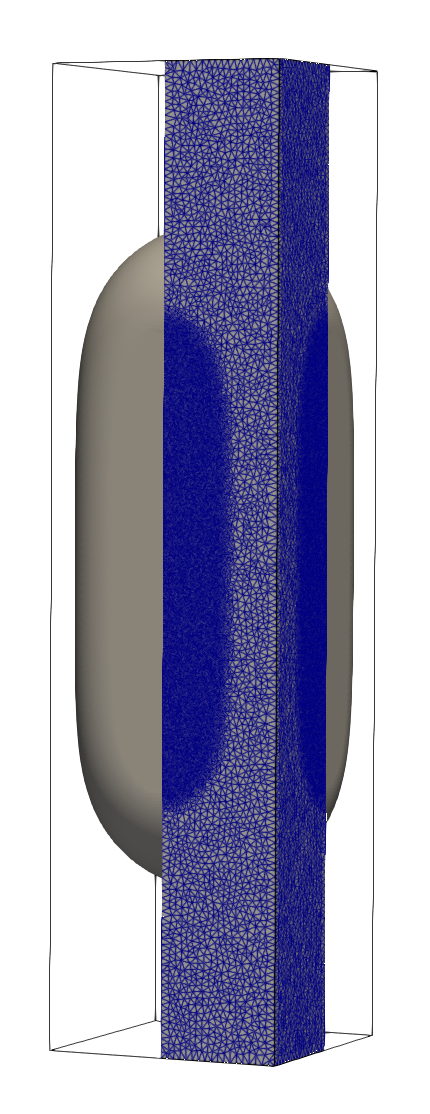}
        \caption{}
        \label{fig:mesh-mesh}
        \end{subfigure}
    \hfill
    \begin{subfigure}[t]{0.195\linewidth}
        \centering
        \includegraphics[width=\linewidth,trim=23 1 23 2,clip]{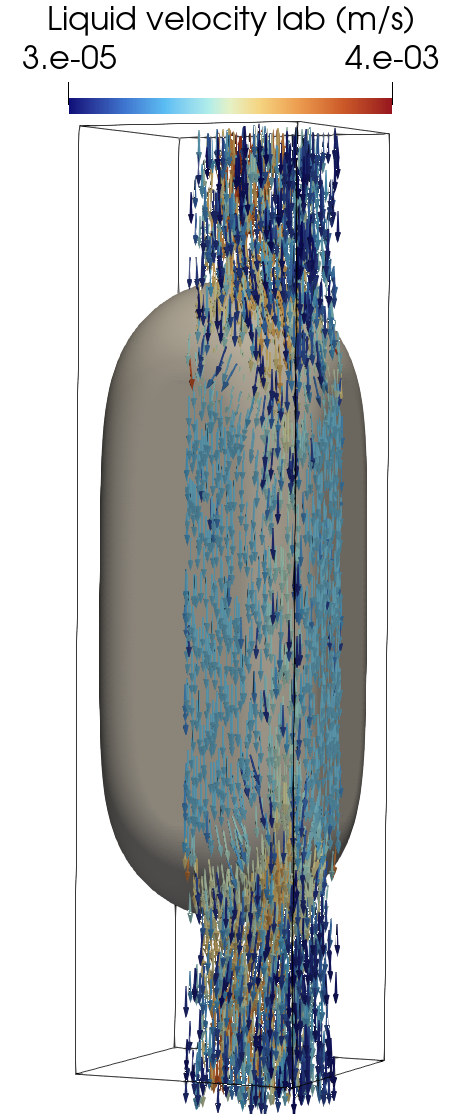}
        \caption{}
        \label{fig:mesh-vel}
        \end{subfigure}
    \hfill
    \begin{subfigure}[t]{0.195\linewidth}
        \centering
        \includegraphics[width=\linewidth,trim=23 1 23 2,clip]{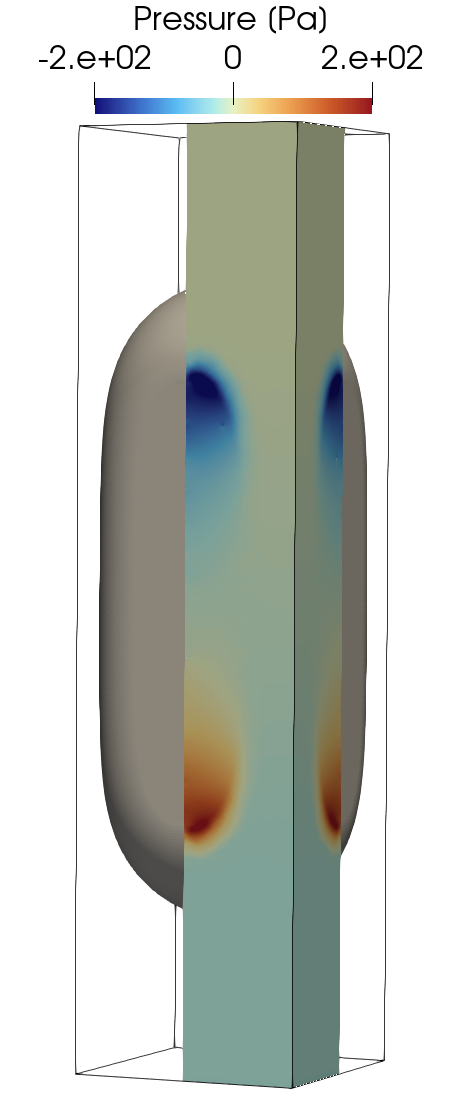}
        \caption{}
        \label{fig:mesh-pres}
        \end{subfigure}
    \end{minipage}
    \caption{Elongated bubbles in a channel with inscribed radius $r_{ij}=25~\mathrm{\mu m}$ and \(l_{ij}=200~\mathrm{\mu m}\). (a) Cross-sectional view of elongated bubbles with equivalent radius $r_b \in \left[15,~26,~32,~36,~40 \right]~\mathrm{\mu m}$. (b) Elongated bubble of \(r_{b}=40~\mathrm{\mu m}\). (c) Fluid mesh between the bubble and channel walls. (d) Liquid velocity around the bubble with $\Delta p=25~\mathrm{Pa}$ and $\Gamma =0$. (e) Pressure field with $\Delta p=25~\mathrm{Pa}$ and $\Gamma =0$.}
    \label{fig:mesh}
\end{figure}

We visually inspect the generated bubble surfaces and fluid meshes before use for visible mesh defects. \autoref{fig:mesh} illustrates how bubbles elongate within the channel, along with a generated pill-shaped bubble surface, its mesh, and the solved liquid velocity and pressure fields.

Expressions for the relative bubble length and bubble curvature radius can be fitted as functions of the equivalent bubble radius $r_b$ and effective throat inscribed radius $r_{\mathrm{eff}}$,

\begin{align}
    l_{b,ij}^* &= 1.823 - 3.094\cdot \left(\frac{r_b}{r_{\mathrm{eff}}}\right) + 2.269\cdot \left(\frac{r_b}{r_{\mathrm{eff}}}\right)^{2}\,, \quad r_b\geq r_{\mathrm{eff}} \,, \label{eqn:bubble_length} \\
    \frac{r_a}{r_{\mathrm{eff}}} &= 0.508 + 0.491 \cdot \left(l_{b,ij}^*\right)^{-3.539}\,, \quad r_b\geq r_{\mathrm{eff}} \,.\label{eqn:curvature_radius}
\end{align}

\begin{figure}[pos=ht!,width=0.48\linewidth]
    \centering
    \begin{minipage}[t]{0.48\linewidth}
        \vspace{0pt}
        \centering
        \includegraphics[width=\linewidth]{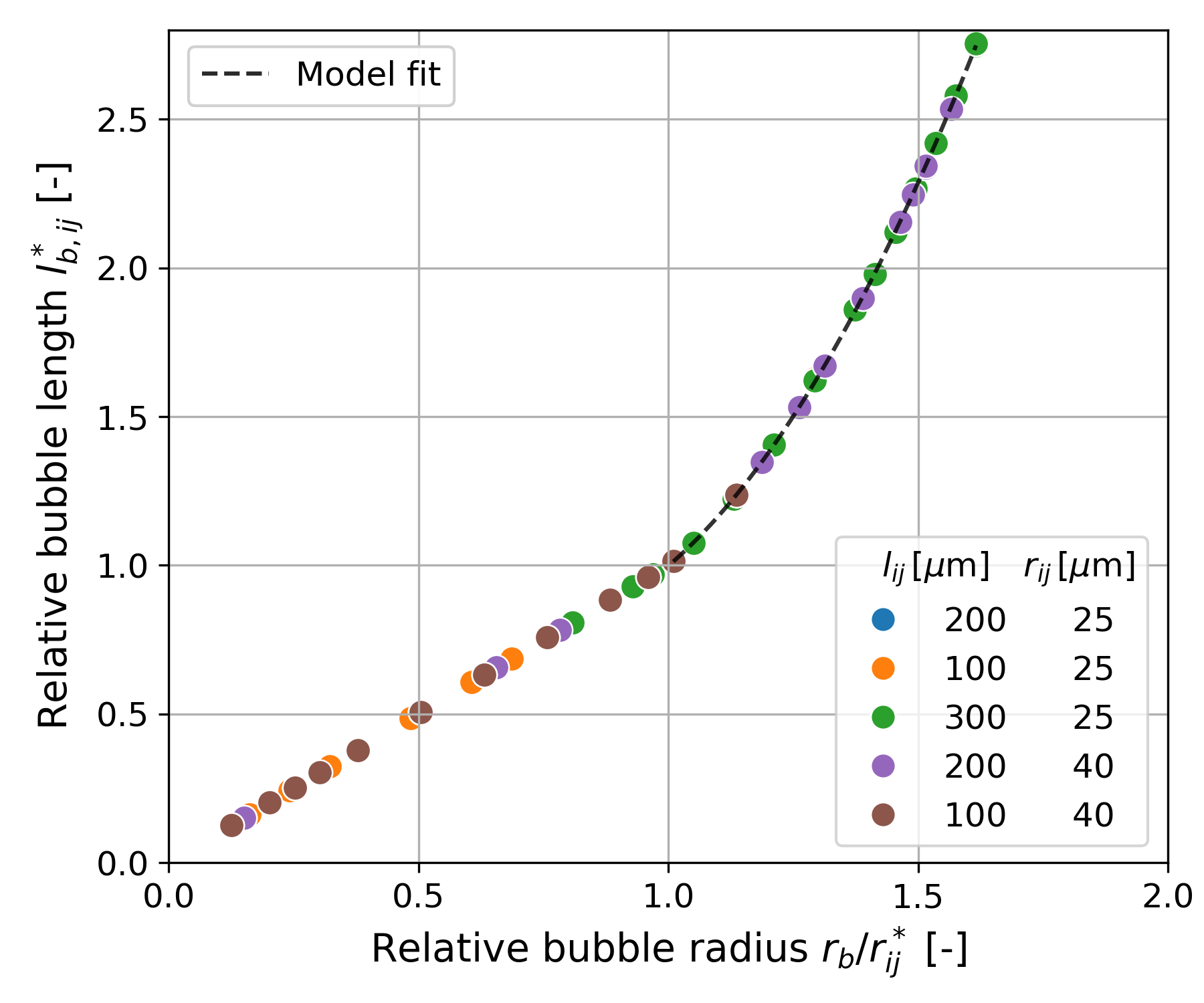}
        \caption{Bubble length fitted as a function of the effective bubble radius, given by \autoref{eqn:bubble_length}.}
        \label{fig:bubble-length-fit}
    \end{minipage}
    \hfill
    \begin{minipage}[t]{0.48\linewidth}
        \vspace{0pt}
        \centering
        \includegraphics[width=\linewidth]{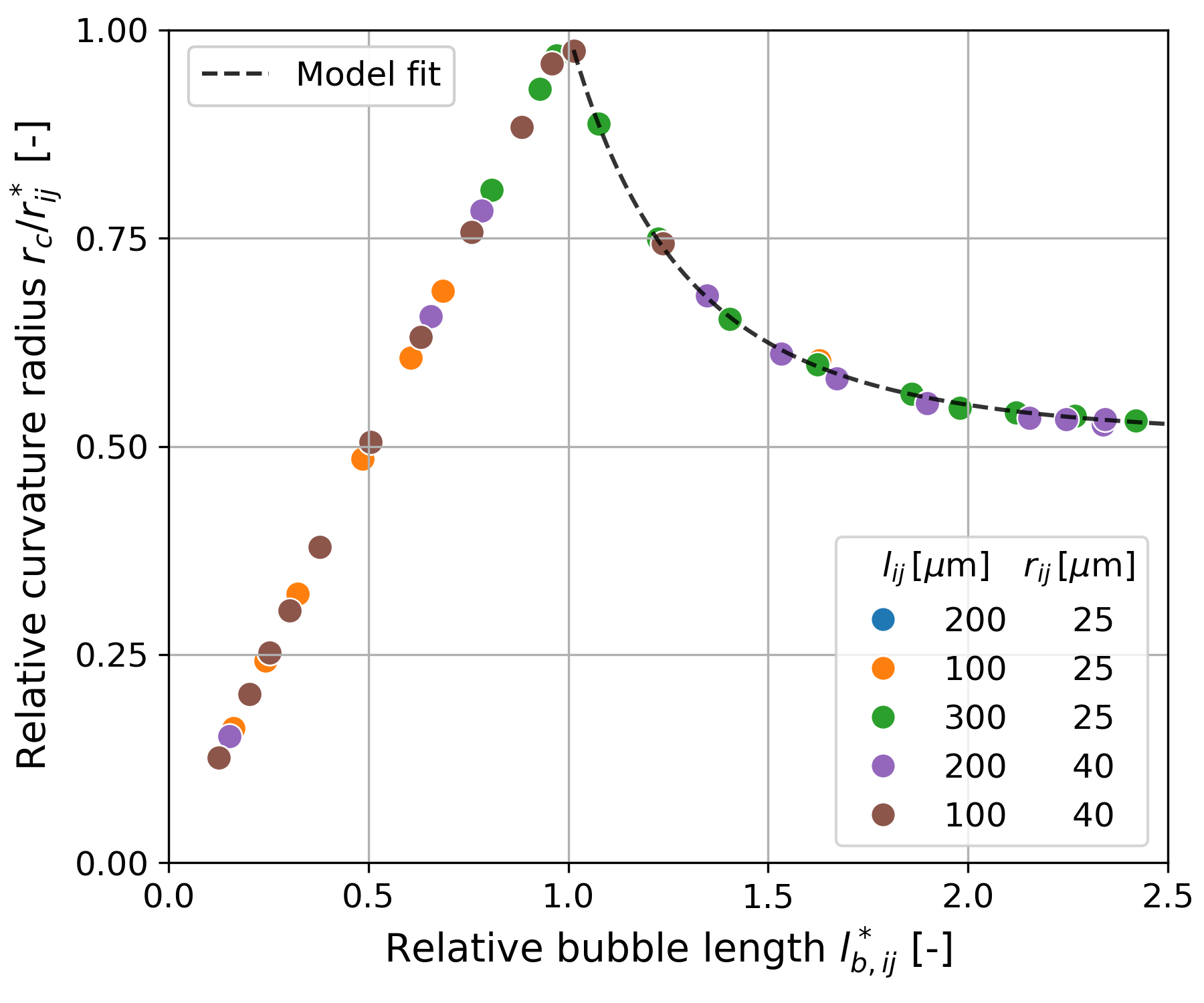}
        \caption{Bubble corner curvature radius fitted as a function of the relative bubble length, given by \autoref{eqn:curvature_radius}.}
        \label{fig:curvature-radius-fit}
    \end{minipage}
\end{figure}

We obtain the model fits using the least-squares method and plot them in \autoref{fig:bubble-length-fit} and \autoref{fig:curvature-radius-fit}.

\subsection{Stokes system and boundary conditions}
At low Reynolds numbers, incompressible flow is described by the Stokes equations,

\begin{subequations}
\begin{align}
    \nabla \cdot \sigma + \rho \,\boldsymbol a_{\mathrm{cf}}& = 0 \, ,\\
     \nabla \cdot \boldsymbol u & = 0 \, ,
\end{align}
\end{subequations}

with the Cauchy stress tensor \(\sigma = -pI + \mu \left(\nabla \boldsymbol u + (\nabla \boldsymbol u)^T\right)\).

\paragraph{Bubble boundary condition.}
Following the discussion in \autoref{sec:bubble-interface}, we choose to model the bubble surface with a no-slip condition.

\paragraph{Inlet and outlet boundary conditions.}
The top and bottom boundaries are assigned Neumann conditions for the momentum problem; the imposed boundary traction is

\begin{equation}
    \boldsymbol F_N = p_B \boldsymbol n - \mu(\nabla\boldsymbol u)^T\boldsymbol n,
\end{equation}

where \(p_B\) denotes the prescribed boundary pressure relative to the top boundary.

\paragraph{Wall boundary condition.}
The outer walls are fixed with a Dirichlet condition. In the lab frame, the velocity at the outer walls is zero, but in the bubble co-moving reference frame the velocity on the outer walls will be the negative of the rising velocity of the bubble.

\subsection{Calculating the bubble velocity}\label{app:bubble-velocity}
The only forces acting on the bubble are drag and buoyancy. When the bubble has reached its terminal velocity the sum of the forces will be zero. The terminal velocity can thus be obtained iteratively by solving the system and then adjusting the velocity at the walls for the next iteration until the sum of the forces is zero. The force exerted by the liquid on the bubble is evaluated as

\begin{equation}
    \boldsymbol T = -\int_{\partial B} \sigma \cdot \boldsymbol n\, dS = \int_{\partial B} \left[ pI - \mu \left(\nabla \boldsymbol u + (\nabla \boldsymbol u)^T \right) \right] \cdot \boldsymbol n\, dS .
\end{equation}

Since the gas phase inside the bubble is not explicitly solved, its weight must be included separately. The total force residual is therefore

\begin{equation}
    \boldsymbol F_{\text{tot}} = \boldsymbol T + \rho_n V_b \, \boldsymbol a_{\mathrm{cf}} .
\end{equation}

The terminal velocity is then obtained iteratively using a Newton method with finite-difference derivatives to converge the velocity on the walls until the vertical force residual is zero,

\begin{enumerate}
    \item The flow problem is solved and the vertical force residual \(F_z(u_z)\) is evaluated. The wall velocity corresponding to the terminal bubble velocity satisfies \(F_z=0\).

    \item Its derivative is approximated by a finite difference,
    \begin{equation}
        F_z'(u_z) \approx \frac{F_z(u_z+\Delta u_z)-F_z(u_z) }{\Delta u_z}.
    \end{equation}

    \item The wall velocity is then updated,
    \begin{equation}
        u_z^{(n+1)} = u_z^{(n)} - \frac{F_z(u_z^{(n)})}{F_z'(u_z^{(n)})}.
    \end{equation}

    \item The procedure is repeated until $|F_z| < 10^{-18} ~\mathrm{N}$.
\end{enumerate}

\subsection{Calculating the viscous pressure drop}\label{app:viscous-pressure}
In the special case where there is no potential difference between the top and bottom and the bubble rises exclusively due to buoyancy, the solver still gives a small flux due to the viscous pressure drop across the bubble. This pressure drop occurs due to the viscous friction the liquid experiences when passing the narrow space between the bubble and the walls of the throat. The hydrostatic pressure gradient is therefore not the correct boundary pressure value when solving the flow around the bubble in the bubble co-moving frame. This issue can be circumvented by converging the pressure \(p\) at the bottom boundary until there is zero wetting phase flux \(Q\) across the boundary. The general method remains the same, but the \(2\times 2\) Jacobian must now be constructed to converge both quantities simultaneously. 

\begin{enumerate}
    \item The flow problem is solved and the resulting force residual \(F_z(u_z,p)\) and flux \(Q(u_z,p)\) are computed. We seek \((u_z, p)\) such that \(|F_z(u_z,p)|<\epsilon\) and \(|Q(u_z,p)|<\epsilon\).

    \item We approximate the derivatives using finite differences. With the Jacobian
    \begin{equation}\label{eqn:jacobian}
        \mathbf{J} =
        \begin{bmatrix}
        \dfrac{\partial F_z}{\partial u_z} & \dfrac{\partial F_z}{\partial p} \\
        \dfrac{\partial Q}{\partial u_z} & \dfrac{\partial Q}{\partial p}
        \end{bmatrix},
    \end{equation}

    and entries approximated using finite differences.
    % and entries approximated as
    % \begin{align}
    %     \frac{\partial F_z}{\partial u_z} &\approx
    %     \frac{F_z(u_z + \Delta u_z, p) - F_z(u_z, p)}{\Delta u_z}, \\
    %     \frac{\partial F_z}{\partial p} &\approx
    %     \frac{F_z(u_z, p + \Delta p) - F_z(u_z, p)}{\Delta p}, \\
    %     \frac{\partial Q}{\partial u_z} &\approx
    %     \frac{Q(u_z + \Delta u_z, p) - Q(u_z, p)}{\Delta u_z}, \\
    %     \frac{\partial Q}{\partial p} &\approx
    %     \frac{Q(u_z, p + \Delta p) - Q(u_z, p)}{\Delta p}.
    % \end{align}

    \item The inlet pressure and wall velocity are then updated,
    \begin{equation}
        \begin{bmatrix}
        u_z^{(n+1)} \\
        p^{(n+1)}
        \end{bmatrix}
        =
        \begin{bmatrix}
        u_z^{(n)} \\
        p^{(n)}
        \end{bmatrix}
        -
        \mathbf{J}^{-1}
        \begin{bmatrix}
        F_z(u_z^{(n)}, p^{(n)}) \\
        Q(u_z^{(n)}, p^{(n)})
        \end{bmatrix}.
    \end{equation}

    \item The procedure is repeated until both the force and flux are sufficiently close to zero, i.e. $\|F_z(u_z,p)\| < 10^{-18}~\mathrm{N}$ and $\|Q(u_z,p)\| < 10^{-18}~\mathrm{m^3/s}$.
\end{enumerate}

\subsection{Mesh independence}\label{app:mesh}
\begin{figure}[pos=htb!]
    \centering
    \includegraphics[width=0.33\linewidth]{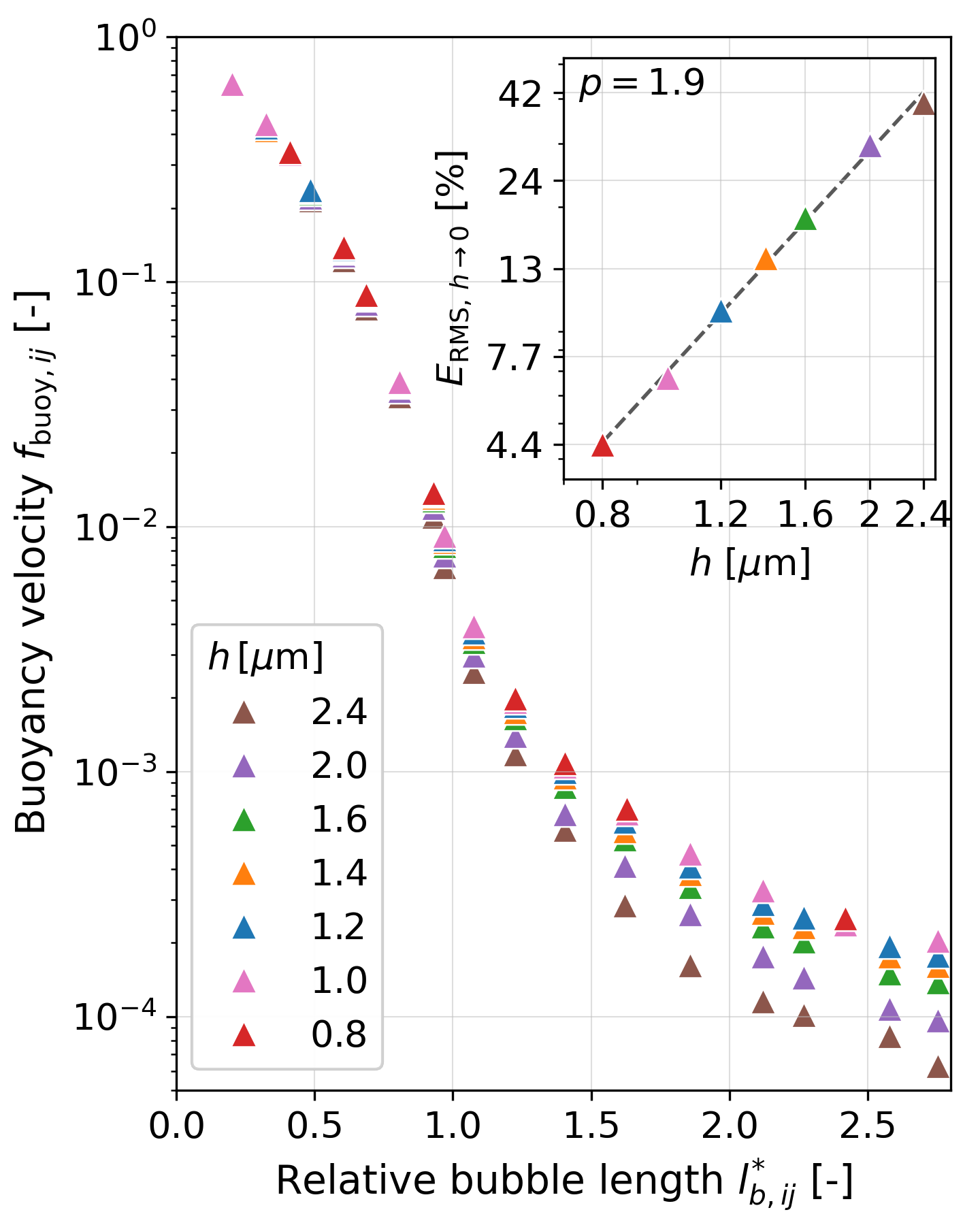}%
    \includegraphics[width=0.33\linewidth]{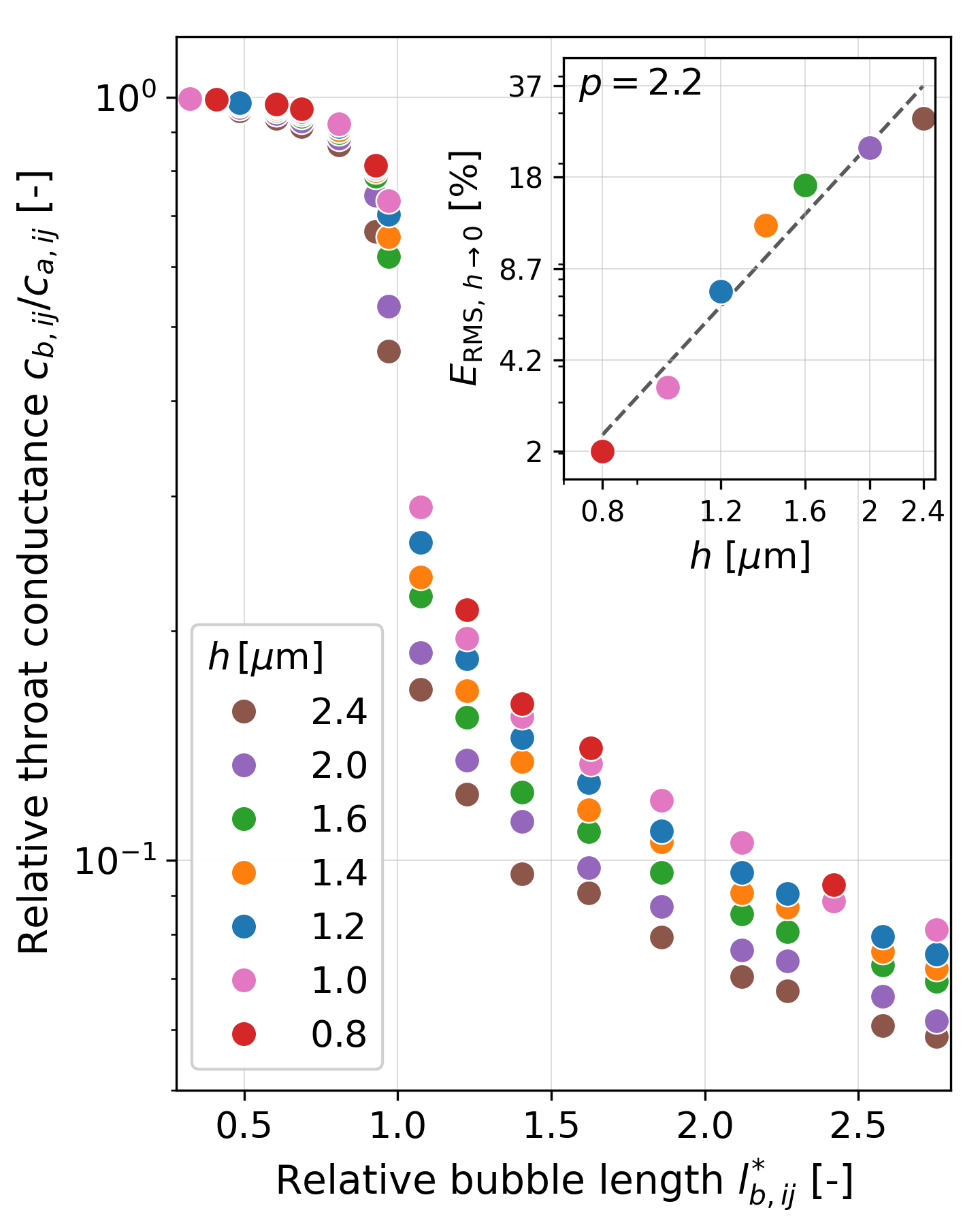}%
    \includegraphics[width=0.33\linewidth]{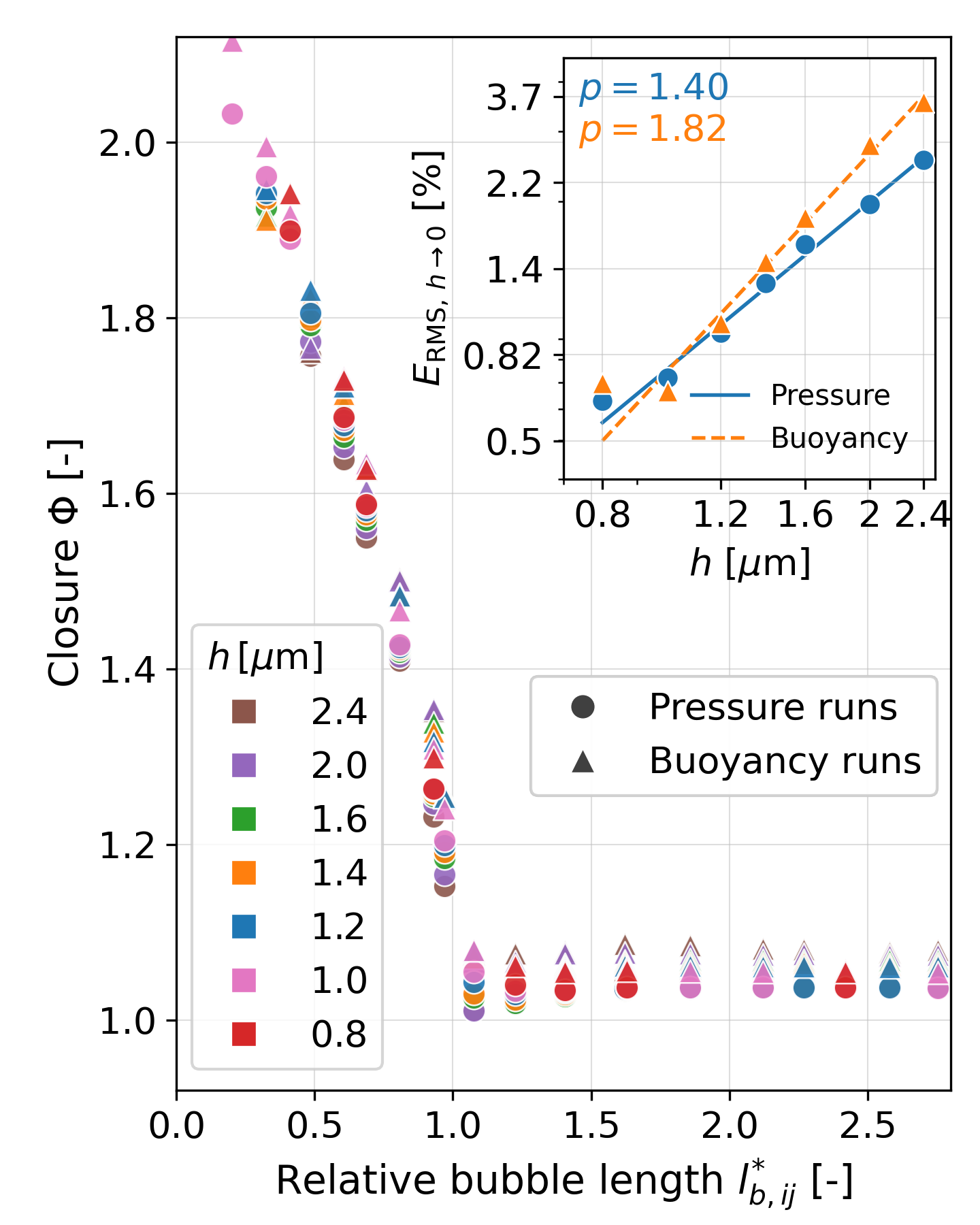}%
    \caption{Mesh-convergence study of the Stokes-flow simulations as the cell size $h$ is reduced. The main panels show the buoyancy velocity (left), relative throat conductance (middle), and closure $\phi$ (right). Each panel is inset with the root-mean-square relative errors with respect to the Richardson-extrapolated $h\rightarrow0$ limits. The throat geometry for all runs is \(r_{ij}=25~\mathrm{\mu m}\) and \(l_{ij}=200~\mathrm{\mu m}\).}
    \label{fig:mesh-convergence}
\end{figure}

We next investigate mesh independence of the results. A throat geometry of \(r_{ij}=25~\mathrm{\mu m}\) and \(l_{ij}=200~\mathrm{\mu m}\) is chosen, and five different values for the element base size \(h\) are selected. We generate a polyhedral mesh in CGAL \citep{cgal:eb-26a} with cell sizes $h=0.8-2.4~\mathrm{\mu m}$, where $h=0.8~\mathrm{\mu m}$ is the reference cell size used for results in \autoref{sec:Stokes_models}. \autoref{fig:mesh-convergence} shows that the root mean squared error of the buoyancy velocity $f_{\mathrm{buoy}}$, relative conductance $c_{b,ij}$, and closure $\phi$, decreases relative to the Richardson-extrapolated $h\rightarrow0$ limit for finer meshes. Finer meshes also yield a smaller error in $\phi$ relative to the analytical limit, with approximately $1\%$ for the finest mesh. 

\begin{figure}[pos=htb!]
    \centering
    \includegraphics[width=0.33\linewidth]{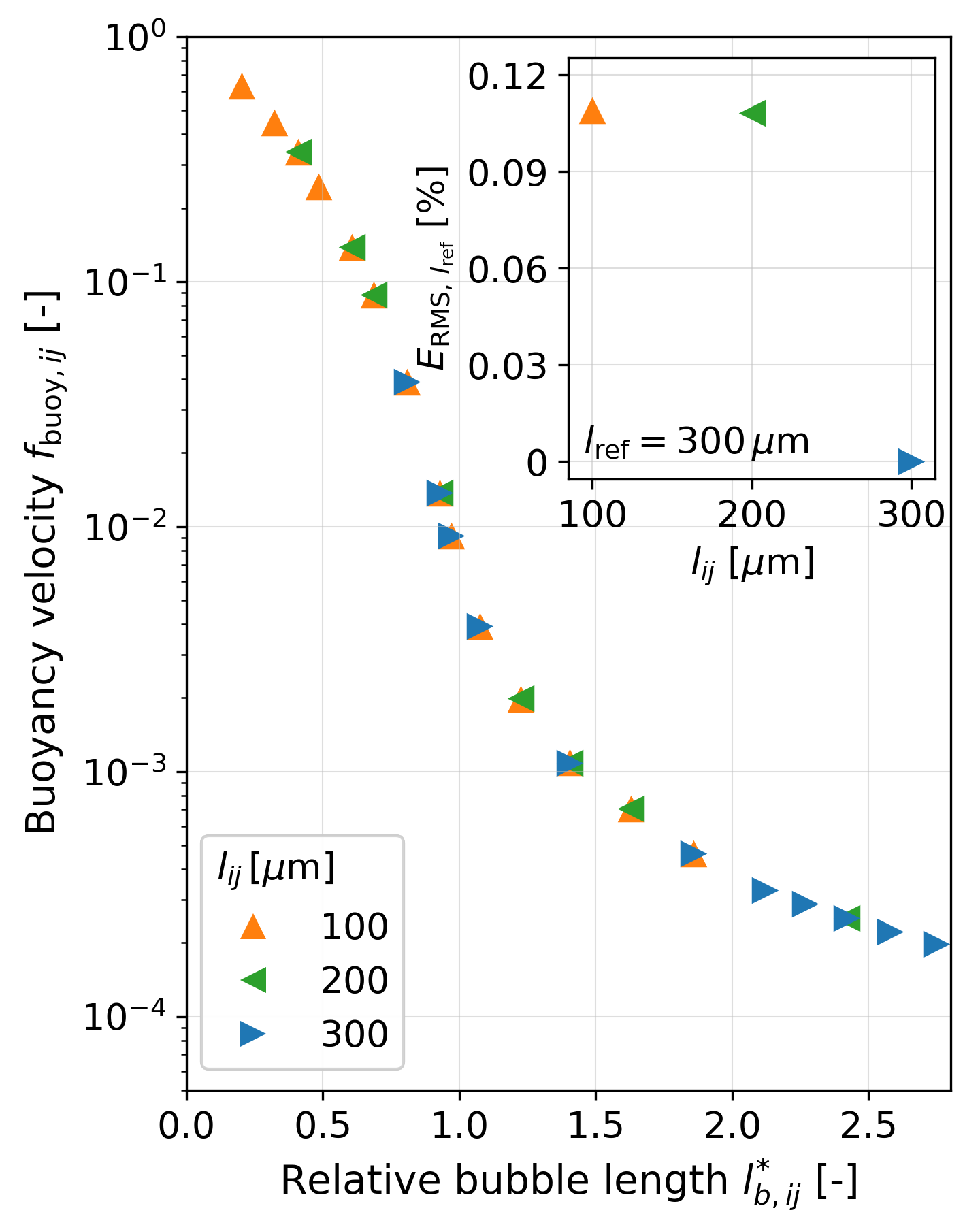}%
    \includegraphics[width=0.33\linewidth]{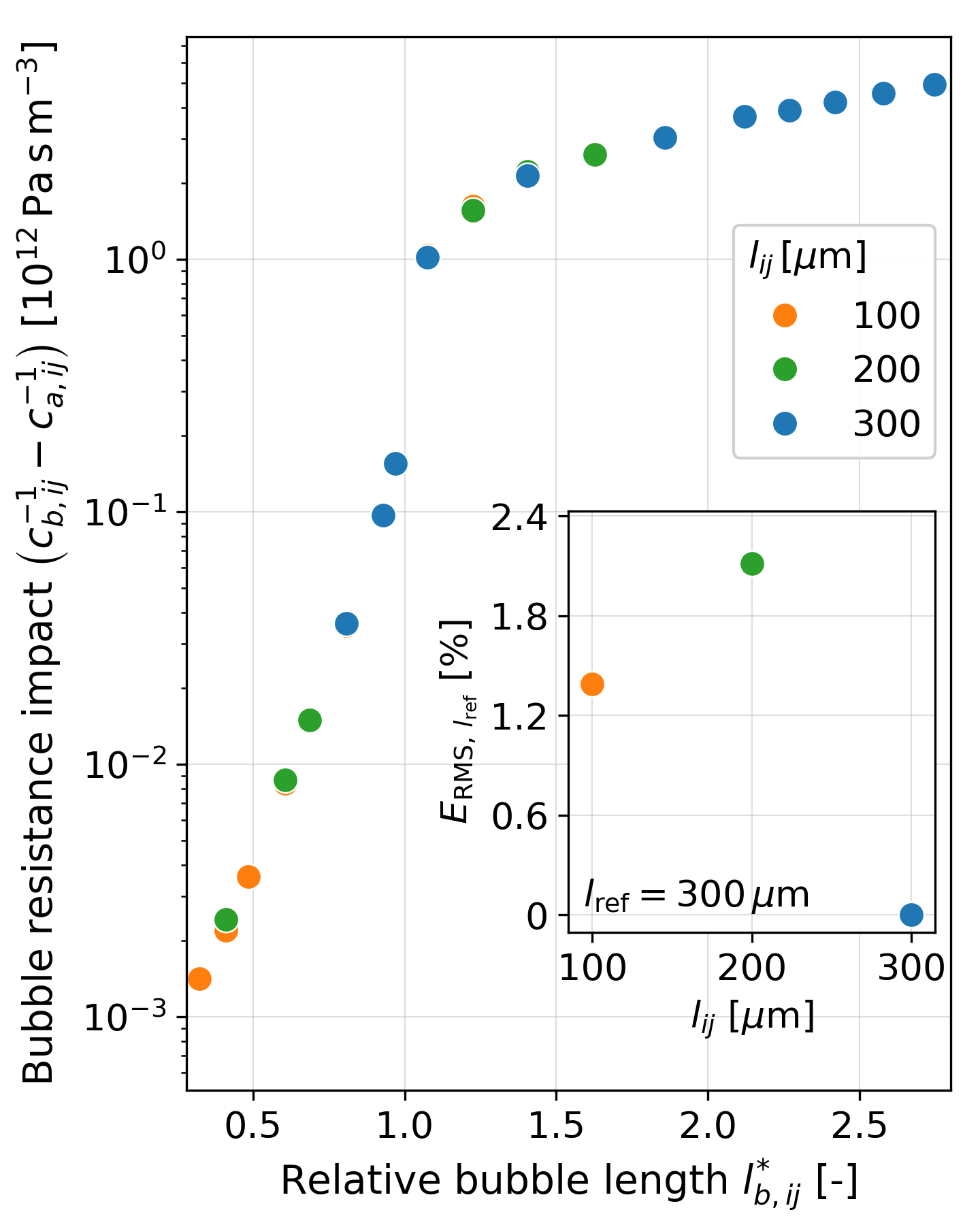}%
    \includegraphics[width=0.33\linewidth]{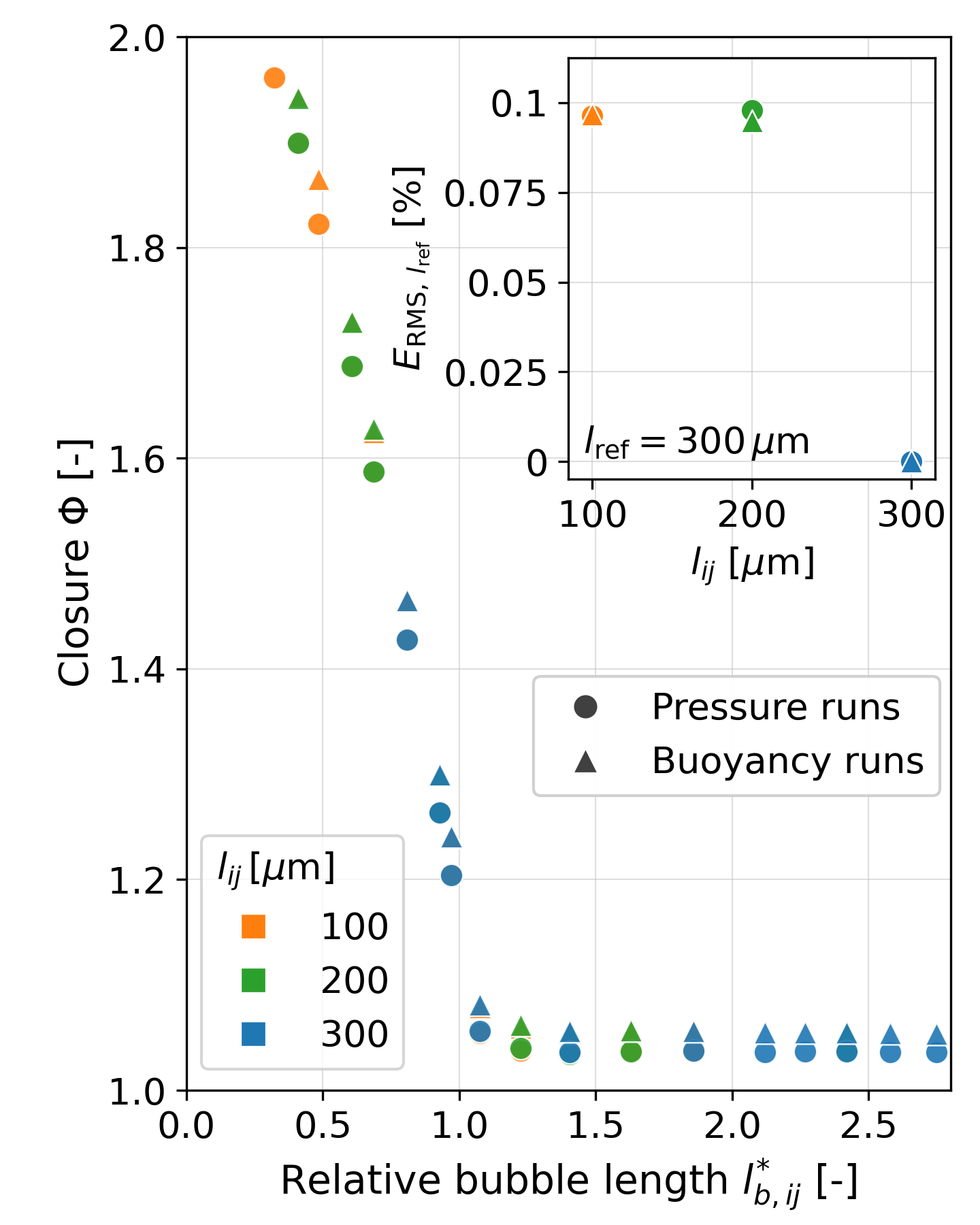}%
    \caption{Throat-length study of the Stokes-flow simulations as the throat length is increased for fixed bubble sizes. The main panels show the Buoyancy velocity (left), bubble resistance impact (middle), and closure $\phi$ (right); each inset shows the root-mean-square error relative to the longest throat, $l_{\mathrm{ref}}=300~\mathrm{\mu m}$. All runs use throat radius \(r_{ij}=25~\mathrm{\mu m}\) and mesh cell size \(h=0.8~\mathrm{\mu m}\).}
    \label{fig:length-convergence}
\end{figure}

Similarly, we want to confirm that the throat is sufficiently long for the flow to stabilise between the end of the bubble and the end of the throat. This is done by adjusting the length of the throat \(l_{ij}\in\{100,~200,~300\}~\mathrm{\mu m}\), while keeping the bubble sizes constant. The observed effect is that changing the throat length did not substantially affect the calculated quantities, as shown in \autoref{fig:length-convergence}.

\FloatBarrier
\section{Corner flow factor $k$}\label{app:corner-k}
\autoref{eqn:phi_inf} contains a geometric factor, $k$, the ratio between the mean liquid velocity in the corners and the bubble velocity when corner flow dominates, fixed by the corner cross-section geometry alone.

Far from the bubble cap, the flow is fully developed, so the axial velocity obeys the two-dimensional Stokes balance $\mu\nabla_\perp^2 u = \partial p/\partial x$, which, by linearity, splits into a pressure-driven part (the conductance $\beta$) and a part driven purely by the bubble's own motion at zero pressure gradient. In the bubble frame, the duct wall translates at $-u_b$ while the no-slip bubble surface is stationary, so that part is a Laplace problem for the drag velocity $W$ on the liquid annulus $A_w$ (corner fillets plus wall films), $\nabla^2 W = 0\,, W = 1$ on the duct wall, and $W = 0$ on the bubble surface.

\begin{figure}[pos=ht!]
    \centering
    \begin{tikzpicture}[scale=4.4, every node/.style={font=\footnotesize}]
        \def\xc{0.90} % bubble half-width / r_{ij} (true value 1-d = 0.99; d widened here to stay legible)
        \def\rc{0.499} % corner radius / r_{ij}, the long-bubble asymptote -- drawn to scale
        \pgfmathsetmacro{\cx}{\xc-\rc}
        \pgfmathsetmacro{\cy}{\xc-\rc}

        % liquid annulus A_w, shaded
        \fill[blue!12] (0,\xc) -- (\cx,\xc) arc (90:0:\rc) -- (\xc,0)
                       -- (1,0) -- (1,1) -- (0,1) -- cycle;
        % bubble A_b, shaded
        \fill[gray!22] (0,0) -- (\xc,0) -- (\xc,\cy) arc (0:90:\rc) -- (0,\xc) -- cycle;

        % duct wall (Dirichlet V=0)
        \draw[very thick] (0,1) -- (1,1) -- (1,0);
        % symmetry planes through the duct centreline (zero flux)
        \draw[dashed,gray] (0,0) -- (0,1);
        \draw[dashed,gray] (0,0) -- (1,0);
        % bubble / meniscus surface (Dirichlet V=1)
        \draw[thick,blue] (0,\xc) -- (\cx,\xc) arc (90:0:\rc) -- (\xc,0);

        % film thickness d, measured across each flat film strip (not to scale, see caption)
        \draw[<->] (0.10,\xc) -- (0.10,1);
        \draw[<->] (\xc,0.10) -- (1,0.10);

        % corner radius r_a
        \fill (\cx,\cy) circle (0.4pt);
        \draw[gray!60!black] (\cx,\cy) -- ({\cx+\rc*cos(50)},{\cy+\rc*sin(50)});

        % boundary-condition leaders, stacked in the right margin: V=0 to the wall, V=1
        % just below it to the meniscus, both short and pointing straight left
        \draw[thin,-{Latex[length=1.6mm]}] (1.16,0.65) -- (1.01,0.65);
        \draw[blue!60!black,thin,-{Latex[length=1.6mm]}] (1.16,0.53) -- (0.885,0.53);

        % --- every text label is placed last, so no line is ever drawn on top of one -------
        \node at (0.25,0.25) {$A_b$};
        \node[blue!60!black] at (0.75,0.80) {$A_w$};
        \node at (0.17,0.95) {$d$};
        \node at (0.95,0.17) {$d$};
        \node[gray!60!black] at ({\cx+0.5*\rc*cos(58)},{\cy+0.5*\rc*sin(58)+0.03}) {$r_a$};
        \node[right] at (1.16,0.65) {$V=0$};
        \node[blue!60!black,right] at (1.16,0.53) {$V=1$};
        \node[gray!70!black,rotate=90] at (0.05,0.5) {symmetry};
        \node[gray!70!black] at (0.5,0.05) {symmetry};
    \end{tikzpicture}
    \caption{One quadrant of the corner cross-section (four-fold symmetry of the square duct), used to solve for $k$. Shaded: bubble $A_b$ ($V=1$, blue boundary) and liquid annulus $A_w$ (light blue). Dashed lines are the zero-flux symmetry planes through the duct centreline; the solid line is the no-slip duct wall ($V=0$). The corner radius $r_a$ is drawn to scale (long-bubble asymptote $r_a/r_{ij}=0.499$); the film thickness $d$ is widened from its true value $0.01\,r_{ij}$ to stay legible.}
    \label{fig:corner-laplace}
\end{figure}
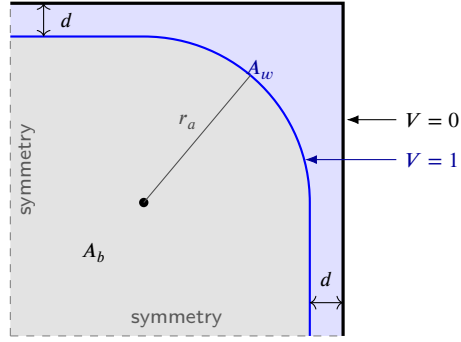

Back in the lab frame, the corner-mean velocity this drag produces is $u_b(1-\langle W\rangle_{A_w})$, so $k = 1 - \langle W\rangle_{A_w}$. Since $1-W$ solves the same Laplace equation with the boundary data swapped, writing $V \equiv 1-W$ turns this into $\nabla^2 V = 0\,, V = 1,$ on the bubble surface, and $V = 0$ on the duct wall. Then, $k$ can be computed as

\begin{equation}
    k = \frac{1}{A_w}\int_{A_w} V ~\mathrm{d}A \,,
\end{equation}

a plain area average: $k$ is fixed purely by the film thickness $d$ and the meniscus radius $r_a$ relative to $r_{ij}$. \autoref{fig:corner-laplace} shows the quadrant domain on which we solve, exploiting the duct's four-fold symmetry.

With $d=0.01\,r_{ij}$ and $r_a/r_{ij}\rightarrow0.499$ (the long-bubble asymptote), solving numerically on a quadrant gives $k=0.428$ (converged to $0.427$--$0.429$ over a $3$--$12$ cell refinement across the film). Because this uses none of the Stokes velocity or pressure data, $\phi_\infty=A_{ij}/(A_b+kA_w)$ is a purely geometric quantity, independent of both datasets.

\FloatBarrier

\section{Flow resistance \(\beta\)}\label{app:beta}
The wetting phase throat conductance from \autoref{eqn:wetting-phase-conductance} depends on a dimensionless flow resistance \(\beta\) derived by \citet{zhouHydrocarbonDrainageCorners1997}. Assuming sharp throat corners and a no-slip condition between the wetting and solid phase, it takes the form

\begin{equation}
    % FULL EQUATION, assuming sharp corners.
    \beta = \frac{12\sin^2(\alpha_\Lambda)(1-B)^2 \left(\psi_1-B\psi_2\right) \left(\psi_3+Bf_2\psi_2\right)^2}{
            (1-\sin\alpha_\Lambda)^2 B^2 \left(\psi_1-B\psi_2\right)^3}
    % SIMPLIFIED EQUATION, when f_2=0 (no stress boundary) and cancelling one \left(\psi_1-B\psi_2\right)
    % \beta = \frac{12\sin^2(\alpha_\Lambda)(1-B)^2 \left(\psi_3\right)^2}{
    % (1-\sin\alpha_\Lambda)^2 B^2 \left(\psi_1-B\psi_2\right)^2} \, ,
\end{equation}

where \(\alpha_\Lambda\) is the half-angle of corner \(\Lambda\) and \(\theta_\Lambda\) is the contact angle in the same corner. The function \(f_2\) represents the boundary condition between the wetting and non-wetting phases, such that \(f_2=1\) for a no-slip and \(f_2=0\) for a free-slip gas-liquid interface. An intermediate boundary can be applied by setting \(0<f_2<1\). The functions \(B\), \(\psi_1\), \(\psi_2\), and \(\psi_3\) are:

\begin{subequations}
\begin{align}
    B &= \left(\frac{\pi}{2}-\alpha_\Lambda\right)\tan(\alpha_\Lambda) \, ,\\
    \psi_1 &= \cos^2(\alpha_\Lambda + \theta_\Lambda)+\cos(\alpha_\Lambda + \theta_\Lambda)\sin(\alpha_\Lambda + \theta_\Lambda)\tan(\alpha_\Lambda) \, ,\\
    \psi_2 &= 1-\frac{\theta_\Lambda}{\pi/2 - \alpha_\Lambda} \, ,\\
    \psi_3 &= \frac{\cos(\alpha_\Lambda+\theta_\Lambda)}{\cos(\alpha_\Lambda)} \; .
\end{align}
\end{subequations}

\clearpage
\includepdf[pages=1-11]{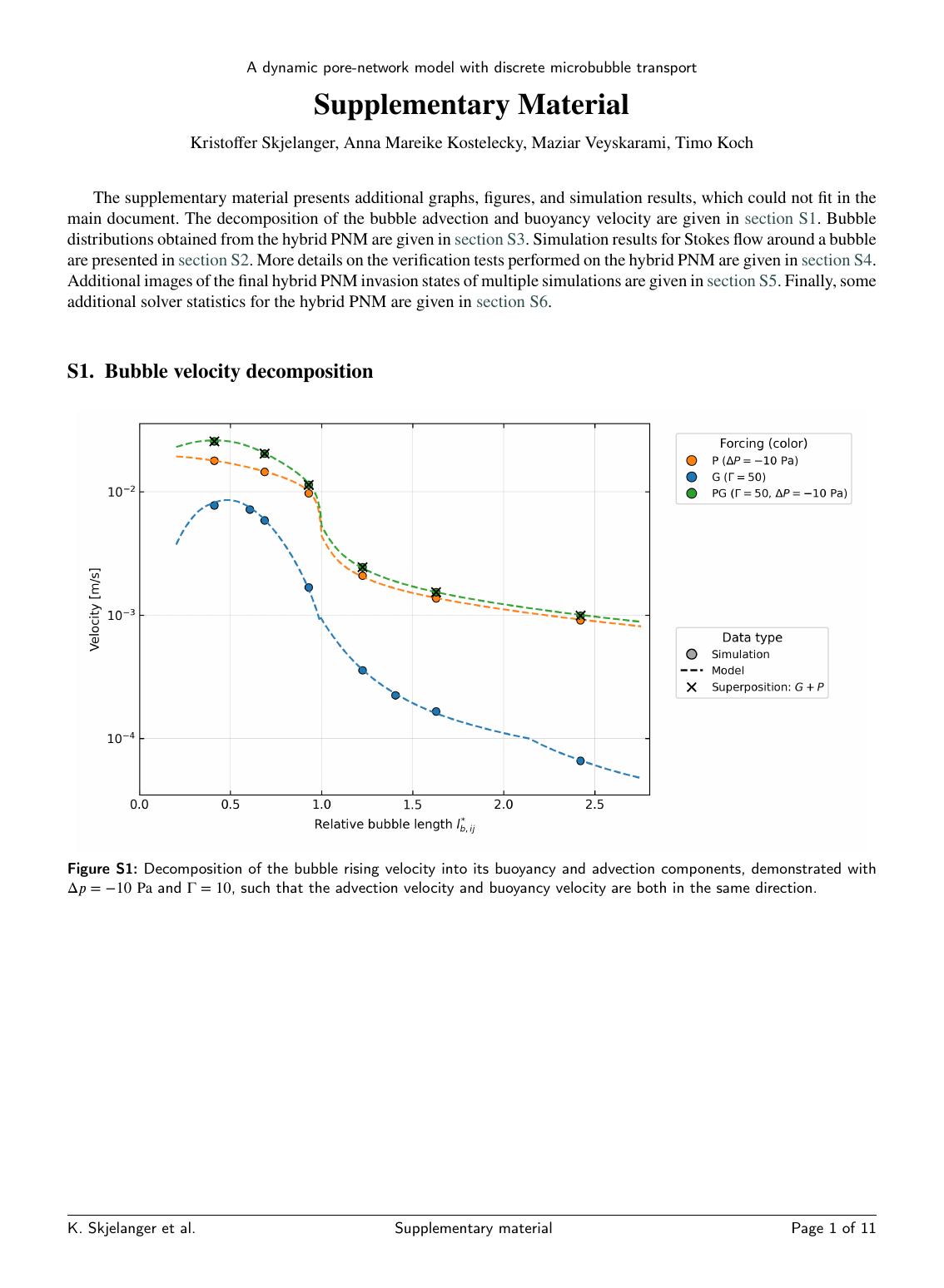}
\addtocounter{page}{-12}

\end{document}